%% file: Final_JHEP.tex
\documentclass[12pt,a4paper]{article}
\usepackage{tikz,lipsum,lmodern}
\usepackage[most]{tcolorbox}

\input{setting.tex}
\def \Mpl {M_{\text{Pl}}}

\def \i{\mathrm{i}}
\newcommand{\m}[1]{\mathrm{#1}}
 
\begin{document}

\begin{titlepage}
\null
\begin{flushright}
CTPU-PTC-23-47, UT-Komaba/23-13
\end{flushright}

\vskip 1cm
\begin{center}
\baselineskip 0.8cm
{\LARGE \bf
Analytic Formulae for Inflationary Correlators with Dynamical Mass
}

\lineskip .75em
\vskip 1cm

\normalsize

{\large } {\large Shuntaro Aoki} $^{1}${\def\thefootnote{\fnsymbol{footnote}}\footnote[1]{E-mail address: shuntaro1230@gmail.com}},
{\large Toshifumi Noumi} $^{2}${\def\thefootnote{\fnsymbol{footnote}}\footnote[2]{E-mail address: tnoumi@g.ecc.u-tokyo.ac.jp}},
{\large Fumiya Sano} $^{3,4}${\def\thefootnote{\fnsymbol{footnote}}\footnote[3]{E-mail address: sanof.cosmo@gmail.com}},

{\large Masahide Yamaguchi} $^{4,3}${\def\thefootnote{\fnsymbol{footnote}}\footnote[4]{E-mail address: gucci@ibs.re.kr}}

\vskip 1.0em

$^1${\small\it Particle Theory and Cosmology Group, Center for Theoretical Physics of the Universe,\\
Institute for Basic Science, Daejeon, 34126, Korea}
\vskip 0.5em

$^2${\small{\it Graduate School of Arts and Sciences, The University of Tokyo, Tokyo 153-8902, Japan}}
\vskip 0.5em

$^3${\small{\it Department of Physics, Tokyo Institute of Technology, Tokyo 152-8551, Japan}}
\vskip 0.5em

$^4${\small{\it Cosmology, Gravity and Astroparticle Physics Group, Center for Theoretical Physics of the Universe, Institute for Basic Science, Daejeon 34126, Korea}}
\vskip 1.0em

{\bf Abstract}\\[5mm]
{\parbox{15cm}{\hspace{5mm} \small
Massive fields can imprint unique oscillatory features on primordial correlation functions or inflationary correlators, which is dubbed the cosmological collider signal. In this work, we analytically investigate the effects of a \textit{time-dependent} mass of a scalar field on inflationary correlators, extending previous numerical studies and implementing techniques developed in the cosmological bootstrap program. The time-dependent mass is in general induced by couplings to the slow-roll inflaton background, with particularly significant effects in the case of non-derivative couplings. By linearly approximating the time dependence, the mode function of the massive scalar is computed analytically, on which we derive analytic formulae for two-, three-, and four-point correlators with the tree-level exchange of the massive scalar. The obtained formulae are utilized to discuss the phenomenological impacts on the power spectrum and bispectrum, and it is found that the scaling behavior of the bispectrum in the squeezed configuration, i.e., the cosmological collider signal, is modified from a time-dependent Boltzmann suppression. By investigating the scaling behavior in detail, we are in principle able to determine the non-derivative couplings between the inflaton and the massive particle.

}}

\end{center}

\end{titlepage}

\tableofcontents
\vspace{35pt}
\hrule

\begin{fmffile}{diagram}

\section{Introduction}\label{intro}

Observations of the Cosmic Microwave Background~\cite{Boomerang:2000jdg,WMAP:2012fli,Planck:2018jri} strongly support the existence of cosmic inflation~\cite{Starobinsky:1980te,Sato:1980yn,Guth:1980zm,Linde:1981mu,Albrecht:1982wi}  which can give solutions to several problems in the standard cosmology as well as is the source of primordial scalar (curvature) and tensor perturbations. Furthermore, given the high energy scale {($\rho^{1/4}_{\text{inf}}\sim 10^{15}$ GeV at most)}, inflation is a unique opportunity to explore high energy physics including models beyond the Standard Model.

In recent year, there has been a growing interest in an approach that exploit higher-point correlation functions, or non-Gaussianity, of scalar and tensor perturbations, called cosmological collider (CC) program (see earlier works~\cite{Chen:2009zp,Baumann:2011nk,Noumi:2012vr,Arkani-Hamed:2015bza} and recent developments~\cite{Chen:2009we,Assassi:2012zq,Sefusatti:2012ye,Norena:2012yi,Chen:2012ge,Pi:2012gf,Cespedes:2013rda,Gong:2013sma,Emami:2013lma,Kehagias:2015jha,Liu:2015tza,Dimastrogiovanni:2015pla,Schmidt:2015xka,Chen:2015lza,Delacretaz:2015edn,Bonga:2015urq,Chen:2016nrs,Flauger:2016idt,Lee:2016vti,Delacretaz:2016nhw,Meerburg:2016zdz,Chen:2016uwp,Meerburg:2016zdz,Chen:2016hrz,Kehagias:2017cym,An:2017hlx,Tong:2017iat,Iyer:2017qzw,An:2017rwo,Kumar:2017ecc,RiquelmeM:2017qhp,Franciolini:2017ktv,RiquelmeM:2017qhp,Tong:2017iat,Tong:2018tqf,Chen:2018sce,Saito:2018omt,Cabass:2018roz,Wang:2018tbf,Chen:2018xck,Bartolo:2018hjc,Dimastrogiovanni:2018uqy,Bordin:2018pca,Chen:2018cgg,Achucarro:2018ngj,Chua:2018dqh,Kumar:2018jxz,Goon:2018fyu,Wu:2018lmx,Anninos:2019nib,Li:2019ves,McAneny:2019epy,Kim:2019wjo,Alexander:2019vtb,Lu:2019tjj,Hook:2019zxa,Hook:2019vcn,Kumar:2019ebj,Liu:2019fag,Wang:2019gbi,Wang:2019gok,Wang:2019gok,Wang:2020uic,Li:2020xwr,Wang:2020ioa,Fan:2020xgh,Bodas:2020yho, Aoki:2020zbj,Maru:2021ezc,Kim:2021pbr,Lu:2021gso,Sou:2021juh, Lu:2021wxu,Wang:2021qez,Pinol:2021aun,Cui:2021iie,Tong:2022cdz,Reece:2022soh,Chen:2022vzh,Qin:2022lva,Cabass:2022rhr,Cabass:2022oap,Niu:2022quw,Niu:2022fki,Aoki:2023tjm,Werth:2023pfl,Tong:2023krn,Jazayeri:2023xcj,Yin:2023jlv,Stefanyszyn:2023qov,Chakraborty:2023qbp,Pinol:2023oux}).  Indeed, these correlation functions can contain information of (possibly new) massive particles created during inflation, and especially in the soft limits, we expect a sharp oscillatory behavior (what we call ``signal'') characterized by the mass of particles around the Hubble scale.

A number of recent developments have been seen in the computational methods of primordial
correlators. In particular, the so-called cosmological bootstrap method~\cite{Arkani-Hamed:2018kmz, Sleight:2019mgd, Sleight:2019hfp, Baumann:2019oyu, Baumann:2020dch, Pajer:2020wnj, Sleight:2020obc, Goodhew:2020hob, Pajer:2020wxk, Jazayeri:2021fvk, Melville:2021lst, Goodhew:2021oqg, Sleight:2021iix, Gomez:2021qfd, Bonifacio:2021azc, Meltzer:2021zin, Hogervorst:2021uvp, DiPietro:2021sjt, Sleight:2021plv, Cabass:2021fnw, Tong:2021wai, Baumann:2021fxj, Gomez:2021ujt, Baumann:2022jpr, Heckelbacher:2022hbq, Pimentel:2022fsc, Jazayeri:2022kjy, Qin:2022fbv, Xianyu:2022jwk, Wang:2022eop, Chen:2023txq, Qin:2023ejc, Qin:2023bjk, Qin:2023nhv, Xianyu:2023ytd, Green:2023ids, DuasoPueyo:2023viy,De:2023xue} (see also  AdS techniques~\cite{Albayrak:2018tam, Albayrak:2019asr, Albayrak:2019yve, Albayrak:2020bso, Albayrak:2020fyp, Albayrak:2023jzl}) has made it possible to compute the correlation functions rigorously and analytically without having to perform the awkward time integrals of special functions in the cosmological in-in calculations (see Ref.~\cite{Wang:2013zva,Chen:2017ryl} for the detail). This allows us to evaluate not only the signal parts of CC but also the non-oscillatory parts (background). This is quite important to understand how large the net signal is.

So far, most of the studies have focused on situations where the masses of massive fields are constant, and consequently, the correlation function has a scale-invariant form. However, interactions with inflaton can produce a non-negligible time dependence on the masses of these massive fields, resulting in a scale-dependent correlation function. For example in Ref.~\cite{Reece:2022soh}, it is numerically shown that a significant deviation from the standard CC signal {in bispectrum} can be obtained {in case of the non-derivative coupling $e^{\alpha\phi/\Mpl}\sigma^2$, where $\phi$ is an inflaton and $\sigma$ is an isocurvaton.}

In this paper, we derive \textit{analytic} formulae for two-, three-, and four-point inflaton correlation functions or ``inflationary correlators''. We focus on a simple model consisting of an inflaton and a massive scalar field, where the inflaton imparts a time dependence on the massive scalar through interactions. We then solve the so-called ``bootstrap equations'' for the correlators~\cite{Arkani-Hamed:2018kmz} by approximating the time-dependence of the mass at the linear order of time. In the constant mass limit, our results consistently reproduce those obtained in Refs.~\cite{Qin:2022fbv,Qin:2023ejc}. The analytic formulae allow us to take into account arbitrary momentum configurations and background parts, so that we can extract various phenomenological aspects of non-Gaussianity in a more precise manner. Specifically, the bispectrum (three-point correlation function of the curvature perturbation) in the squeezed limit contains information on the time dependence of the mass of $\sigma$, which is useful to distinguish couplings between inflaton and the massive field.

The paper is organized as follows. In section~\ref{setup}, we introduce the above setup and solve the mode equations for a massive scalar field with time-dependent mass. We also introduce integral formulae for the two-, three-, and four-point correlators of interest. We then derive and solve the bootstrap equations satisfied by the correlators in section~\ref{sec:boot_eq}. In section~\ref{PNG}, we discuss the observational impact by taking a particular interaction between the inflaton and the massive scalar. Section~\ref{summary} is devoted to the summary. Appendices~\ref{MB} and~\ref{CML} provide boundary conditions necessary to solve the bootstrap equations, and a consistency check in the constant mass limit. 

\subsection*{Notations}
The spacetime metric is Friedman--Lema{\^i}tre--Robertson--Walker (FLRW) metric: $\mathrm{d}s^2=-\mathrm{d}t^2 +a^2(t)\mathrm{d}\mathbf{x}^2=a^2(\tau)(-\mathrm{d}\tau^2+\mathrm{d}\mathbf{x}^2)$, where $t$ and $\tau$ are physical time and conformal time respectively. In (quasi) de Sitter space, a scale factor is given by $a(\tau)=-1/H\tau$ with the Hubble parameter $H$. Regarding the derivative with respect to physical time (conformal time), we use a dot (prime) for operation, i.e.,  $\cdot=\mathrm{d}/\mathrm{d}t$ and $\prime=\mathrm{d}/\mathrm{d}\tau$.

\section{Setup}\label{setup}
In this work, we consider a simple system consisting of only an inflaton $\phi$ and a massive scalar $\sigma$,
and assume that their interactions include the following non-derivative coupling:\footnote{In this work, we do not consider back-reaction effects on the inflation dynamics due to Eq.~\eqref{e_mass}, to keep our setup as simple as possible. But we only remark that these kinds of terms breaking a shift symmetry of inflaton affects the inflation dynamics in general at the classical and quantum level, which requires some non-trivial extension of the setup. For example, to relax the quantum effects, one may consider supersymmetric embedding~\cite{Baumann:2014nda} or a system with a discrete symmetry~\cite{Deshpande:2020lmf,Lee:2022fkd,Lee:2023dcy}.} 
\begin{align}
\mc{L}_{\text{int}}\supset -\frac{1}{2}g(\phi)\sigma^2.    \label{e_mass}
\end{align}
During inflation, the interaction~\eqref{e_mass} gives an effective mass of $\sigma$, $m_{\rm{eff}}^2=g(\phi_0(t))$ with $\phi_0(t)$ being the inflaton background, and thus the mass of $\sigma$ becomes time-dependent. We will first investigate the effects 
of time dependence with Eq.~\eqref{e_mass} on the mode function of $\sigma$. Then, based on the mode function derived there, we study its impact on the inflationary correlation functions in the next sections.

\subsection{Mode functions and propagators with time-dependent mass}

\subsubsection*{Canonical quantization}
We quantize the inflaton fluctuation $\delta\phi$ and the massive scalar $\sigma$ with the mode expansion
\begin{align}
  \delta\phi(\tau,\mathbf{x})&= \int \frac{\mathrm{d}^3 \mathbf{k}}{(2 \pi)^3}\left(u_k(\tau) a_{\mathbf{k}}+u_k^*(\tau) a_{-\mathbf{k}}^{\dagger}\right)e^{\i{\mathbf{k} \cdot \mathbf{x}}}, \label{Q_phi} \\
  \sigma(\tau,\mathbf{x})&=\int \frac{\mathrm{d}^3 \mathbf{k}}{(2 \pi)^3}\left(v_k(\tau) b_{\mathbf{k}}+v_k^{ *}(\tau) b_{-\mathbf{k}}^{ \dagger} \right)e^{\i{\mathbf{k} \cdot \mathbf{x}}}, \label{Q_s} 
\end{align}
and the commutation relations for the annihilation and the creation operators
\begin{align}
[a_{\mathbf{k}},a^{\dagger}_{\mathbf{k}'}]=(2\pi)^3\delta^3({\mathbf{k}}-{\mathbf{k}'}), \quad [b_{\mathbf{k}},b^{\dagger}_{\mathbf{k}'}]=(2\pi)^3\delta^3({\mathbf{k}}-{\mathbf{k}'}).
\end{align}
Note that $\mathbf{k}$ expresses three-dimensional vectors, and $k \equiv\left|\mathbf{k}\right|$ is the absolute value. Under the slow-roll approximation, the mode function of inflaton fluctuations, $u_k$, satisfies an equation of motion for a massless scalar field in de Sitter spacetime,
\begin{align}
u_k^{\prime \prime}-\frac{2}{\tau} u_k^{\prime}+k^2u_k=0. \label{eom_u}    
\end{align}
Assuming the Bunch–Davies vacuum, we obtain the canonically normalized mode function
\begin{align}
u_k=\frac{H}{\sqrt{2 k^3}}(1+\i k \tau) e^{-\i k \tau}. \label{sol_u}
\end{align}
On the other hand, the mode function of the massive field, $v_k$, satisfies the following equation:
\begin{align}
v_k^{\prime \prime}-\frac{2}{\tau}v_k^{\prime}+\left(k^2+\frac{m^2_{\rm{eff}}}{H^2 \tau^2}\right) v_k=0.    \label{eom_v}
\end{align}
Here $ m^2_{\rm{eff}}\equiv g(\phi_0(t))$ is an effective mass of $\sigma$, which in general depends on time.

\subsubsection*{Approximation of effective mass for analytic computations}

In the forthcoming analysis, we impose several assumptions on the effective mass $m_{\rm eff}$ and employ approximations to perform analytic computations. To explain them, we begin by recalling that, within the framework of the slow-roll approximation, the inflaton background is described by
\begin{align}
\phi_{0}(\tau)=\sqrt{2\epsilon}\Mpl \log \left(\frac{\tau}{\tau_0}\right),   \end{align}
where we used $\phi_0^{\prime}= \sqrt{2\epsilon}\Mpl/\tau $ (assuming $\dot{\phi}_0<0$ without loss of generality) and introduced a reference time $\tau_0$ as an integration constant. In this context, the value of the background inflaton field $\phi_{0*}$ at the horizon crossing time $\tau_*$ for a mode $k$ is given by
\begin{align}
\phi_{0*}=\sqrt{2\epsilon}\Mpl \log \left(\frac{\tau_*}{\tau_0}\right)=-\sqrt{2\epsilon}\Mpl \log v(k),
\quad
v(k)\equiv \frac{k}{k_0},
\label{phi_s} 
\end{align}
where we used the condition $k\tau_*=-1$ and also introduced a reference scale $k_0$ such that $k_0\tau_0=-1$. The effective mass at the horizon crossing of the mode $k$ reads
\begin{align}
m_{\rm eff}^2(\tau_*)=g(\phi_{0*})\equiv g_*(v).
\end{align}
We are interested in the situation where the variation of the effective mass is not negligible, i.e., $|\Delta m_{\rm eff}^2|\gtrsim m_{\rm eff}^2$ during observable inflation {with e-folding number roughly $50\lesssim N\lesssim 60$ for CMB,} where $\Delta m_{\rm eff}^2$ represents the change in the square of the effective mass from the beginning to the end of observable inflation. It is worth emphasizing that, under this condition, the time dependence of the effective mass produces significant effects that dominate over other slow-roll suppressed effects such as the inflaton mass. Moreover, we mainly focus on the situation that the effective mass remains at the Hubble scale throughout inflation, $m_{\rm eff}\sim H$, {and undergoes several times larger or smaller changes relative to the initial mass}. This assumption is made to avoid exponential suppression of the CC signal $\sim e^{-\mathcal{O}(m_{\rm eff}/H)}$.

When we solve the equation of motion~\eqref{eom_v} for the massive field $\sigma$, it is also necessary to consider the time evolution of the effective mass around the time of the horizon crossing. In order to estimate the effect, let us perform an expansion of the inflaton background $\phi_0(\tau)$ around the horizon crossing time $\tau_*$ as
\begin{align}
 \phi_0(\tau)=\phi_{0*} - \sqrt{2 \epsilon} \Mpl\left(1-\frac{\tau}{\tau_*}\right) +\cdots, \label{exp_phi}   
\end{align}
where the dots stand for higher order terms in $1-(\tau/\tau_*)$. Correspondingly, we expand the effective mass as
\begin{align}
m^2_{\rm{eff}}(\tau)= g_*-g_{\phi,*}\sqrt{2 \epsilon} \Mpl\left(1-\frac{\tau}{\tau_*}\right) +\cdots, \label{m_eff}     
\end{align}
where $g_{\phi,*}\equiv \mathrm{d}g/\mathrm{d}\phi|_{\phi=\phi_{0*}}$. The expansion works well at least within a span of several e-foldings around the horizon crossing if the following condition is satisfied:
\begin{align}
\frac{|g_{\phi,*} |\Mpl}{g_*}\lesssim \epsilon^{-1/2}.
\end{align}
In the following analysis, we will take into account the leading order correction, specifically, the second term of Eq.~\eqref{m_eff}. On the other hand, the previously mentioned condition $|\Delta m_{\rm eff}^2|\gtrsim m_{\rm eff}^2$ for non-negligible time dependence during inflation can be rephrased as
{$|g_{\phi,*} |\Mpl/g_*\gtrsim \epsilon^{-1/2}\delta N^{-1}$
with $\delta N \sim 10$ being the change of the e-folding number over observable part of inflation} and $m_{\rm eff}^2$ being evaluated at the horizon crossing time. To sum up, there exists a parameter regime where both conditions are simultaneously satisfied:{
\begin{align}
\epsilon^{-1/2}\delta N^{-1}\lesssim\frac{|g_{\phi,*} |\Mpl}{g_*}\lesssim \epsilon^{-1/2}.
\label{eq:analytic_app}
\end{align}
}We work in this regime and study the effects of the time-dependent mass analytically.

\subsubsection*{Analytical mode function for massive field}

With the leading order correction in Eq.~\eqref{m_eff}, we simplify the equation of motion~\eqref{eom_v} for the massive field $\sigma$ to
\begin{align}
v_k^{\prime \prime}-\frac{2}{\tau}v_k^{\prime}+\left(k^2+\frac{\mu^2+9/4}{\tau^2}+\frac{2k\kappa}{\tau}\right) v_k=0\label{Eq_v}    
\end{align}
where 
\begin{align}
\mu^2\equiv \frac{ g_*}{H^2}\left(1-\frac{\sqrt{2 \epsilon}g_{\phi,*} \Mpl}{g_*}\right)-\frac{9}{4}, \quad \kappa \equiv  -\frac{g_*}{2H^2}\frac{\sqrt{2 \epsilon}g_{\phi,*} \Mpl}{g_*}. \label{def_mu}
\end{align}
It is important to note that $g_*$ and $g_{\phi,*}$ are functions of $v(k)=k/k_0$, thus $\mu$ and $\kappa$ have scale dependence accordingly. We can solve Eq.~\eqref{Eq_v} analytically as
\begin{align}
v_k=\frac{e^{\pi \kappa /2}}{\sqrt{2 k}}H(-\tau) W_{-\i \kappa, \i \mu}(2 \i k \tau),  \label{sol_v}  
\end{align}
where $W_{a,b}(\cdot)$ is the Whittaker function.\footnote{In this paper, we focus on the case with $\mu^2>0$ to see how the oscillatory features of the cosmological collider signal are affected by the time-dependent mass.
}
Note that when the mass of $\sigma$ is constant, $g(\phi)=m_0^2$ (and hence $\kappa=0$), Eq.~\eqref{sol_v} is reduced to
\begin{align}
v_k= e^{-\frac{\pi}{2} \mu+\i \frac{\pi}{4}} \frac{\sqrt{\pi}}{2} H(-\tau)^{3 / 2} H_{\i \mu}^{(1)}(-k \tau)\label{hankel}
\end{align}  
with $\mu=\sqrt{(m_0/H)^2-9/4}$, which reproduces the mode function for a constant mass~\cite{Chen:2009zp}.

\subsubsection*{Propagators}
Based on the mode functions, \eqref{sol_u} and~\eqref{sol_v}, one can construct Schwinger--Keldysh (SK) propagators as follows (see Ref.~\cite{Chen:2017ryl} for details).

\begin{itemize}
\item Bulk-to-boundary propagators for the inflaton fluctuation $\delta\phi$:
\begin{align}
G_{\mathrm{a}}(k ; \tau)=\frac{H^2}{2 k^3}(1-\i\mathrm{a} k \tau) e^{\i\mathrm{a}k \tau}, \label{G_a}   
\end{align}
where ${\mathrm{a}}=\pm$.

\item Bulk-to-bulk propagators $D_{\mathrm{ab}}$ with ${\mathrm{a},\mathrm{b}}=\pm$ for the massive field $\sigma$:
\begin{align}
& D_{\pm \mp}\left(k ; \tau_1, \tau_2\right)=D_{\lessgtr}\left(k ; \tau_1, \tau_2\right), \label{D_pm}\\
& D_{\pm \pm}\left(k ; \tau_1, \tau_2\right)=D_{\gtrless}\left(k ; \tau_1, \tau_2\right) \theta\left(\tau_1-\tau_2\right)+D_{\lessgtr}\left(k ; \tau_1, \tau_2\right) \theta\left(\tau_2-\tau_1\right), \label{D_pp}
\end{align}
where $\theta(\cdot)$ is a unit step function and
\begin{align}
&D_{>}\left(k ; \tau_1, \tau_2\right)=v_k\left( \tau_1\right) v_k^*\left( \tau_2\right)=\frac{e^{\pi \kappa}}{2k}H^2(\tau_1\tau_2) W_{-\i \kappa, \i \mu}(2 \i k \tau_1)W_{\i \kappa, -\i \mu}(-2 \i k \tau_2), \\
&D_{<}\left(k ; \tau_1, \tau_2\right)=D_{>}^*\left(k ; \tau_1, \tau_2\right).
\end{align}

\end{itemize}

\subsection{Inflationary correlators}
Here, we specify inflationary correlation functions to be investigated in this paper.
The correlation functions can be computed from the formula~\cite{Wang:2013zva}, 
\begin{align}
\left\langle\mathcal{O}\right\rangle \equiv\langle 0|\left[\bar{\mathrm{T}} \exp \left(\i \int_{-\infty}^0 \mathrm{d} \tau^{\prime} H_I\left(\tau^{\prime}\right)\right)\right] \mathcal{O}_I\left[\mathrm{T} \exp \left(-\i \int_{-\infty}^0 \mathrm{d}\tau^{\prime} H_I\left(\tau^{\prime}\right)\right)\right]  |0\rangle,\label{master}
\end{align}
where $\mathcal{O}$ is the operator of our interest, and $\mathrm{T}\ (\bar{\mathrm{T}})$ is a (anti) time-ordering operator. $H_I$ is an interaction Hamiltonian, and all quantities on the right-hand side are understood in the interaction picture.

\subsubsection*{Interactions}
\begin{figure}[t]
\centering
\includegraphics[width=16.0cm]{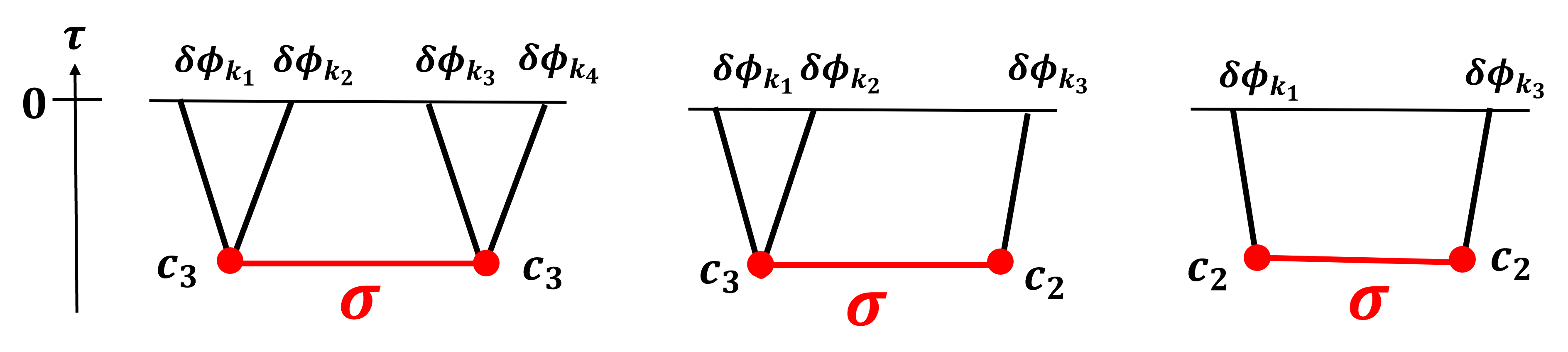}
\caption{Diagrams with a massive scalar field $\sigma$ with time-dependent mass.}
\label{fig0}
\end{figure}

In addition to Eq.~\eqref{e_mass}, we assume the presence of the following derivative (shift-symmetric with respect to $\delta\phi$) interactions involving the inflaton fluctuation $\delta\phi$ and the massive scalar~$\sigma$,
\begin{align}
&\mathcal{L}_{2,{\rm{int}}}= c_2(-H\tau)^{-3}  \sigma  \delta\phi^{\prime},\label{L_2}\\
&\mathcal{L}_{3,{\rm{int}}}= c_3(-H\tau)^{-2} \sigma ( \delta\phi^{\prime})^2, \label{L_3}
\end{align}
where $c_2$ and $c_3$ are coupling constants with mass dimension $1$ and $-1$, respectively.
With the presence of the interactions and the formula~\eqref{master}, we can calculate inflationary correlators $\left\langle\delta\phi\cdots \delta\phi\right\rangle$. In particular, this paper focuses on analyzing the two-, three-, and four-point correlation functions of the inflaton fluctuations as shown in Fig.~\ref{fig0}.  

Before proceeding, let us comment on other types of $\phi$-$\sigma$ interactions which arise from the non-derivative coupling~\eqref{e_mass} and have unavoidable contribution to inflaton correlators. The interaction~\eqref{e_mass} can be expanded as  
\begin{align}
g(\phi)\sigma^2=m^2_{\rm{eff}}\sigma^2+g_\phi\delta \phi \sigma^2 +\cdots,  
\end{align}
and the second term, in conjunction with two transfer vertices~\eqref{L_2}, contributes to the three-point inflaton correlator as a tree, double $\sigma$-exchange diagram, in addition to the diagram of our interest (the middle diagram with single $\sigma$-exchange of Fig.~\ref{fig0}). The contribution from the double-exchange diagram is estimated as 
\begin{align}
\langle\delta\phi^3\rangle_{\text{DE}}  \sim g_\phi c_2^2,
\end{align}
while the one from the single-exchange diagram is 
\begin{align}
\langle\delta\phi^3\rangle_{\text{SE}}  \sim H^3c_2c_3.
\end{align}
Here, in order to compare these contributions, let us further assume that the interactions, \eqref{L_2} and~\eqref{L_3}, are generated from a single covariant interaction $\sigma (\partial \phi)^2/\Lambda$ with a cutoff scale $\Lambda$. Then, the coupling constants are fixed as $c_2\sim \sqrt{\epsilon} \Mpl H/\Lambda$ and $c_3\sim 1/\Lambda$, and thus the ratio of these two correlators is evaluated as
\begin{align}
\frac{1}{\delta{N}}\lesssim\frac{\langle\delta\phi^3\rangle_{\text{DE}} }{\langle\delta\phi^3\rangle_{\text{SE}}}  \sim \frac{\sqrt{\epsilon}\Mpl g_\phi}{H^2}\lesssim 1.\label{eq:double}
\end{align}
The inequalities are from the lower and the upper bound of Eq.~\eqref{eq:analytic_app} with the quasi-single field regime $m_{\rm{eff}}^2=g(\phi_0)\sim H^2$. Therefore, we find the contribution from the double-exchange diagram is comparable to or smaller than the single-exchange one in the cosmological collider setup. 
In section~\ref{PNG} where some phenomenological aspects of the time-dependent mass are discussed, we choose the model parameters within the region~\eqref{eq:double}, in particular near the lower bound. In this case, the contribution from the double-exchange is negligible (about $10\%$ of the single-exchange contribution). Furthermore, 
each massive propagator is {naively} expected to be suppressed by the Boltzmann factor $e^{-\pi\mu}$, leading to the prediction that the double exchange process {can} receive stronger suppression by one more Boltzmann factor than the single-exchange process\footnote{{Note that it was numerically shown that the triple-exchange diagram exhibits only a single Boltzmann suppression factor in Ref.~\cite{Chen:2017ryl}, so we need more detailed analysis for double-exchange one.}}. Taking the above discussions into account, we ignore the contribution from the double-exchange diagram in this work\footnote{An analytic approach to multiple exchange diagrams was recently discussed in Ref.~\cite{Xianyu:2023ytd}. As in this work, it would be interesting to consider the effects of time-dependent couplings on these more general diagrams. We leave it as  non-trivial future works.}.

\subsubsection*{4-point correlators}
Let us begin with the correction to the four-point correlation functions since its specific limit produces the three- and two-point correlation functions. 

By employing the in-in formula~\eqref{master} (or SK diagrammatic rule~\cite{Chen:2017ryl}), we obtain 
\begin{align}
\nonumber &\langle\delta\phi_{\bf{k}_1} \delta\phi_{\bf{k}_2}\delta\phi_{\bf{k}_3}\delta\phi_{\bf{k}_4}\rangle '\\
&=-4\frac{c_3^2}{H^4}\sum_{\mathrm{a}, \mathrm{b}=\pm} \mathrm{ab} \int  \frac{\mathrm{d} \tau_1}{\left(-\tau_1\right)^2} \frac{\mathrm{d} \tau_2}{\left(-\tau_2\right)^2} G^{\prime}_{\mathrm{a}}(k_1;\tau_1)G^{\prime}_{\mathrm{a}}(k_2;\tau_1) G^{\prime}_{\mathrm{b}}(k_3;\tau_2)G^{\prime}_{\mathrm{b}}(k_4;\tau_2)D_{\mathrm{ab}}\left(k_s ; \tau_1, \tau_2\right)+{\rm{5 per.}} ,
\end{align}
where the prime on the left-hand side means that a momentum conservation factor $(2\pi)^3\delta^{(3)}({\bf{k}}_1 + \cdots+ {\bf{k}}_4)$ is extracted, and ``5 per.'' represents permutations of the external momenta ${\bf{k}}_i\ (i=1,\cdots, 4)$. The propagators $G_{\m{a}}$ and $D_{\m{ab}}$ are given in Eqs.~\eqref{G_a}, \eqref{D_pm} and \eqref{D_pp} respectively, and $k_s\equiv |\mathbf{k}_1+\mathbf{k}_2|$ is the ``$s$-channel'' momentum. This can be further simplified as 
\begin{align}
\nonumber &\langle\delta\phi_{\bf{k}_1} \delta\phi_{\bf{k}_2}\delta\phi_{\bf{k}_3}\delta\phi_{\bf{k}_4}\rangle '\\
\nonumber &= -4c_3^2H^4\cdot \frac{1}{16 k_1 k_2 k_3k_4} \sum_{\mathrm{a}, \mathrm{b}=\pm} \mathrm{ab} \int  \mathrm{d} \tau_1\mathrm{d} \tau_2 e^{\mathrm{i}\mathrm{a} k_{12} \tau_1+\i\mathrm{b} k_{34} \tau_2} D_{\mathrm{ab}}\left(k_s ; \tau_1, \tau_2\right)+{\rm{5 per.}} \\
&=4c_3^2H^4\cdot \frac{1}{16 k_1 k_2 k_3k_4k_s^5}\sum_{\mathrm{a}, \mathrm{b}=\pm}\mathcal{I}_{\mathrm{ab}}^{0 , 0}+{\rm{5 per.}},    \label{4pt}
\end{align}
where  $k_{12}\equiv k_1+k_2$ and $k_{34}\equiv k_3+k_4$. In the first line we inserted the explicit expression of $G_{\m{a}}$, and in the second line we introduced a ``seed integral''~\cite{Qin:2022fbv,Qin:2023ejc},
\begin{align}
&\mathcal{I}_{\mathrm{ab}}^{p_1 p_2} \equiv-\mathrm{ab} k_s^{5+p_{12}} \int_{-\infty}^0 \mathrm{~d} \tau_1 \mathrm{~d} \tau_2\left(-\tau_1\right)^{p_1}\left(-\tau_2\right)^{p_2} e^{\i\mathrm{a}  k_{12} \tau_1+\i\mathrm{b} k_{34} \tau_2} D_{\mathrm{ab}}\left(k_s ; \tau_1, \tau_2\right),  \label{seed}  
\end{align}
with $p_{1}$ and $p_{2}$ being constant numbers with $p_{1,2}>-5/2$, and $p_{12}\equiv p_1+p_2$. Therefore, once the seed integral~\eqref{seed} is computed, the four-point correlation function can be automatically obtained from Eq.~\eqref{4pt}, and this is also true for three- and two-point functions, as shown below.

\subsubsection*{3-point correlators}
In the same way, by utilizing the seed integral~\eqref{seed}, the three-point function is given by
\begin{align}
\nonumber &\langle\delta\phi_{\bf{k}_1} \delta\phi_{\bf{k}_2}\delta\phi_{\bf{k}_3}\rangle '\\
\nonumber &=-2\frac{c_2c_3}{H^5}\sum_{\mathrm{a}, \mathrm{b}=\pm} \mathrm{ab} \int  \frac{\mathrm{d} \tau_1}{\left(-\tau_1\right)^2}  \frac{\mathrm{d} \tau_2}{\left(-\tau_2\right)^3} G^{\prime}_{\mathrm{a}}(k_1;\tau_1)G^{\prime}_{\mathrm{a}}(k_2;\tau_1) G^{\prime}_{\mathrm{b}}(k_3;\tau_2)D_{\mathrm{ab}}\left(k_3 ; \tau_1, \tau_2\right)+{\rm{2 per.}} \\ 
\nonumber &= 2c_2c_3H\cdot \frac{1}{8 k_1 k_2 k_3} \sum_{\mathrm{a}, \mathrm{b}=\pm} \mathrm{ab} \int  \mathrm{d} \tau_1\frac{\mathrm{d} \tau_2}{\left(-\tau_2\right)^2} e^{\mathrm{i}\mathrm{a} k_{12} \tau_1+\i\mathrm{b} k_{3} \tau_2} D_{\mathrm{ab}}\left(k_3 ; \tau_1, \tau_2\right)+{\rm{2 per.}} \\
&= -2c_2c_3H\cdot \frac{1}{8 k_1 k_2 k_3^4} \lim_{k_4 \to 0}\sum_{\mathrm{a}, \mathrm{b}=\pm}\mathcal{I}_{\mathrm{ab}}^{0 ,-2}+{\rm{2 per.}}. \label{3pt}
\end{align}
The three-point correlator corresponds to a soft limit ($k_4 \rightarrow 0$) of the four-point one or the seed integral.
\subsubsection*{2-point correlators}
Finally, the leading order correction to the two-point function is described as follows:
\begin{align}
\nonumber \langle\delta\phi_{\bf{k}_1} \delta\phi_{\bf{k}_3}\rangle '&= -\frac{c_2^2}{H^6}\sum_{\mathrm{a}, \mathrm{b}=\pm} \mathrm{ab} \int  \frac{\mathrm{d} \tau_1}{\left(-\tau_1\right)^3}  \frac{\mathrm{d} \tau_2}{\left(-\tau_2\right)^3} G^{\prime}_{\mathrm{a}}(k_1;\tau_1)G^{\prime}_{\mathrm{b}}(k_1;\tau_2)D_{\mathrm{ab}}\left(k_1 ; \tau_1, \tau_2\right)\\ 
\nonumber &= -\frac{c_2^2}{H^2}\cdot \frac{1}{4k_1^2} \sum_{\mathrm{a}, \mathrm{b}=\pm} \mathrm{ab} \int  \frac{\mathrm{d} \tau_1}{\left(-\tau_1\right)^2} \frac{\mathrm{d} \tau_2}{\left(-\tau_2\right)^2} e^{\mathrm{i}\mathrm{a} k_{1} \tau_1+\i\mathrm{b} k_{1} \tau_2} D_{\mathrm{ab}}\left(k_1 ; \tau_1, \tau_2\right) \\
&= \frac{c_2^2}{H^2}\cdot \frac{1}{4k_1^3} \lim_{k_2,k_4 \to 0}\sum_{\mathrm{a}, \mathrm{b}=\pm}\mathcal{I}_{\mathrm{ab}}^{-2 ,-2}.\label{2pt}
\end{align}
Here, we need an expression for double soft limits ($k_2,k_4\rightarrow 0$) of the seed integral. Note that the above correlator is a correction to the one in free theory, $\langle\delta\phi_{\bf{k}_1} \delta\phi_{\bf{k}_3}\rangle '=H^2/2k_1^3$.

\section{Bootstrapping Seed Integrals with Time-dependent Mass}\label{sec:boot_eq}
Our task is reduced to evaluate the seed integral~\eqref{seed}, 
\begin{align}
&\mathcal{I}_{\mathrm{ab}}^{p_1 p_2} \equiv-\mathrm{ab} k_s^{5+p_{12}} \int_{-\infty}^0 \mathrm{~d} \tau_1 \mathrm{~d} \tau_2\left(-\tau_1\right)^{p_1}\left(-\tau_2\right)^{p_2} e^{\i\mathrm{a}  k_{12} \tau_1+\i\mathrm{b} k_{34} \tau_2} D_{\mathrm{ab}}\left(k_s ; \tau_1, \tau_2\right).  \label{seed2}  
\end{align}
Performing the time-integral in Eq.~\eqref{seed2} is challenging for general momentum configurations. Therefore, we employ another approach developed in Refs.~\cite{Arkani-Hamed:2018kmz,Qin:2022fbv,Qin:2023ejc}, where we can obtain analytic expressions for the seed integrals by solving differential equations (``bootstrap equations'') for Eq.~\eqref{seed2}. We will see that this method is also applicable to our case. 
Note that this section describes technical details, so readers who are not interested in the derivation can skip to the final results, Eqs.~\eqref{I_4_pm}, \eqref{I_4_pp} for the seed integral, Eqs.~\eqref{I_3_pm}, \eqref{I_3_pp} for its single soft limit, and Eqs.~\eqref{I_pm_2}, \eqref{I_pp_2} for the double soft limit.
\subsection{Bootstrap equations}
The SK propagators in Eqs.~\eqref{D_pm} and~\eqref{D_pp} satisfy the following differential equations
\begin{align}
&\left(\tau_1^2 \partial_{\tau_1}^2-2 \tau_1 \partial_{\tau_1}+k_s^2 \tau_1^2+\mu^2+\frac{9}{4}+2\kappa k_s\tau_1\right) D_{ \pm \mp}\left(k_s ; \tau_1, \tau_2\right)=0,\label{Green1}\\
&\left(\tau_1^2 \partial_{\tau_1}^2-2 \tau_1 \partial_{\tau_1}+k_s^2 \tau_1^2+\mu^2+\frac{9}{4}+2\kappa k_s\tau_1\right) D_{ \pm \pm}\left(k_s ; \tau_1, \tau_2\right)=\mp \mathrm{i} H^2\tau_1^2 \tau_2^2 \delta\left(\tau_1-\tau_2\right),\label{Green2}
\end{align}
where we recall $k_s\equiv |\mathbf{k}_1+\mathbf{k}_2|$. They satisfy the same differential equations with respect to $\tau_2$. By introducing 
\begin{align}
z_1 =-k_{12}\tau_1, \quad  z_2 =-k_{34} \tau_2,
\end{align}
with $k_{12}\equiv k_1+k_2$ and $k_{34}\equiv k_3+k_4$, and 
\begin{align}
r_1\equiv \frac{k_s}{k_{12}}, \quad r_2\equiv \frac{k_s}{k_{34}},
\end{align}
Eqs.~\eqref{Green1} and \eqref{Green2} can be rewritten as
\begin{align}
&\left(z_1^2 \partial_{z_1}^2-2 z_1 \partial_{z_1}+r_1^2z_1^2+\mu^2+\frac{9}{4}-2\kappa r_1z_1\right) \widehat{D}_{ \pm \mp}\left(r_1z_1, r_2z_2\right)=0,\label{Green1_z}\\
&\left(z_1^2 \partial_{z_1}^2-2 z_1 \partial_{z_1}+r_1^2z_1^2+\mu^2+\frac{9}{4}-2\kappa r_1z_1\right) \widehat{D}_{ \pm \pm}\left(r_1z_1, r_2z_2\right)=\mp \mathrm{i} H^2r_1^2z_1^2 r_2^2z_2^2 \delta\left(r_1z_1-r_2z_2\right),\label{Green2_z}
\end{align}
where we defined $\widehat{D}_{\m{ab}}=k_s^3D_{\m{ab}}$, or explicitly, 
\begin{align}
&\widehat{D}_{\pm\mp}\left(r_1 z_1, r_2 z_2\right)=  \frac{e^{\pi \kappa}}{2}H^2r_1z_1r_2z_2 W_{\pm\i \kappa, \mp\i \mu}(\pm 2 \i r_1z_1)W_{\mp\i \kappa, \pm\i \mu}(\mp 2 \i r_2z_2),\\
&\widehat{D}_{\pm\pm}\left(r_1 z_1, r_2 z_2\right)=\theta(r_2z_2-r_1z_1)\widehat{D}_{\mp\pm}\left(r_1 z_1, r_2 z_2\right)+\theta(r_1z_1-r_2z_2)\widehat{D}_{\pm\mp}\left(r_1 z_1, r_2 z_2\right).
\end{align}
An important observation is that $\widehat{D}$ depends on $z_i\ (i=1,2)$ with a specific combination $r_iz_i$. Thus, Eqs.~\eqref{Green1_z} and \eqref{Green2_z} can be regarded as the differential equations with respect to $r_i$, i.e.,
\begin{align}
&\left(r_1^2 \partial_{r_1}^2-2 r_1 \partial_{r_1}+r_1^2z_1^2+\mu^2+\frac{9}{4}-2\kappa r_1z_1\right) \widehat{D}_{ \pm \mp}\left(r_1z_1, r_2z_2\right)=0,\label{Green1_r}\\
&\left(r_1^2 \partial_{r_1}^2-2 r_1 \partial_{r_1}+r_1^2z_1^2+\mu^2+\frac{9}{4}-2\kappa r_1z_1\right) \widehat{D}_{ \pm \pm}\left(r_1z_1, r_2z_2\right)=\mp \mathrm{i} H^2r_1^2z_1^2 r_2^2z_2^2 \delta\left(r_1z_1-r_2z_2\right) ,\label{Green2_r}
\end{align}
and similarly for $r_2$.

The subsequent task is identifying differential equations (bootstrap equations) for the seed integral. In terms of $z_i$ and $r_i$, Eq.~\eqref{seed2} is expressed as 
\begin{align}
&\mathcal{I}_{\mathrm{ab}}^{p_1 p_2} =(-\mathrm{ab}) r_1^{1+p_1} r_2^{1+p_2} \int_0^{\infty} \mathrm{d} z_1 \mathrm{~d} z_2 z_1^{p_1} z_2^{p_2} e^{-\mathrm{ia} z_1-\mathrm{ib} z_2} \widehat{D}_{\mathrm{ab}}\left(r_1 z_1, r_2 z_2\right). \label{seed_z}
\end{align}
Let us start from the opposite sign seed integral $\mathcal{I}_{\pm \mp}^{p_1 p_2}$. By applying Eq.~\eqref{Green1_r}, $r_1^{-1-p_1}r_2^{-1-p_2}\mathcal{I}_{\pm \mp}^{p_1 p_2}$ satisfies
\begin{align}
\nonumber &\left(r_1^2 \partial_{r_1}^2-2 r_1 \partial_{r_1}+\mu^2+\frac{9}{4}\right)\left(r_1^{-1-p_1}r_2^{-1-p_2}\mathcal{I}_{\pm \mp}^{p_1 p_2}\right) \\
\nonumber &= \int_0^{\infty} \mathrm{d} z_1 \mathrm{~d} z_2 z_1^{p_1} z_2^{p_2} e^{\mp\mathrm{i} z_1\pm \mathrm{i} z_2} \left(r_1^2 \partial_{r_1}^2-2 r_1 \partial_{r_1}+\mu^2+\frac{9}{4}\right)\widehat{D}_{\pm \mp}\left(r_1 z_1, r_2 z_2\right)\\
\nonumber &= \int_0^{\infty} \mathrm{d} z_1 \mathrm{~d} z_2 z_1^{p_1} z_2^{p_2} e^{\mp\mathrm{i} z_1\pm \mathrm{i} z_2} \left(-r_1^2z_1^2+2\kappa r_1z_1\right)\widehat{D}_{\pm \mp}\left(r_1 z_1, r_2 z_2\right)\\
&=\left[r_1^2\left(r_1 \partial_{r_1}+p_1+2\right)\left(r_1 \partial_{r_1}+p_1+1\right) \mp  2\i \kappa r_1\left(r_1 \partial_{r_1}+p_1+1\right)\right] \left(r_1^{-1-p_1}r_2^{-1-p_2}\mathcal{I}_{\pm \mp}^{p_1 p_2}\right), \label{eq_rI}  
\end{align}
where we used the following formulae for an arbitrary function $f(rz)$,
\begin{align}
&\int_0^{\infty} \mathrm{d} z\ z^{p+1} e^{-\i \mathrm{a} z} f(r z)=-\i \mathrm{a}(r \partial_r+p+1) \int_0^{\infty} \mathrm{d} z\ z^p e^{-\i \mathrm{a} z} f(r z),\\
&\int_0^{\infty} \mathrm{d} z\ z^{p+2} e^{-\i \mathrm{a} z}  f(r z)=-\left(r \partial_r+p+2\right)\left(r \partial_r+p+1\right) \int_0^{\infty}\mathrm{d} z\ z^p e^{-\i \mathrm{a} z} f(r z),
\end{align}
from the third to the fourth line. Thus, we find 
\begin{align}
\nonumber &\left[\left(r_1^2-r_1^4\right) \partial_{r_1}^2-2\left( r_1+\left(p_1+2\right) r_1^3\mp \i \kappa r_1^2\right) \partial_{r_1}+\left(\mu^2+\frac{9}{4}-(p_1+1)(p_1+2) r_1^2\pm 2(p_1+1)\i \kappa r_1\right)\right] \\
&\times  \left(r_1^{-1-p_1}r_2^{-1-p_2}\mathcal{I}_{\pm \mp}^{p_1 p_2}\right)=0,
\end{align}
or equivalently,
\begin{align}
\mathcal{D}_{\pm, r_1}^{p_1} \mathcal{I}_{ \pm \mp}^{p_1 p_2}=0,\label{B_r1}
\end{align}
where
\begin{align}
\mathcal{D}_{\pm,r}^p \equiv\left(r^2-r^4\right) \partial_r^2-2\left[(p+2) r+ r^3\mp \i \kappa r^2\right] \partial_r+\mu^2+\frac{(5+2 p)^2}{4}.
\end{align}
Eq.~\eqref{B_r1} gives the bootstrap equations for $\mathcal{I}_{\pm \mp}^{p_1 p_2}$ with respect to $r_1$. 
Those with respect to~$r_2$ are obtained in the same way, $\mathcal{D}_{\mp, r_2}^{p_1} \mathcal{I}_{ \pm \mp}^{p_1 p_2}=0$.

Let us shift our focus to the seed integral with the same sign, $\mathcal{I}_{\pm \pm}^{p_1 p_2}$. The derivation is basically parallel to the previous case, except for the presence of the ``source'' term on the right-hand side of Eq.~\eqref{Green2_z}. In fact, we find 
\begin{align}
\nonumber &\left(r_1^2 \partial_{r_1}^2-2 r_1 \partial_{r_1}+\mu^2+\frac{9}{4}\right)\left(r_1^{-1-p_1}r_2^{-1-p_2}\mathcal{I}_{\pm \pm}^{p_1 p_2} \right) \\
\nonumber &=\left[r_1^2\left(r_1 \partial_{r_1}+p_1+2\right)\left(r_1 \partial_{r_1}+p_1+1\right) \mp  2\i \kappa r_1\left(r_1 \partial_{r_1}+p_1+1\right)\right] \left(r_1^{-1-p_1}r_2^{-1-p_2}\mathcal{I}_{\pm \pm}^{p_1 p_2} \right)\\
&\ \ \ \ \pm \i H^2 r_1^2r_2^2 \int_0^{\infty} \mathrm{d} z_1 \mathrm{~d} z_2 z_1^{p_1+2} z_2^{p_2+2} e^{\mp\mathrm{i} z_1\mp\mathrm{i} z_2} \delta (r_1z_1-r_2z_2),  
\end{align}
where the second term of the right-hand side is the new contribution. After performing the one-dimensional integral, we obtain the bootstrap equations
\begin{align}
\mathcal{D}_{\pm,r_1}^{p_1} \mathcal{I}_{ \pm \pm}^{p_1 p_2}=H^2e^{\mp \mathrm{i} p_{12} \frac{\pi}{2}} \Gamma\left(5+p_{12}\right)\left(\frac{r_1 r_2}{r_1+r_2}\right)^{5+p_{12}}.    
\end{align}
The same equation for $r_2$ is provided with a replacement $\mathcal{D}_{\pm, r_1}^{p_1} \rightarrow \mathcal{D}_{\pm, r_2}^{p_1}$.

As pointed out in Refs.~\cite{Qin:2022fbv,Qin:2023ejc}, it is useful to introduce new variables 
\begin{align}
u_i\equiv \frac{2 r_i}{1+r_i}, \quad (i=1,2).   
\end{align}
In terms of $u_i$, the set of bootstrap equations is summarized as 
\begin{align}
&\mathcal{D}_{\pm,u_1}^{p_1} \mathcal{I}_{ \pm \mp}^{p_1 p_2}=0,\label{Homo}\\
&\mathcal{D}_{\pm,u_1}^{p_1} \mathcal{I}_{ \pm \pm}^{p_1 p_2}=H^2e^{\mp \mathrm{i} p_{12} \frac{\pi}{2}} \Gamma\left(5+p_{12}\right)\left(\frac{u_1 u_2}{2\left(u_1+u_2-u_1 u_2\right)}\right)^{5+p_{12}},\label{inHomo}
\end{align}
and
\begin{align}
&\mathcal{D}_{\mp,u_2}^{p_2} \mathcal{I}_{ \pm \mp}^{p_1 p_2}=0,\label{Homo2}\\
&\mathcal{D}_{\pm,u_2}^{p_2} \mathcal{I}_{ \pm \pm}^{p_1 p_2}=H^2e^{\mp \mathrm{i} p_{12} \frac{\pi}{2}} \Gamma\left(5+p_{12}\right)\left(\frac{u_1 u_2}{2\left(u_1+u_2-u_1 u_2\right)}\right)^{5+p_{12}},\label{inHomo2}
\end{align}
where 
\begin{align}
\mathcal{D}_{\pm,u}^p \equiv\left(u^2-u^3\right) \partial_u^2-\left[(4+2 p) u-(1+p\pm \i \kappa) u^2\right] \partial_u+\left[\mu^2+\left(p+\frac{5}{2}\right)^2\right].    
\end{align}

\subsection{Solutions}
Here we solve the bootstrap equations, \eqref{Homo}, \eqref{inHomo}, \eqref{Homo2}, and \eqref{inHomo2}. 

\subsubsection*{Opposite sign seed $\mathcal{I}_{ \pm \mp}^{ p_1 p_2}$}
The opposite sign seeds $\mathcal{I}_{ \pm \mp}^{ p_1 p_2}$ satisfy the homogeneous differential equations, \eqref{Homo} and \eqref{Homo2}. Considering each equation has two independent solutions, a general solution for $\mathcal{I}_{ \pm \mp}^{ p_1 p_2}$ can be expressed as the following combination:
\begin{align}
\mathcal{I}_{ \pm \mp}^{ p_1 p_2}=\sum_{\mathrm{a}, \mathrm{b}= \pm} \mathcal{C}_{ \pm \mp \mid \mathrm{ab}}\ \mathcal{U}_{ \pm \mid \mathrm{a}}^{p_1}\left(u_1\right) \mathcal{U}_{\mp \mid \mathrm{b}}^{p_2}\left(u_2\right), \label{Gsol_pm}
\end{align}
where $\mathcal{C}_{ \pm \mp \mid \mathrm{ab}} $ are coefficients fixed from boundary conditions later, and 
$\mathcal{U}_{\mathrm{a} \mid \mathrm{b}}(u)$ with $\m{b}=\pm 1$ are the two independent solutions
\begin{align}
\mathcal{U}_{\mathrm{a} \mid \mathrm{b}}^p(u)=\mathrm{iab} 2^{\mathrm{iab} \mu} \pi \operatorname{csch}(2 \pi \mu)\left(\frac{u}{2}\right)^{5/ 2+p+\mathrm{iab} \mu} { }_2\mathcal{F}_1\left[\begin{array}{c|c}
\frac{5}{2}+p+\operatorname{iab} \mu, \frac{1}{2}-\mathrm{ia} \kappa +\operatorname{iab} \mu \\
1+2 \mathrm{iab} \mu
\end{array} \  u\right]. \label{def_U}
\end{align}
Here, we inserted some numerical factors for later convenience, and ${ }_p\mathcal{F}_q$ is defined by
\begin{align}
{ }_p\mathcal{F}_q\left[\begin{array}{c|c}
a_1, \cdots, a_p \\
b_1, \cdots, b_q
\end{array} \ z\right]\equiv \Gamma\left[\begin{array}{c}
a_1, \cdots, a_p \\
b_1, \cdots, b_q
\end{array}\right]{ }_p \mathrm{F}_q\left[\begin{array}{c|c}
a_1, \cdots, a_p \\
b_1, \cdots, b_q
\end{array} \ z\right] , \label{def_mF}
\end{align}
where ${ }_p\mathrm{F}_q$ is a generalized hypergeometric function, and the products of 
gamma function is abbreviated by
\begin{align}
&\Gamma\left[z_1, \cdots, z_m\right]  \equiv \Gamma\left(z_1\right) \cdots \Gamma\left(z_m\right), \\
&\Gamma\left[\begin{array}{c}
z_1, \cdots, z_m \\
w_1, \cdots, w_n
\end{array}\right]  \equiv \frac{\Gamma\left(z_1\right) \cdots \Gamma\left(z_m\right)}{\Gamma\left(w_1\right) \cdots \Gamma\left(w_n\right)} .
\end{align}

\subsubsection*{Same sign seed $\mathcal{I}_{ \pm \pm }^{ p_1 p_2}$}
The same sign seeds $\mathcal{I}_{ \pm \pm}^{ p_1 p_2}$ satisfy the inhomogeneous differential equations, \eqref{inHomo} and \eqref{inHomo2}. The general solution is obtained by combining the general solution for the corresponding homogeneous equation with the particular solution for the inhomogeneous one. Therefore, we can take
\begin{align}
\mathcal{I}_{ \pm \pm}^{ p_1 p_2}=\mathcal{G}_{ \pm \pm}^{ p_1 p_2}+\sum_{\mathrm{a}, \mathrm{b}= \pm} \mathcal{C}_{ \pm \pm \mid \mathrm{ab}}\ \mathcal{U}_{ \pm \mid \mathrm{a}}^{p_1}\left(u_1\right) \mathcal{U}_{\pm \mid \mathrm{b}}^{p_2}\left(u_2\right),  \label{Gsol_pp}      
\end{align}
where $\mathcal{G}_{ \pm \pm}^{ p_1 p_2}$ is a particular solution and the second term is the solution for homogeneous equations consisting of undetermined coefficients $\mathcal{C}_{ \pm \pm \mid \mathrm{ab}} $ and Eq.~\eqref{def_U}. 

Regarding the particular solution, we rewrite the right-hand side of Eq.~\eqref{inHomo} by using
\begin{align}
\left(\frac{u_1 u_2}{2\left(u_1+u_2-u_1 u_2\right)}\right)^{5+p_{12}}   = \left(\frac{u_1}{2}\right)^{p_{12}+5}\sum_{n=0}^{\infty}\left(\begin{array}{c}
n+p_{12}+4 \\
n
\end{array}\right)u_1^n\left(1-\frac{1}{u_2}\right)^n,\label{expand}
\end{align}
where the array on the right-hand side means the binomial coefficient. This motivates us to take the following ansatz for $\mathcal{G}_{ \pm \pm}^{ p_1 p_2}$:
\begin{align}
\mathcal{G}_{ \pm \pm}^{ p_1 p_2}=\frac{H^2e^{\mp \frac{\pi}{2} \i p_{12}} \Gamma\left(p_{12}+5\right) }{2^{p_{12}+5}}\sum_{m, n=0}^{\infty} \chi_{m, n}^{\pm} u_1^{m+n+p_{12}+5}\left(1-\frac{1}{u_2}\right)^n. \label{ansatz}
\end{align}
Inserting Eqs.~\eqref{expand} and \eqref{ansatz} to Eq.~\eqref{inHomo}, we find that $ \chi_{m, n}^{\pm}$ satisfies the following recurrence relations,
\begin{align}
&\chi_{0, n}^{\pm}= \frac{1}{\mu^2+\left(n+\frac{5}{2}+p_2\right)^2}\left(\begin{array}{c}
n+p_{12}+4 \\
n
\end{array}\right),\label{ini_chi}\\
&\chi_{m+1, n}^{\pm}=\frac{\left(m+n+p_2+3\mp\i \kappa\right)\left(m+n+p_{12}+5\right)}{\mu^2+\left(m+n+\frac{7}{2}+p_2\right)^2} \chi_{m, n}^{\pm},
\end{align}
which can be solved as
\begin{align}
\chi_{m, n}^{\pm}=\frac{\left(5+n+p_{12}\right)_m\left(3+n+p_2\mp \i \kappa\right)_m}{\left(\frac{5}{2}+n+p_2-\i \mu\right)_{m+1}\left(\frac{5}{2}+n+p_2+\i \mu\right)_{m+1}}\left(\begin{array}{c}
n+p_{12}+4 \\
n
\end{array}\right)\label{sol_chi}
\end{align}
where $(z)_n\equiv \Gamma[z+n]/\Gamma[z]$ is the Pochhammer symbol. Therefore, we obtain the particular solution from Eq.~\eqref{sol_chi} and Eq.~\eqref{ansatz}. While resolving the double summation in Eq.~\eqref{ansatz} is a non-trivial challenge, we manage to perform at least one of the summations, which results in
\begin{align}
\nonumber \mathcal{G}_{ \pm \pm}^{ p_1 p_2}=&\ \frac{H^2e^{\mp \frac{\pi}{2} \i p_{12}} \Gamma\left(p_{12}+5\right) }{2^{p_{12}+5}} \sum_{n=0}^{\infty} u_1^{n+p_{12}+5}\left(1-\frac{1}{u_2}\right)^n\left(\begin{array}{c}
n+p_{12}+4 \\
n
\end{array}\right)\\
&\times \frac{1}{\mu^2+\left(\frac{5}{2}+n+p_2\right)^2}{ }_3 \mathrm{F}_2\left[\begin{array}{c|c}
1,3+n+p_2 \mp \mathrm{i} \kappa, 5+n+p_{12} \\
\frac{7}{2}+n+p_2-\mathrm{i} \mu, \frac{7}{2}+n+p_2+\mathrm{i} \mu
\end{array} \ u_1\right]. \label{g_pp}
\end{align}

Finally, we note that the particular solution~\eqref{g_pp} automatically satisfies the second inhomogeneous differential equation with respect to $u_2$~\eqref{inHomo2}. This is confirmed by substituting the ansatz~\eqref{ansatz} into Eq.~\eqref{inHomo2}. Then, in the same way as above, we obtain recurrence relations for $\chi_{m,n}$, 
\begin{align}
&\chi_{0, n}^{\pm}= \frac{1}{\mu^2+\left(n+\frac{5}{2}+p_2\right)^2}\left(\begin{array}{c}
n+p_{12}+4 \\
n
\end{array}\right),\\
&\chi_{m+1,n}^{\pm}=\frac{(n+1)\left(n+3+p_2\mp\i \kappa\right)}{\mu^2+\left(p_2+n+5 / 2\right)^2}\chi_{m,n+1}^{\pm}.
\end{align}
The first relation is exactly the same as Eq.~\eqref{ini_chi} and one can check the second relation is automatically satisfied by our solution~\eqref{sol_chi}. Therefore, Eq.~\eqref{g_pp} gives a particular solution for both inhomogeneous differential equations~\eqref{inHomo} and~\eqref{inHomo2}.

\subsubsection*{Determining coefficients}
We will specify boundary conditions for the seed integrals~\eqref{seed2} to determine the coefficients $\mathcal{C}$ in Eqs.~\eqref{Gsol_pm} and \eqref{Gsol_pp}. In Appendix~\ref{MB}, we evaluated the seed integrals in the hierarchical collapsed limit ($u_1 \ll u_2 \ll 1$) by employing another method based on the Mellin--Barnes representation of the Whittaker function~\cite{Qin:2022fbv,Qin:2023ejc}. The results are given by
\begin{align}
&\lim _{u_1 \ll u_2 \ll 1} \mathcal{I}_{\pm \mp}^{p_1 p_2} =\sum_{\m{a}, \m{b}= \pm} \tilde{\mathcal{C}}_{ \pm \mp \mid \m{a b}}\  \tilde{\mathcal{U}}_{ \pm \mid \m{a}}^{p_1}\left(u_1\right) \tilde{\mathcal{U}}_{\mp \mid \mathrm{b}}^{p_2}\left(u_2\right),\label{sq_pm}\\
&\lim _{u_1 \ll u_2 \ll 1} \mathcal{I}_{\pm \pm}^{p_1 p_2}=\sum_{\m{a}, \m{b}= \pm}\tilde{\mathcal{C}}_{ \pm \pm \mid \m{a b}}\  \tilde{\mathcal{U}}_{ \pm \mid \m{a}}^{p_1}\left(u_1\right) \tilde{\mathcal{U}}_{\pm \mid \mathrm{b}}^{p_2}\left(u_2\right), \label{sq_pp} 
\end{align}
where 
\begin{align}
\tilde{\mathcal{U}}_{\mathrm{a} \mid \mathrm{b}}^p(u)  =\operatorname{iab} 2^{\mathrm{iab} \mu} \pi \operatorname{csch}(2 \pi \mu)\left(\frac{u}{2}\right)^{5 / 2+p+\mathrm{iab} \mu} \Gamma\left[\begin{array}{c}
\frac{5}{2}+p+\operatorname{iab} \mu, \frac{1}{2}-\operatorname{ia}\kappa+\operatorname{iab} \mu \\
1+2 \operatorname{iab} \mu
\end{array}\right], \label{u_til}
\end{align}
and
\begin{align}
&\tilde{\mathcal{C}}_{ \pm \mp \mid ++} =\tilde{\mathcal{C}}_{ \pm \mp \mid +-}=\tilde{\mathcal{C}}_{ \pm \mp \mid -+}=\tilde{\mathcal{C}}_{ \pm \mp \mid --}=\frac{e^{\pi \kappa}}{2 \pi^2} H^2 \left[\cosh \left(2\pi \kappa\right)+\cosh (2 \pi \mu)\right]  e^{\mp \mathrm{i}\pi \bar{p}_{12}  / 2} ,\label{Ctil_pm}\\
&\tilde{\mathcal{C}}_{ \pm \pm \mid ++} =\tilde{\mathcal{C}}_{ \pm \pm \mid +-}= \frac{\pm\i e^{\pi \left(\kappa+\mu\right)}\cosh \pi\left(-\mu+\kappa\right)H^2}{\pi \Gamma\left[\frac{1}{2}-\mathrm{i} \mu\mp\i\kappa, \frac{1}{2}+\mathrm{i} \mu\mp\i\kappa\right]}e^{\mp\mathrm{i} \pi p_{12} / 2} ,\label{Ctil_pp1}\\
&\tilde{\mathcal{C}}_{ \pm \pm \mid -+} =\tilde{\mathcal{C}}_{ \pm \pm \mid --}= \frac{\pm\i e^{\pi \left(\kappa-\mu\right)}\cosh \pi\left(\mu+\kappa\right)H^2}{\pi \Gamma\left[\frac{1}{2}-\mathrm{i} \mu\mp\i\kappa, \frac{1}{2}+\mathrm{i} \mu\mp\i\kappa\right]}e^{\mp\mathrm{i} \pi p_{12} / 2} \label{Ctil_pp2}
\end{align}
with $\bar{p}_{12}\equiv p_1-p_2$.
On the other hand, in the limit $u_1 \ll u_2 \ll 1$, $\mathcal{U}_{\mathrm{a} \mid \mathrm{b}}^p(u)$ defined in Eq.~\eqref{def_U} reduced to $\tilde{\mathcal{U}}_{\mathrm{a} \mid \mathrm{b}}^p(u)$, and thus Eqs.~\eqref{Gsol_pm} and \eqref{Gsol_pp} have to coincide with Eqs.~\eqref{sq_pm} and \eqref{sq_pp}. Therefore, the coefficient is determined as $\mathcal{C}_{ \pm \mp \mid \mathrm{ab}} =\tilde{\mathcal{C}}_{ \pm \mp \mid \mathrm{ab}} $ and $\mathcal{C}_{ \pm \pm \mid \mathrm{ab}} =\tilde{\mathcal{C}}_{ \pm \pm \mid \mathrm{ab}} $. In the derivation, we used the fact that, in the limit $u_1 \ll u_2 \ll 1$, the particular solution~\eqref{g_pp} is negligible compared to the second term in Eq.~\eqref{Gsol_pp}. This is because the particular solution behaves as $\sim u_1^{5+p_{12}}$ in the collapsed limit, whereas scales for the dominant term is $\sim u_1^{5/2+p_{1}\pm \i\mu}u_2^{5/2+p_{2}\pm \i\mu}$.

\subsubsection*{Summary}
In summary, the analytic expressions of the seed integrals~\eqref{seed2} are obtained as follows:
\begin{tcolorbox}[arc=10pt, colback=gray!10, boxrule=1pt]
\begin{align}
\nonumber &\mathcal{I}_{ \pm \mp}^{ p_1 p_2}/H^2\\
\nonumber&= \frac{- e^{\mp \i \frac{\pi}{2}\bar{p}_{12}} e^{\pi \kappa} \left[\cosh (2\pi\kappa)+\cosh (2\pi\mu) \right]}{2\operatorname{sinh}^2(2 \pi \mu)}\\
\nonumber &\times \left\{ 2^{ \pm \i \mu}\left(\frac{u_1}{2}\right)^{\frac{5}{2}+p_1 \pm \i \mu}{}_2\mathcal{F}_1\left[\begin{array}{c|c}
\frac{5}{2}+p_1\pm\operatorname{i} \mu, \frac{1}{2}\mp\mathrm{i} \kappa \pm\operatorname{i} \mu \\
1\pm 2 \mathrm{i} \mu
\end{array} \ u_1\right]-(\mu\rightarrow-\mu)\right\}  \\
&\times \left\{ 2^{ \pm \i \mu}\left(\frac{u_2}{2}\right)^{\frac{5}{2}+p_2 \pm \i \mu}{}_2\mathcal{F}_1\left[\begin{array}{c|c}
\frac{5}{2}+p_2\pm\operatorname{i} \mu, \frac{1}{2}\pm\mathrm{i} \kappa \pm\operatorname{i} \mu \\
1\pm 2 \mathrm{i} \mu
\end{array} \ u_2\right]-(\mu\rightarrow-\mu)\right\}, 
 \label{I_4_pm} \\
\nonumber &\mathcal{I}_{ \pm \pm }^{ p_1 p_2}/H^2\\
\nonumber &=\frac{\mp \i e^{\mp\i\frac{\pi}{2}  p_{12}} e^{\pi\kappa} \pi }{\Gamma\left[\frac{1}{2} \mp \i \kappa-\i \mu, \frac{1}{2} \mp \i \kappa+\i \mu\right]\sinh^2(2\pi\mu)}\\
\nonumber &\times \left\{ \frac{e^{\pi\mu}\cosh \left[\pi(-\mu+\kappa)\right]}{2^{ \mp \i \mu}}\left(\frac{u_1}{2}\right)^{\frac{5}{2}+p_1 \pm \i \mu}{}_2\mathcal{F}_1\left[\begin{array}{c|c}
\frac{5}{2}+p_1\pm\operatorname{i} \mu, \frac{1}{2}\mp\mathrm{i} \kappa \pm\operatorname{i} \mu \\
1\pm 2 \mathrm{i} \mu
\end{array} \ u_1\right]-(\mu\rightarrow-\mu)\right\}  \\
\nonumber &\times \left\{ 2^{ \pm \i \mu}\left(\frac{u_2}{2}\right)^{\frac{5}{2}+p_2 \pm \i \mu}{}_2\mathcal{F}_1\left[\begin{array}{c|c}
\frac{5}{2}+p_2\pm\operatorname{i} \mu, \frac{1}{2}\mp\mathrm{i} \kappa \pm\operatorname{i} \mu \\
1\pm 2 \mathrm{i} \mu
\end{array} \ u_2\right]-(\mu\rightarrow-\mu)\right\}\\
\nonumber &+\frac{e^{\mp \i\frac{\pi}{2}  p_{12}} \Gamma\left(p_{12}+5\right) }{2^{p_{12}+5}} \sum_{n=0}^{\infty} u_1^{n+p_{12}+5}\left(1-\frac{1}{u_2}\right)^n\left(\begin{array}{c}
n+p_{12}+4 \\
n
\end{array}\right)\\
&\times \frac{1}{\mu^2+\left(\frac{5}{2}+n+p_2\right)^2}{ }_3 \mathrm{F}_2\left[\begin{array}{c|c}
1,3+n+p_2 \mp \mathrm{i} \kappa, 5+n+p_{12} \\
\frac{7}{2}+n+p_2-\mathrm{i} \mu, \frac{7}{2}+n+p_2+\mathrm{i} \mu
\end{array} \ u_1\right].
\label{I_4_pp}
\end{align}
\end{tcolorbox}
\noindent{}where the definition of ${}_2\mathcal{F}_1$ is found in Eq.~\eqref{def_mF} and we remind $p_{12}\equiv p_1+p_2$ and $\bar{p}_{12}\equiv p_1-p_2$.
These are the generalization from the results for constant mass~\cite{Qin:2022fbv,Qin:2023ejc} to those with time-dependent mass. In fact, they reproduce the results for constant mass when $g(\phi)=m_0^2$, which formally corresponds to take $\kappa=0$ (see Appendix~\ref{CML}).  

Before going to the three- and two-point correlation functions, let us comment on the momentum dependence of the seed integrals. These are the functions of $u_1$ and $u_2$, and the same is true for constant mass.
However, as already mentioned, they also depend on $v$ defined in Eq.~\eqref{phi_s} through $\mu$ and $\kappa$ (see Eq.~\eqref{def_mu}) because of the time-dependent $\sigma$-mass. Incorporating this effect into correlation functions is one of the main results of this work, and its impact on cosmological observables will be investigated in the next section.  
In the following, this additional momentum dependence of the seed integrals is explicitly denoted by $\mathcal{I}_{ \m{ab}}^{ p_1 p_2}\left(u_1, u_2, v(k_s))\right)$. 

\subsection{Soft limit}
The three-point correlation function~\eqref{3pt} is obtained by taking the soft limit $k_4\rightarrow 0 \ (u_2\rightarrow 1)$ of the seed integrals.
For the two-point correlation function~\eqref{2pt}, we further take $k_2\rightarrow 0\ (u_1\rightarrow 1)$ corresponding to the double soft limit. These limits are non-trivial at first sight because of the cancellation of some apparent divergences~\cite{Qin:2022fbv,Qin:2023ejc}, so we look at it in detail in the following.

\subsubsection*{Single soft limit}
In the single soft limit $k_4\rightarrow 0\ (u_2\rightarrow 1)$, the opposite sign seed becomes 
\begin{align}
 \mathcal{I}_{ \pm \mp}^{ p_1 p_2}\left(u_1, 1,v(k_3)\right)=\tilde{\mathcal{C}}_{ \pm \mp \mid ++} \left(\mathcal{U}_{ \pm \mid +}^{p_1}\left(u_1\right)+\mathcal{U}_{ \pm \mid -}^{p_1}\left(u_1\right)\right)\left(U_{ \mp \mid +}^{p_2}+U_{ \mp \mid -}^{p_2}\right),\label{SF_limit_pm}
\end{align}
where $\tilde{\mathcal{C}}_{ \pm \mp \mid ++}$ is given in Eq.~\eqref{Ctil_pm}, and we defined 
\begin{align}
U_{\mathrm{a} \mid \mathrm{b}}^p \equiv \operatorname{Fin.}\left\{\mathcal{U}_{\mathrm{a} \mid \mathrm{b}}^p(1)\right\} =\frac{\mathrm{i a b} \pi}{2^{5 / 2+p}}\operatorname{csch}(2 \pi \mu) \Gamma\left[\begin{array}{l}
p+\frac{5}{2}+\mathrm{i a b} \mu, \frac{1}{2}+\mathrm{i a b} \mu-\i  \mathrm{a}\kappa,-2-p+\mathrm{i a} \kappa \\
-\frac{3}{2}-p+\mathrm{i a b} \mu, \frac{1}{2}+\mathrm{i a b} \mu+\mathrm{i a} \kappa
\end{array}\right].    
\end{align}
The $\operatorname{Fin.}\{\cdots \}$ denotes the finite parts of $\mathcal{U}_{\mathrm{a} \mid \mathrm{b}}^p(1)$, and the divergence is canceled out in the combination $\mathcal{U}_{ \mp \mid +}^{p_2}\left(1\right)+\mathcal{U}_{ \mp \mid -}^{p_2}\left(1\right)$.
In the same way, for the same sign seed, we have
\begin{align}
\nonumber &\mathcal{I}_{ \pm \pm}^{ p_1 p_2}\left(u_1, 1,v(k_3)\right)\\
&=\mathcal{G}_{ \pm \pm}^{ p_1 p_2}\left(u_1, 1,v(k_3)\right)+   \left( \tilde{\mathcal{C}}_{ \pm \pm \mid ++}\ \mathcal{U}_{ \pm \mid +}^{p_1}\left(u_1\right) + \tilde{\mathcal{C}}_{ \pm \pm \mid -+}\ \mathcal{U}_{ \pm \mid -}^{p_1}\left(u_1\right) \right)\left(U_{ \pm \mid +}^{p_2}+U_{ \pm \mid -}^{p_2}\right),\label{SF_limit_pp}
\end{align}
in the limit $k_4\rightarrow 0\ (u_2\rightarrow 1)$, where $\tilde{\mathcal{C}}_{ \pm \pm \mid ++}$ and $\tilde{\mathcal{C}}_{ \pm \pm \mid -+}$ are given in Eqs.~\eqref{Ctil_pp1} and \eqref{Ctil_pp2}. For the first term $\mathcal{G}_{ \pm \pm}^{ p_1 p_2}$, only $n=0$ of the $n$-summation in Eq.~\eqref{g_pp} survives in the limit $u_2\rightarrow 1$, and we obtain the expression without the infinite summation,  
\begin{align}
&\mathcal{G}_{ \pm \pm}^{ p_1 p_2}\left(u_1, 1,v(k_3)\right)=\frac{H^2e^{\mp \frac{\pi}{2} \i p_{12}}\Gamma\left(p_{12}+5\right) u_1^{p_{12}+5}}{2^{p_{12}+5}\left[\mu^2+\left(p_2+\frac{5}{2}\right)^2\right]}{ }_3 \mathrm{F}_2\left[\begin{array}{c|c}
1,5+p_{12}, 3+p_2\mp\i \kappa \\
\frac{7}{2}+p_2-\i \mu, \frac{7}{2}+p_2+\i \mu &
\end{array}u_1\right] .   
\end{align}

In summary, after some simplification, we obtain
\begin{tcolorbox}[arc=10pt, colback=gray!10, boxrule=1pt]
\begin{align}
\nonumber &\mathcal{I}_{ \pm \mp}^{ p_1 p_2}\left(u_1, 1,v(k_3)\right)/H^2\\
\nonumber &= \left\{\frac{e^{\mp \i \frac{\pi}{2}\bar{p}_{12}} e^{\pi \kappa} \cosh \left[\pi(\mu-\kappa) \right]}{2^{\frac{5}{2}+p_2}\operatorname{sinh}(2 \pi \mu)} \Gamma\left[\begin{array}{c}
-2-p_2 \mp \mathrm{i} \kappa, \frac{5}{2}+p_2 \pm \mathrm{i} \mu \\
\frac{1}{2}+i \mu \mp \i \kappa, \frac{1}{2}-\i \mu \mp \i \kappa,-\frac{3}{2}-p_2 \pm \mathrm{i} \mu
\end{array}\right]  +(\mu\rightarrow-\mu)\right\}  \\
&\ \ \ \ \times \left\{-2^{ \pm \i \mu}\left(\frac{u_1}{2}\right)^{\frac{5}{2}+p_1 \pm \i \mu} \pi \operatorname{csch}(2 \pi \mu){}_2\mathcal{F}_1\left[\begin{array}{c|c}
p_1+\frac{5}{2} \pm \i \mu, \frac{1}{2} \pm \i \mu \mp \i \kappa \\
1\pm 2\i\mu
\end{array} \ u_1\right]+(\mu\rightarrow-\mu)\right\}, \label{I_3_pm}\\
\nonumber &\mathcal{I}_{ \pm \pm}^{ p_1 p_2}\left(u_1, 1,v(k_3)\right)/H^2\\
\nonumber &= \frac{ e^{\mp \i\frac{\pi}{2}  p_{12}} \Gamma\left(5+p_{12}\right)}{2^{5+p_{12}}\left[\mu^2+\left(p_2+\frac{5}{2}\right)^2\right]} u_1^{5+p_{12}}{ }_3 \mathrm{F}_2\left[\begin{array}{c|c}
1,3+p_2 \mp \mathrm{i} \kappa, 5+p_{12} \\
\frac{7}{2}+p_2-\mathrm{i} \mu, \frac{7}{2}+p_2+\mathrm{i} \mu
\end{array} \ u_1\right]\\
\nonumber &\ \ \ \ \mp\i\left\{\frac{e^{\mp \i \frac{\pi}{2}p_{12}} e^{\pi \kappa} \cosh \left[\pi(\mu+\kappa) \right]}{2^{\frac{5}{2}+p_2}\operatorname{sinh}(2 \pi \mu)} \Gamma\left[\begin{array}{c}
-2-p_2 \pm \mathrm{i} \kappa, \frac{5}{2}+p_2 \pm \mathrm{i} \mu \\
-\frac{3}{2}-p_2 \pm \mathrm{i} \mu
\end{array}\right]  +(\mu\rightarrow-\mu)\right\}  \\
&\ \ \ \ \times\left\{\left(\frac{u_1}{2}\right)^{\frac{5}{2}+p_1 \pm \i \mu} \frac{e^{\pi \mu}\cosh \left[\pi(\mu-\kappa) \right]}{2^{ \mp \i \mu}\operatorname{sinh}(2 \pi \mu)}{}_2\mathcal{F}_1\left[\begin{array}{c|c}
p_1+\frac{5}{2} \pm \i \mu, \frac{1}{2} \pm \i \mu \mp \i \kappa \\
1\pm 2\i\mu
\end{array} \ u_1\right]+(\mu\rightarrow-\mu)\right\},\label{I_3_pp}
\end{align}
    \end{tcolorbox}
\noindent{}in the single soft limit $k_4\rightarrow 0$. Here $\bar{p}_{12} \equiv p_1-p_2$ and $p_{12} \equiv p_1+p_2$.
These reproduces the results for constant mass when $g(\phi)=m_0^2$ or  $\kappa=0$ (see Appendix~\ref{CML}).

\subsubsection*{Double soft limit}
To consider the double soft limit, we take a limit $k_2\rightarrow 0\ (u_1\rightarrow 1)$ of the expressions in single soft limit Eqs.~\eqref{SF_limit_pm} and \eqref{SF_limit_pp}.
For the opposite sign seed, we have
\begin{align}
\mathcal{I}_{ \pm \mp}^{ p_1 p_2}\left(1, 1,v(k_3)\right)=\tilde{\mathcal{C}}_{ \pm \mp \mid ++} \left(U_{ \pm \mid +}^{p_1}+U_{ \pm \mid -}^{p_1}\right)\left(U_{ \mp \mid +}^{p_2}+U_{ \mp \mid -}^{p_2}\right).
\end{align}
In the same way as the single soft limit, divergences are canceled in a combination, $\mathcal{U}_{ \pm \mid +}^{p_1}(1)+\mathcal{U}_{ \pm \mid -}^{p_1}(1)$.
For the same sign seed, we have
\begin{align}
\nonumber &\mathcal{I}_{ \pm \pm}^{ p_1 p_2}\left(1, 1,v(k_3)\right)\\
&=\ \operatorname{Fin.}\left\{\mathcal{G}_{ \pm \pm}^{ p_1 p_2}\left(1, 1,v(k_3)\right)\right\}+  \left( \tilde{\mathcal{C}}_{ \pm \pm \mid ++}\ U_{ \pm \mid +}^{p_1} +\tilde{\mathcal{C}}_{ \pm \pm \mid -+}\ U_{ \pm \mid -}^{p_1} \right)\left(U_{ \pm \mid +}^{p_2}+U_{ \pm \mid -}^{p_2}\right),    
\end{align}
where 
\begin{align}
\operatorname{Fin.}\left\{\mathcal{G}_{ \pm \pm}^{ p_1 p_2}\left(1, 1,v(k_3)\right)\right\}=&\  \frac{H^2e^{\mp \frac{\pi}{2} \i p_{12}}\Gamma\left(p_{12}+5\right) }{2^{p_{12}+5}} {}_3\mathcal{F}_2\left[\begin{array}{c|c}
\frac{5}{2}+p_2-\i \mu, \frac{5}{2}+p_2+\i \mu,-2-p_1 \pm \i \kappa & \\
3+p_2 \pm \i \kappa, 1-p_1+p_2 
\end{array} 1\right].
\end{align}
Here we used 
\begin{align}
\operatorname{Fin.}\left\{\lim _{u \rightarrow 1}{ }_3 \mathrm{F}_2\left[\begin{array}{c|c}
a, b, c \\
d, e
\end{array} \ u\right]\right\}=\Gamma\left[\begin{array}{c}
d, e, s \\
c, a+s, b+s
\end{array}\right]{ }_3 \mathrm{F}_2\left[\begin{array}{c|c}
d-c, e-c, s  \\
a+s, b+s
\end{array}\ 1\right]  ,  
\end{align}
where $s\equiv d+e-a-b-c$. In this case, cancellation of the divergence is not within the combination of $\tilde{\mathcal{C}}_{ \pm \pm \mid ++}\ \mathcal{U}_{ \pm \mid +}^{p_1}\left(1\right) + \tilde{\mathcal{C}}_{ \pm \pm \mid -+}\ \mathcal{U}_{ \pm \mid -}^{p_1}\left(1\right) $, but with the one from $\mathcal{G}_{ \pm \pm}^{ p_1 p_2}\left(1, 1,v(k_3)\right)$.

In summary, we obtain the double soft limit ($k_4,k_2\rightarrow 0$) of the seed integrals as
\begin{tcolorbox}[arc=10pt, colback=gray!10, boxrule=1pt]
\begin{align}
\nonumber &\mathcal{I}_{ \pm \mp}^{ p_1 p_2}\left(1, 1,v(k_3)\right)/H^2\\
&=\ \frac{e^{\mp\mathrm{i} \frac{\pi}{2}\bar{p}_{12} } e^{\pi \kappa}}{2^{5+p_{12}}} \Gamma\left[\begin{array}{c}
\frac{5}{2}+p_1-\mathrm{i} \mu, \frac{5}{2}+p_1+\mathrm{i}\mu, \frac{5}{2}+p_2-\mathrm{i} \mu, \frac{5}{2}+p_2+\mathrm{i} \mu \\
3+p_1 \mp \mathrm{i} \kappa, 3+p_2 \pm \mathrm{i} \kappa
\end{array}\right] ,\label{I_pm_2} \\
\nonumber &\mathcal{I}_{ \pm \pm}^{ p_1 p_2}\left(1, 1,v(k_3)\right)/H^2\\
\nonumber &=\ \Gamma\left[\begin{array}{c}
\frac{5}{2}+p_1-\mathrm{i} \mu, \frac{5}{2}+p_1+\mathrm{i}\mu, \frac{5}{2}+p_2-\mathrm{i} \mu, \frac{5}{2}+p_2+\mathrm{i} \mu, \frac{1}{2}+\i \mu \mp \i \kappa, \frac{1}{2}-\i \mu \mp \i \kappa \\
3+p_1 \mp \mathrm{i} \kappa, 3+p_2 \pm \mathrm{i} \kappa
\end{array}\right]\\
\nonumber &\ \ \ \ \times  \frac{\pm \i e^{\mp\mathrm{i}\frac{\pi}{2} p_{12} } \left(e^{-2 \pi \mu}+e^{2 \pi \kappa}\right)}{2^{6+p_{12}}\pi} \\
&\ \ \ \ -\frac{e^{\mp\mathrm{i} \frac{\pi}{2}p_{12} } }{2^{5+p_{12}}}\Gamma\left[\begin{array}{c}
\frac{5}{2}+p_1 \pm \i \mu, \frac{5}{2}+p_2 \pm \i \mu \\
\frac{1}{2} \pm \i \mu \pm \i \kappa
\end{array}\right] {}_3\mathcal{F}_2\left[\begin{array}{c|c}
\frac{1}{2} \pm \i \mu \pm \i \kappa,5+p_{12},1 \\
\frac{7}{2}+p_1\pm\i\mu, \frac{7}{2}+p_2\pm\i\mu
\end{array}\ 1\right] .\label{I_pp_2}
\end{align}
    \end{tcolorbox}
\noindent{}As we expected, they reproduce the results for constant mass when $g(\phi)=m_0^2$ or $\kappa=0$ (see Appendix~\ref{CML}).

\section{Impact on Primordial non-Gaussianity}\label{PNG}
With the analytical expressions for the seed integrals~\eqref{I_4_pm} and \eqref{I_4_pp}, as well as their soft limits~\eqref{I_3_pm}, \eqref{I_3_pp}, \eqref{I_pm_2}, and \eqref{I_pp_2}, we are now ready to compute the inflationary correlators by using Eqs.~\eqref{4pt}, \eqref{3pt}, and \eqref{2pt}. This section will discuss phenomenology observed in the scale-dependent power spectrum and bispectrum.

To provide a clear and concrete framework for the subsequent discussion, we assume a specific interaction
\begin{align}
g(\phi)=m_0^2\left(1+\alpha\frac{\phi}{\Mpl}\right),    \label{g_ex}
\end{align}
unless explicitly stated otherwise. The first term represents a time-independent bare mass of $\sigma$, and the second term introduces an inflaton (or time) dependence. In the expression, $\alpha$ is a time-independent coupling constant, and the limit $\alpha\rightarrow 0$ realizes a constant mass scenario. We also note that Eq.~\eqref{g_ex} can be considered as the leading expansion of $e^{\alpha\phi/\Mpl}$, whose impact on the bispectrum was explored numerically in Ref.~\cite{Reece:2022soh}. In this context, the parameters $\mu$ and $\kappa$ in Eq.~\eqref{def_mu} are
\begin{align}
\mu^2(v)= \frac{m_0^2}{H^2}\left(1-\sqrt{2\epsilon}\alpha \log v-\sqrt{2\epsilon}\alpha\right)-\frac{9}{4}, \quad \kappa =  -\frac{\sqrt{2 \epsilon}\alpha m_0^2}{2H^2}. \label{def_mu_3}
\end{align}
Furthermore, to avoid a tachyonic mass for $\sigma$, we focus on the following parameter region,
\begin{align}
|\log v|\lesssim \frac{1}{\sqrt{2\epsilon}|\alpha|}.  \label{eq:cond_tachyon}  
\end{align}

\subsection{Power spectrum}
The power spectrum $P_{\zeta}$ is defined by
\begin{align}
\left\langle\zeta_{\bf{k}_1} \zeta_{\bf{k}_3} \right\rangle ' \equiv \frac{2 \pi^2}{k_1^3}  P_{\zeta},
\end{align}
where the prime denotes the omission of the momentum conservation factor. The curvature perturbation $\zeta$ is related to the inflaton fluctuation $\delta\phi$ by\footnote{In our analysis, we consider a simple scenario in which the curvature perturbation solely arises from inflaton fluctuation, without involving more complicated situations such as curvaton or modulated reheating scenarios.} 
\begin{align}
\zeta=-\frac{H}{\dot{\phi}_0} \delta\phi =\frac{1}{\sqrt{2\epsilon}\Mpl}\delta\phi.    \label{zeta}
\end{align}
{As is well established, the standard expression for the power spectrum at leading order is given by
\begin{align}
P_{\zeta}^{(0)}=\frac{H^4}{4 \pi^2 \dot{\phi}_0^2}=\frac{H^2}{8 \pi^2 \epsilon \Mpl^2},
\end{align}
and scale dependence appears as slow-roll corrections.}
In our scenario, Eq.~\eqref{2pt} combined with Eq.~\eqref{zeta} introduces a correction to the power spectrum, 
\begin{align}
P_{\zeta}^{(1)}=\frac{1}{16 \pi^2} \cdot \frac{c_2^2 }{H^2\Mpl^2}    \sum_{\m{a}, \m{b}= \pm}  \mathcal{I}_{\m{ab}}^{-2,-2}(1,1,v(k_1)),\label{P_1}
\end{align}
where $\mathcal{I}_{\m{ab}}^{p_1,p_2}$ with double soft limit are shown in Eqs.~\eqref{I_pm_2} and~\eqref{I_pp_2}. The correction from a constant mass scalar field was initially computed analytically in Ref.~\cite{Chen:2012ge} through direct integration. The same outcome was subsequently derived by solving the bootstrap equations in Ref.~\cite{Qin:2022fbv,Qin:2023ejc}.  
As a consistency check, our result~\eqref{P_1} reproduces the result in the constant mass limit, i.e., $\alpha=0$. Note that, in contrast to the constant mass case where $P_{\zeta}^{(1)}$ is scale-invariant, Eq.~\eqref{P_1} depends on the specific momentum ratio $v(k_1)=k_1/k_0$, due to the time-dependent nature of the $\sigma$-mass.

\begin{figure}[t]
\centering
\includegraphics[width=9.0cm]{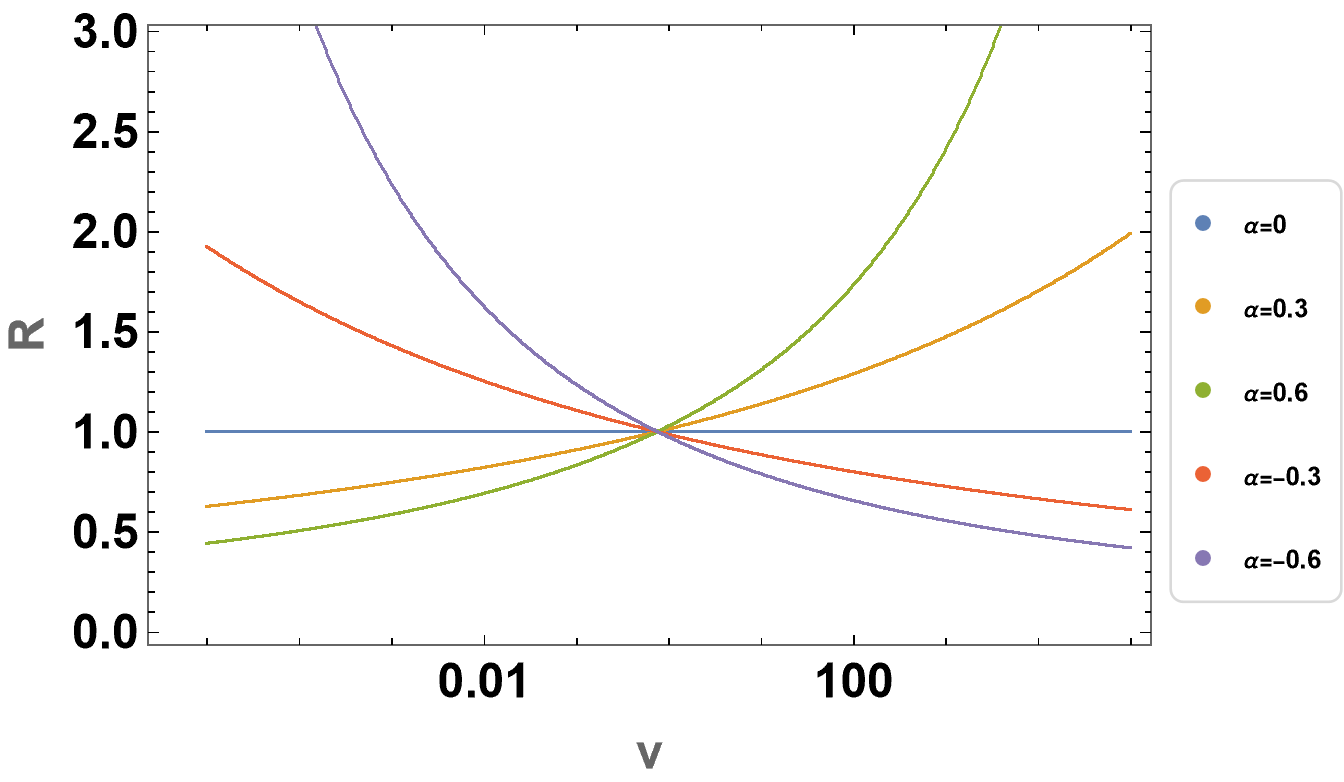}
\caption{$v(k_1)(=k_1/k_0)$-dependence of $R$. We fix $m_0=2H$ and $\epsilon =0.05/16$, and change the parameter $\alpha$ from $-0.6$ to $0.6$.}
\label{fig5}
\end{figure}

To see these effects, we introduce a ratio,
\begin{align}
R\equiv \frac{ P_{\zeta}^{(1)}}{P_{\zeta}^{(1)}(\alpha=0)},   
\end{align}
with the denominator representing the correction to the power spectrum in case of a constant mass. In Fig.~\ref{fig5}, we plot the variation of $R$ as a function of $v(k_1)$. Our choice of parameter includes a fixed $m_0=2H$ and $\epsilon =0.05/16$, with $\alpha$ ranging from $-0.6$ to $0.6$. 
Depending on the values of $v$ and $\alpha$, the correction to the power spectrum can either be increased or decreased.
{Furthermore, we can analytically estimate the leading $v$-dependence of the power spectrum $P_\zeta^{(1)}$ as
\begin{align}
    \pdiff{P_\zeta^{(1)}}{v}=f_P(m_0)\frac{\sqrt{\epsilon}\alpha}{v}+\order{\epsilon},
    \label{scale_P}
\end{align}
where $f_P(m_0)$ is a function of $m_0$. 
This $v$ dependence coincides with the behavior derived in general single field inflation scenarios (see e.g.,~\cite{Chen:2006nt}). Note that, although the $v$-dependence is identical, the leading order of slow-roll parameter dependence is different from the general single field case where the leading order is $\order{\epsilon,\eta}$. 
We also note that the $v$-dependence in Eq.~\eqref{scale_P} is universal as slow-roll correction, while the effect of $\sigma$-field on the inflation background is highly model-dependent. For instance, the power spectrum can often be expanded as $P_\zeta=P_{\zeta *}+\varepsilon(t_\mathrm{exit}-t_*)+\order{\varepsilon^2}\sim P_{\zeta *}+\varepsilon\log{(k/k_*)}$ where $\varepsilon$ represents the leading slow-roll order of the theory, e.g., $\varepsilon=\order{\sqrt{\epsilon}}$ in our case.}

\subsection{Bispectrum}
The bispectrum is characterized by the so-called shape function $S$ defined by
\begin{align}
\left\langle\zeta_{\bf{k}_1} \zeta_{\bf{k}_2} \zeta_{\bf{k}_3}\right\rangle ' \equiv (2 \pi)^4  \frac{P_{\zeta}^2}{\left(k_1 k_2 k_3\right)^2} S.
\end{align}
Combining Eq.~\eqref{3pt} with Eq.~\eqref{zeta}, we derive the shape function
\begin{align}
S =&\ \frac{1}{(2 \pi)^4} \cdot \frac{1}{P_{\zeta}^2} \cdot(2 \epsilon \Mpl^2)^{-\frac{3}{2}} \cdot\left(-2 c_2 c_3\right) \cdot \frac{H}{8} \sum_{\m{a}, \m{b}= \pm} \left[\frac{k_1 k_2}{k_3^2} \mathcal{I}_{\m{a b}}^{0,-2}\left(\frac{2 k_3}{k_{123}}, 1,v(k_3)\right)+{\rm{2 per.}}\right],\label{eq:s}
\end{align}
where the single soft limit of $\mathcal{I}_{\m{ab}}^{p_1p_2}$ is described in Eqs.~\eqref{I_3_pm} and~\eqref{I_3_pp}. In the subsequent subsections, we explore the behavior of the shape function in detail.

\subsubsection{Cosmological collider signal at squeezed configuration}

Let us consider the configuration $k_1=k_2$, and define $x\equiv k_3/k_{1,2} $ and $v\equiv k_1/k_0$. Consequently, we can express Eq.~\eqref{eq:s} as
\begin{align}
{S} =\frac{1}{(2 \pi)^4} \cdot \frac{1}{P_{\zeta}^2} \cdot&(2 \epsilon \Mpl^2)^{-\frac{3}{2}} \cdot\left(-2 c_2 c_3\right) \cdot \frac{H}{8}\notag\\
&\times\sum_{\m{a}, \m{b}= \pm}\left[\frac{1}{x^2}\mathcal{I}_{\m{a b}}^{0,-2}\left(\frac{2 x}{2+x}, 1,vx\right)+2 x \mathcal{I}_{\m{a b}}^{0,-2}\left(\frac{2}{2+x}, 1,v\right)\right] .  \label{Stil2} 
\end{align}
In this expression, the squeezed limit $k_3\ll k_{1,2}$ corresponds to $x\ll 1$, while the equilateral one $k_1\simeq k_2\simeq k_3$ to $x\simeq 1$. 

\begin{figure}[t]
 \begin{minipage}{0.5\hsize}
  \begin{center}
   \includegraphics[width=70mm]{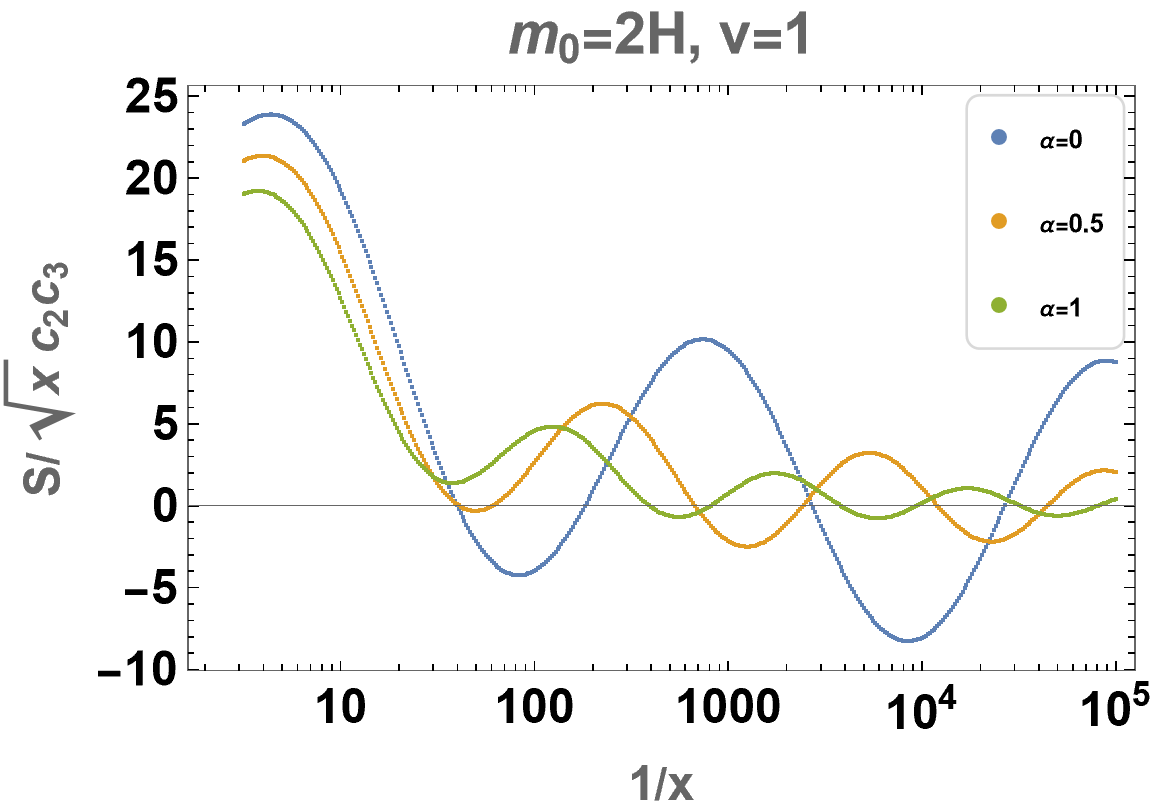}
  \end{center}
 \end{minipage}
 \begin{minipage}{0.5\hsize}
  \begin{center}
   \includegraphics[width=70mm]{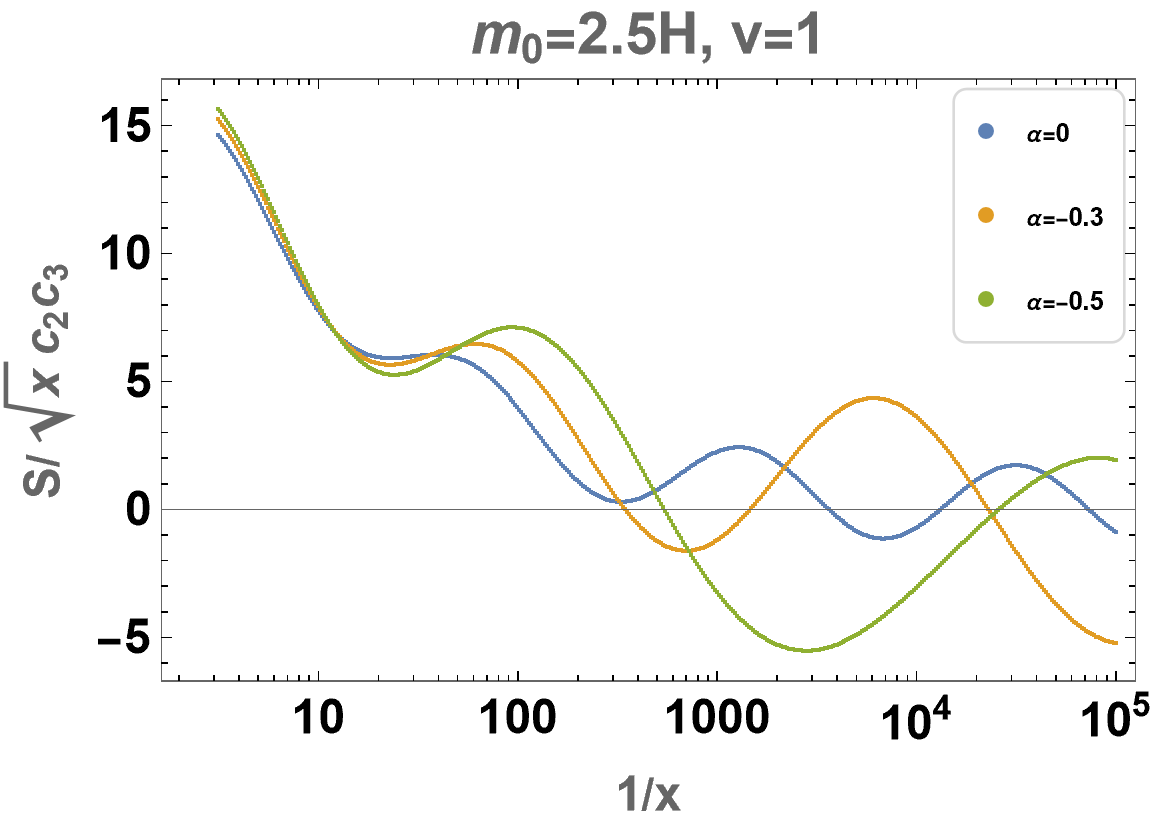}
  \end{center}
 \end{minipage}
\caption{${S}/c_2c_3\sqrt{x}$ as a function of $x^{-1}$ with several choices of $\alpha$. {\it{Left}} : Case with $\alpha \geq 0$, that is changed as  $0$ (blue), $0.5$ (orange), and $1$ (green). {\it{Right}} : Case with $\alpha \leq 0$, that is changed as $0$ (blue), $-0.3$ (orange), and $-0.5$ (green).}
  \label{fig4}
\end{figure}

The squeezed limit of the bispectrum is of particular interest in cosmological collider physics, as specific oscillatory patterns can sharply appear in this limit. In Fig.~\ref{fig4}, we present ${S}/c_2c_3\sqrt{x}$ as a function of $x^{-1}$, with varying values of $\alpha$. Our parameter choices are $v=1$, $m_0=2H\ (2.5H)$ for $\alpha>0 (<0)$, $\epsilon=0.05/16$, and $\Mpl/H=10^{5}$.\footnote{We use these values of $\epsilon$ and $\Mpl/H$ in all figures below.
}
It is readily apparent that both the amplitude and the oscillation frequency exhibit strong dependence on $k_1/k_3$. For $\alpha>0$, the amplitude experiences damping and the frequency diminish in the squeezed configuration ($x\ll 1$). The opposite behavior is observed for $\alpha<0$. This feature is more prominent for an increased value of $|\alpha|$, and the deviation from a constant mass signal is sufficiently significant to observe.
Note that these results are consistent with previous works, including calculations within super-horizon approximation~\cite{Wang:2019gok} and numerical simulation~\cite{Reece:2022soh}.

The physical interpretation of this signal is clear.
In the squeezed limit $x\ll 1$, we are essentially observing the correlation between a long wavelength mode $k_3$ and short ones $k_{1,2}$. The long mode exits the horizon earlier than the short ones and evolves on the super-horizon scale $|k_3\tau|\ll 1$, schematically represented by
\begin{align}
v_{k_3}(\tau) \sim
(-\tau)^{\frac{3}{2}}
\left[
(-\tau)^{\i\mu}
+
e^{-\pi\mu}(-\tau)^{-\i\mu}
\right]
, 
\end{align}
where we showed only $\tau$-dependence of the mode function, disregarding the relative phase of the positive and negative frequency modes. The relative amplitude $e^{-\pi\mu}$ of the positive and negative frequency modes is nothing but (the square root of) the Boltzmann factor responsible for on-shell particle creation relevant to the CC signal. In the present case, $\mu$ exhibits scale dependence due to the horizon crossing time difference.
For $\alpha>0\ (<0)$, the mass of $\sigma $ becomes heavier (lighter) than the constant mass scenario due to its time-dependent nature. Consequently, the mode function is more (less) suppressed by the Boltzmann factor $e^{-\pi \mu}$ compared to the constant mass case, leading to a smaller (larger) amplitude and a shorter (longer) wavelength in the oscillatory signal of the bispectrum. Furthermore, the oscillation frequency is scale-dependent accordingly.

\subsubsection{Distinction of couplings with inflaton}

Recall that the time-dependent mass appears because the non-derivative couplings break the shift symmetry; it is not easy to obtain similar effects from derivative couplings. Therefore, when we observe the specific behavior in the CC signal, such as the damping/enhancement of amplitude and the shrinking/spreading of wavelength in the squeezed limit, it is attributed to these non-derivative (direct) interactions. Moreover, detailed observations of the time-dependent Boltzmann factor associated with non-derivative couplings between the inflaton and the $\sigma$-field allow us to distinguish the form of the interactions. This is obtained from distinctive characteristics in the tail of the ``scaling'' behavior in the CC signal. We will delve deeper into this aspect in the following discussion.

\begin{figure}[t]
 \begin{minipage}{0.5\hsize}
  \begin{center}
   \includegraphics[width=70mm]{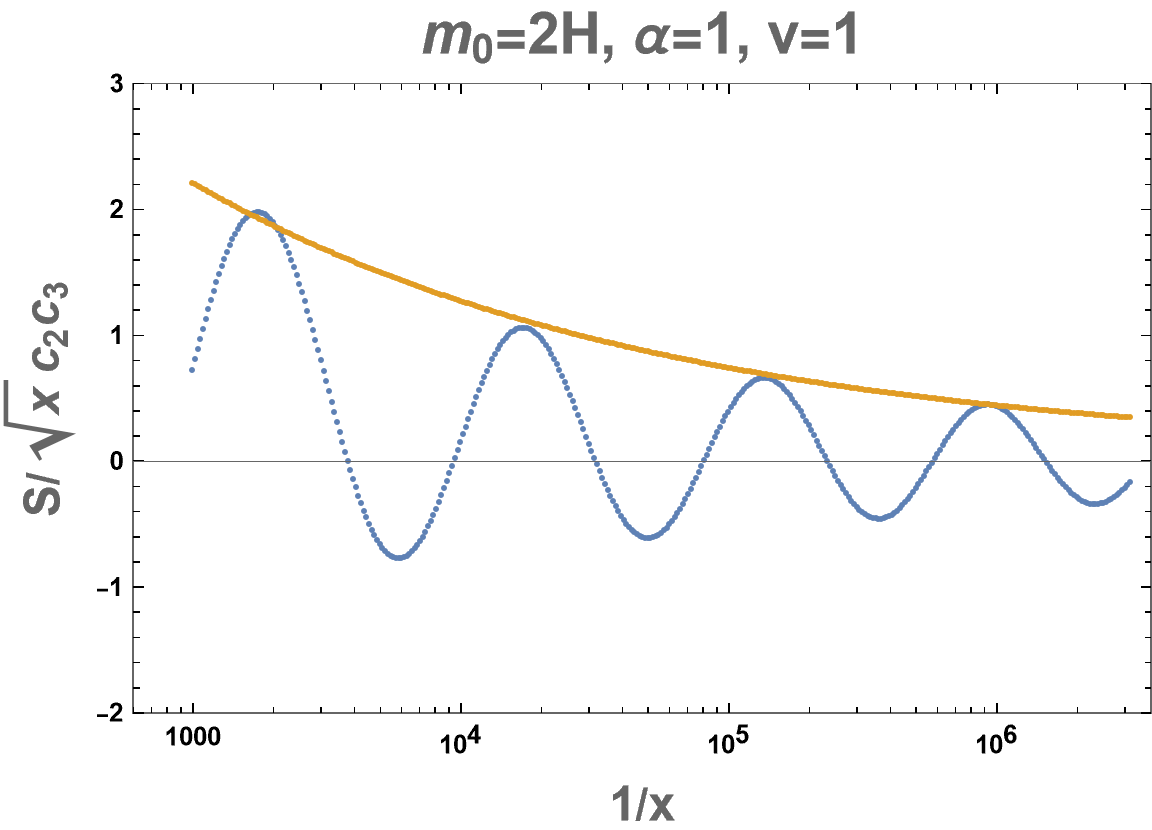}
  \end{center}
 \end{minipage}
 \begin{minipage}{0.5\hsize}
  \begin{center}
   \includegraphics[width=70mm]{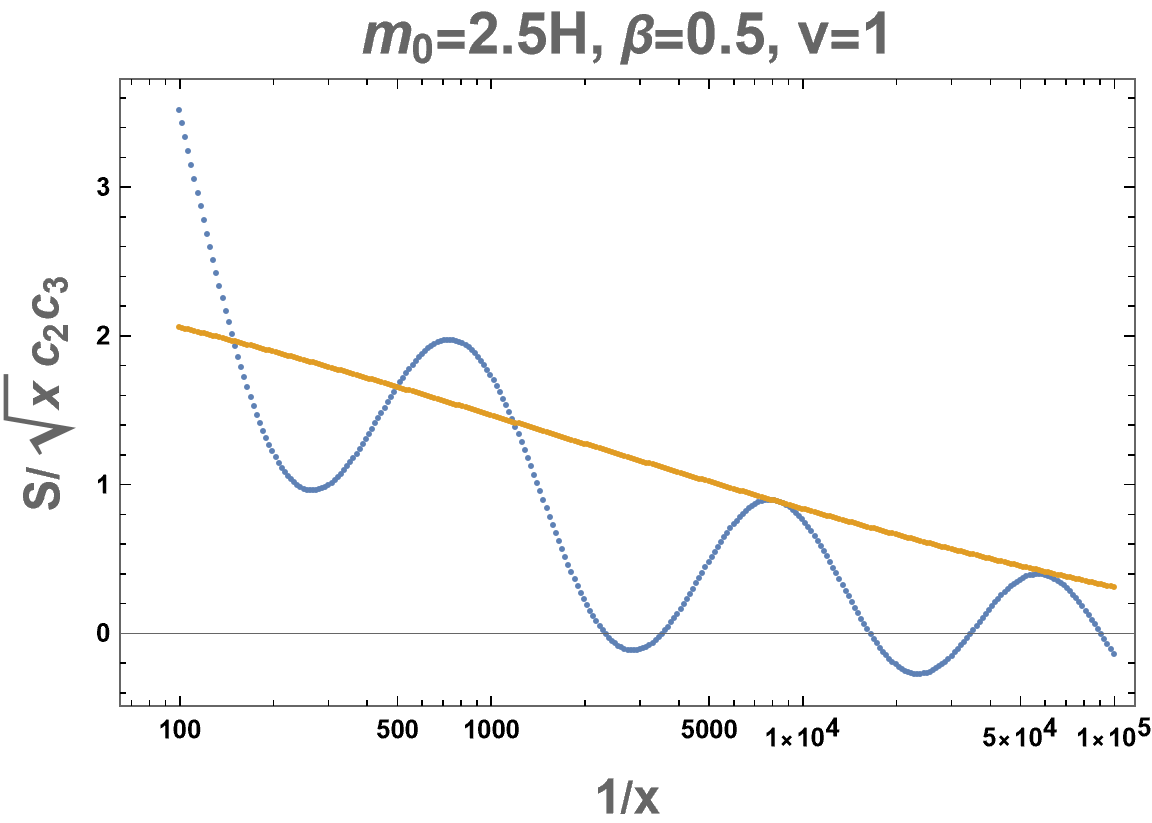}
  \end{center}
 \end{minipage}
 \caption{{Fitting of the full shape~\eqref{Stil2} (blue line) with scaling function $e^{-\pi\mu(x)}$ (orange line).} {\it{Left}} : Linear case~\eqref{g_ex} with Eq.~\eqref{s}. {\it{Right}} : Quadratic case~\eqref{quad} with \eqref{s_q}. 
Parameters are chosen as described in the figure.}
 \label{fig:scaling}
\end{figure}

In the shape function~\eqref{Stil2} with Eqs.~\eqref{I_3_pm} and~\eqref{I_3_pp}, one can extract the Boltzmann factor for $\mu\gtrsim 1$ with 
\begin{align}
|\Gamma(a+\i b)| \sim e^{-\pi|b| / 2}, \quad|b|\gtrsim 1, \quad a, b \in \mathbb{R},    
\end{align}
This reveals that the shape function behaves as
\begin{align}
S/\sqrt{x} \sim e^{-\pi \mu (x)} x^{\pm \i \mu}+{\rm{c.c.}},    
\label{eq:squi}
\end{align}
in the squeezed limit, $x\ll1$\footnote{In Eq.~\eqref{Stil2}, the dominance of the first term over the second one holds true for $x\ll1$.}. Regarding the Boltzmann factor $e^{-\pi \mu (x)}$, it is crucial to emphasize that $\mu$ exhibits $x$-dependence due to the time-dependent mass arising from the coupling to the inflaton, unlike the constant mass situation\footnote{{One may expect the existence of so-called chemical potential~\cite{Wang:2019gbi,Bodas:2020yho} from the expression of the mode function~\eqref{sol_v}. However, the approximation~\eqref{exp_phi} is valid only in the parameter region where the effect of chemical potential is negligible, $e^{\pi(\kappa-\mu)}\sim e^{-\pi\mu}$, which is our focus throughout this paper.}}. Consequently, the amplitude of the CC signal exhibits a scaling behavior in addition to the well-known oscillation, as observed in the previous subsection
(see Fig.~\ref{fig4}).
In case of the linear coupling~\eqref{g_ex}, we can estimate the scaling \begin{align}
 e^{-\pi \mu (x)}\sim e^{-\pi m_0/H}\cdot \left(\frac{v}{x}\right)^{-  \pi\frac{m_0}{H} \sqrt{\frac{\epsilon}{2}} \alpha}, \label{s}    
\end{align}
as a function of $x$. The left figure of Fig.~\ref{fig:scaling} illustrates {the scaling tail of the oscillatory part (orange line) estimated in Eq.~\eqref{s} on the top of the full shape including the background part~\eqref{Stil2}.} Eq.~\eqref{s} essentially characterizes the scaling behavior in the squeezed limit ($x\ll 1$). 

Remarkably, the scaling behavior exhibits dependence on the coupling between the inflaton and the massive field. For example, in the case of a quadratic coupling,
\begin{align}
g(\phi)=m_0^2\left(1+\beta \frac{\phi^2}{\Mpl^2}\right),  \label{quad}
\end{align}
we obtain
\begin{align}
 e^{-\pi \mu (x)}\sim e^{-\pi m_0/H}\cdot e^{- \pi \frac{m_0}{H} \cdot \beta \epsilon\left(\log  v x\right)\left(\log v x+2\right)} \label{s_q}   
\end{align}
instead of Eq.~\eqref{s}. This new formulation represents the amplitude of the corresponding CC signal, as demonstrated in the right panel of Fig.~\ref{fig:scaling}, and the scaling behavior ($x$-dependence) differs significantly from the linear case.
Therefore, in principle, we can distinguish the coupling $g(\phi)$ through careful observations and analysis of these scaling behaviors.

It is straightforward to obtain a similar formula for more generic couplings. For example, for a coupling,
\begin{align}
g(\phi)=m_0^2\left(1+\alpha_{(n)}\frac{\phi^n}{\Mpl^{n}}\right),
\end{align}
with $n$ being positive integers, we have the following scaling function,
\begin{align}
 e^{-\pi \mu (x)}\sim e^{-\pi m_0/H}\cdot e^{- \pi \frac{m_0}{H} \cdot \frac{\alpha_{(n)}}{2} (-\sqrt{2\epsilon})^n\left(\log  v x\right)^{n-1}\left(\log v x+n\right)},  \label{BF_n} 
\end{align}
which predicts a different scaling depending on $n$.

\subsubsection{Equilateral configuration}
While the CC signal in the squeezed configuration is quite essential to extract information about $\sigma$-field, the bispectrum~\eqref{Stil2} exhibits a peak (dominant contribution) at the equilateral limit $x\simeq 1$. Hence, it is equally valuable to investigate the behavior around the equilateral configuration. In our scenario, the magnitude of the shape function or the non-Gaussianity parameter~$f_{\text{NL}}$ in the equilateral limit is estimated by using
\begin{align}
f_{\text{NL}}^{\text{equi}}\sim c_2 c_3 \times \order{10},
\end{align}
for $\alpha\sim \order{1}$ and $m_0\sim \order{H}$. 

\begin{figure}[t]
 \begin{minipage}{0.5\hsize}
  \begin{center}
   \includegraphics[width=7.0cm]{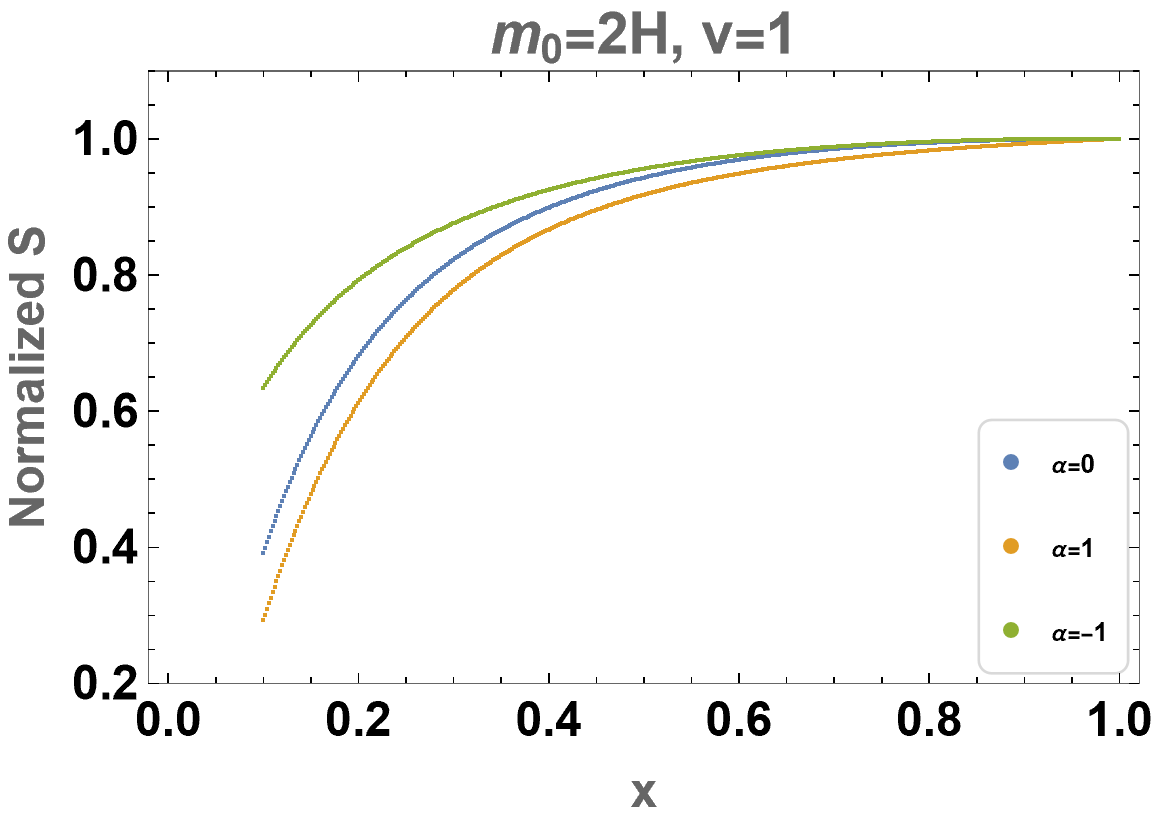}
  \end{center}
 \end{minipage}
 \begin{minipage}{0.5\hsize}
  \begin{center}
   \includegraphics[width=70mm]{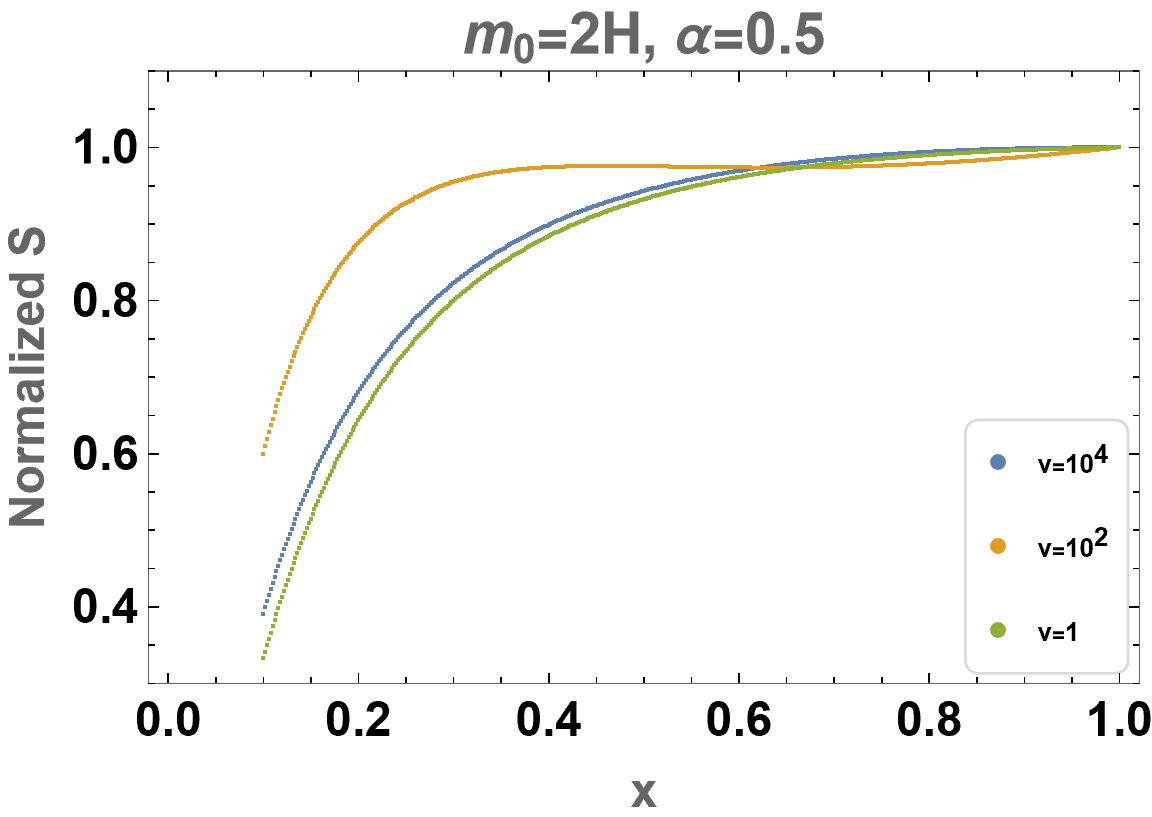}
  \end{center}
 \end{minipage}
 \caption{The shape function $S$ as the function of $x$. Here ${S}$ is normalized to have value $1$ at the equilateral limit $x=1$. \textit{Left:} $\alpha$-dependence with $m_0=2H$ and $v=1$ fixed.  \textit{Right:} $v$-dependence with $m_0=2H$ and $\alpha=0.5$ fixed.}
 \label{fig7}
\end{figure}

In Fig.~\ref{fig7}, we present plots of ${S}$ for $x\gtrsim 0.1$ (equilateral configuration) varying the interaction strength $\alpha$ (left figure) and the additional scale dependence $v$ (right figure). 
We find that the values of $\alpha$ and $v$ change the scaling behavior of the bispectrum towards the equilateral configuration. {Furthermore, the leading $v$-dependence of the shape function $S$ at $x=1$ is obtained in the same way as the power spectrum. Specifically, $\partial{S}/\partial{v} = f_S(m_0) \sqrt{\epsilon}\alpha/v +\order{\epsilon}$ where $f_S(m_0)$ is a function of $m_0$. Unfortunately, for reasons discussed in the subsection of the power spectrum, it is difficult to extract information about the theory from the scale dependence at the exact equilateral configuration.}

\section{Summary}\label{summary}
Our paper provided analytical formulae for the inflationary correlators, specifically two-, three-, and four-point correlation functions, in the presence of a massive scalar field with time-dependent mass. The results are summarized in Eqs.~\eqref{I_4_pm} and \eqref{I_4_pp} for four-point correlators, Eqs.~\eqref{I_3_pm} and \eqref{I_3_pp} for three-point, and Eqs.~\eqref{I_pm_2} and \eqref{I_pp_2} for two-point. The time dependence of mass naturally arises from couplings to the inflaton, and its effects can be significant when the couplings do not respect the shift symmetry of the inflaton, i.e., in the case of non-derivative couplings. The mode function of the massive field~\eqref{sol_v} is analytically given by the Whittaker function under the linear order expansion around the time of horizon crossing for the time-dependence inflaton background, which characterizes the scale dependence of the signal. The analytical formulae for the correlators obtained from the bootstrap method include both the signal (oscillatory) parts and the background (non-oscillatory) parts of the bispectrum as graphically shown in Fig.~\ref{fig4} and \ref{fig7}. The non-oscillatory part of the signal is also crucial for achieving precision in the cosmological collider (CC) program.

As an application, the phenomenological impact of the time dependence on the primordial curvature power spectrum and bispectrum were investigated. Our analysis revealed that the CC signal exhibits the damping/enhancement of amplitude due to the time-dependent (or scale-dependent) Boltzmann factor. The change in amplitude in the CC signal can be readily observed and provide evidence for the time-dependent mass of the massive field. Although this behavior has been qualitatively confirmed in previous numerical studies such as Ref.~\cite{Reece:2022soh}, we quantitatively parameterized this scaling of the amplitude in a simple form~\eqref{BF_n} based on our analytical formulae. By probing the scaling behavior in the tail of the signal, we are able to distinguish couplings between the inflaton and the massive field, which provides a pathway to explore the unknown inflaton sector through the CC signal.  

It should be noted that we made an approximation by linearizing the time dependence of the inflaton and omitting the higher-order time dependence. This is merely the approximation made in our study. While extensions beyond linear order present technical challenges in obtaining analytical representations of non-Gaussianity, there are interesting scenarios beyond the scope of our setup, such as CC signals from tachyonic phase transitions~\cite{Wang:2018tbf}. We plan to explore these topics in future work.


\subsection*{Acknowledgments}
{We thank Lucas Pinol, Yi Wang and Zhong-Zhi Xianyu for valuable discussions and their hospitalities when F.S. stayed at their groups, Dong-Gang Wang and S\'ebastien Renaux-Petel for valuable discussions in Gravity 2023, YITP and COSMO 2023, IFT, and Yuhang Zhu for valuable discussions.}
S.A. is supported by IBS under the project code, IBS-R018-D1.
T.N. is supported in part by JSPS KAKENHI Grant No. 20H01902 and No. 22H01220, and MEXT KAKENHI Grant No. 21H05184 and No. 23H04007. 
F.S. is supported by financial aid from the Center for the Theoretical Physics of the Universe, Institute for Basic Science, financial aid from the Advanced Research Center for Quantum Physics and Nanoscience, Tokyo Institute of Technology, and JSPS Grant-in-Aid for Scientific Research No. 23KJ0938. 
M.Y. is supported by IBS under the project code IBS-R018-D3 and JSPS Grant-in-Aid for Scientific Research No. JP21H01080.
\begin{appendix}

\section{Seed integral with the Mellin--Barnes representation}\label{MB}
In this appendix, we present a direct integration method for the seed integral~\eqref{seed}, based on the technique utilized in Ref.~\cite{Qin:2022fbv,Qin:2023ejc} which makes use of the Mellin--Barnes representation of Whittaker function,  
\begin{align}
 \mathrm{W}_{\kappa, \nu}(z)&=e^{ z / 2} \int_{-\mathrm{i} \infty}^{+\mathrm{i} \infty} \frac{\mathrm{d} s}{2 \pi \mathrm{i}} \Gamma\left[\begin{array}{c}
s-\nu, s+\nu \\
s-\kappa+\frac{1}{2}
\end{array}\right] z^{-s+1 / 2} ,\label{W_1}\\
 \mathrm{~W}_{\kappa, \nu}( z)&=e^{- z / 2} \int_{-\mathrm{i} \infty}^{+\mathrm{i} \infty} \frac{\mathrm{d} s}{2 \pi \mathrm{i}} \Gamma\left[\begin{array}{c}
s-\nu, s+\nu,-s-\kappa+\frac{1}{2} \\
\frac{1}{2}-\kappa-\nu, \frac{1}{2}-\kappa+\nu
\end{array}\right]z^{-s+1 / 2} .\label{W_2}
\end{align}
As a reminder, the notation used in the paper is as follows:
\begin{align}
{ }_p\mathcal{F}_q\left[\begin{array}{c|c}
a_1, \cdots, a_p \\
b_1, \cdots, b_q
\end{array} \ z\right]\equiv \Gamma\left[\begin{array}{c}
a_1, \cdots, a_p \\
b_1, \cdots, b_q
\end{array}\right]{ }_p \mathrm{F}_q\left[\begin{array}{c|c}
a_1, \cdots, a_p \\
b_1, \cdots, b_q
\end{array} \ z\right] , 
\end{align}
where ${ }_p\mathrm{F}_q$ represents the (generalized) hypergeometric function, and the products of gamma functions are abbreviated as
\begin{align}
&\Gamma\left[z_1, \cdots, z_m\right]  \equiv \Gamma\left(z_1\right) \cdots \Gamma\left(z_m\right), \\
&\Gamma\left[\begin{array}{c}
z_1, \cdots, z_m \\
w_1, \cdots, w_n
\end{array}\right]  \equiv \frac{\Gamma\left(z_1\right) \cdots \Gamma\left(z_m\right)}{\Gamma\left(w_1\right) \cdots \Gamma\left(w_n\right)} .
\end{align}

\subsubsection*{Opposite sign seed $\mathcal{I}_{\pm \mp}^{ p_1 p_2}$}
To compute the opposite sign seed integral $\mathcal{I}_{\pm \mp}^{ p_1 p_2}$, we first express the propagators $D_{\pm \mp}$ using Eq.~\eqref{W_2}:
\begin{align}
\nonumber D_{\pm \mp}\left(k ; \tau_1, \tau_2\right) =&\ \frac{e^{\pi \kappa}}{2k}H^2(\tau_1\tau_2) W_{\pm\i \kappa, \mp\i \mu}(\mp2 \i k \tau_1)W_{\mp \i \kappa, \pm \i \mu}(\pm 2 \i k \tau_2)\\
\nonumber =&\ \frac{e^{\pi\kappa}}{2 \pi^2} H^2 e^{\pm\i k\left(\tau_1-\tau_2\right)}\left[\cosh \left(2\pi \kappa\right)+\cosh (2 \pi \mu)\right] \\
\nonumber&\times \int_{-\mathrm{i} \infty}^{+\mathrm{i} \infty} \frac{\mathrm{d} s_1}{2 \pi \i} \frac{\mathrm{d} s_2}{2 \pi \i} e^{\mp \mathrm{i} \pi\left(s_1-s_2\right) / 2}\left(2 k\right)^{-s_{12}}\left(-\tau_1\right)^{-s_1+3/ 2}\left(-\tau_2\right)^{-s_2+3 / 2}\\
&\times \Gamma\left[-s_1+\frac{1}{2} \mp\i \kappa,-s_2+\frac{1}{2} \pm \i \kappa, s_1-\mathrm{i} \mu, s_1+\mathrm{i} \mu, s_2-\mathrm{i} \mu, s_2+\mathrm{i} \mu\right],\label{MB_D_pm}
\end{align}
where $s_{12}=s_1+s_2$. Substituting this expression into $\mathcal{I}_{\pm \mp}^{ p_1 p_2}$ enables us to carry out the $\tau_{1,2}$-integral, resulting in
\begin{align}
\nonumber  \mathcal{I}_{\pm \mp}^{ p_1 p_2}=   &\ \frac{e^{\pi\kappa(k_s)}}{2 \pi^2} H^2 \left[\cosh \left(2\pi \kappa(k_s)\right)+\cosh (2 \pi \mu)\right] \left(\frac{u_1}{2}\right)^{5 / 2+p_1}\left(\frac{u_2}{2}\right)^{5 / 2+p_2} e^{\mp \mathrm{i}\pi\left(p_1-p_2\right)  / 2}\\
\nonumber & \times \int_{-\mathrm{i} \infty}^{\mathrm{i} \infty} \frac{\mathrm{d} s_1}{2 \pi \mathrm{i}} \frac{\mathrm{d} s_2}{2 \pi \mathrm{i}} u_1^{-s_1} u_2^{-s_2}\Gamma\left[p_1+\frac{5}{2}-s_1, p_2+\frac{5}{2}-s_2\right]\\
 &\times \Gamma\left[-s_1+\frac{1}{2} \mp \i \kappa(k_s),-s_2+\frac{1}{2} \pm \i \kappa(k_s), s_1-\mathrm{i} \mu, s_1+\mathrm{i} \mu, s_2-\mathrm{i} \mu, s_2+\mathrm{i} \mu\right],\label{Ipm}
\end{align}
where 
\begin{align}
u_1\equiv \frac{2 r_1}{1+r_1}, \quad  u_2\equiv \frac{2 r_2}{1+r_2},   
\end{align}
and we explicitly denote $k_s$ dependence in $\kappa(k_s)$ (see Eq.~\eqref{def_mu} with Eq.~\eqref{phi_s}). 
The integration over $s_i$ can be computed using the residue theorem. We close the contours from the left with a large semicircle and pick up the poles $s_i=-n_i\pm \i \mu $ with $n_i$ to be non-negative integers. This process allows us to express the seed integral as
\begin{align}
\nonumber \mathcal{I}_{\pm \mp}^{ p_1 p_2}=&\ \frac{e^{\pi\kappa(k_s)}}{2 \pi^2} H^2 \left[\cosh \left(2\pi \kappa(k_s)\right)+\cosh (2 \pi \mu)\right] \left(\frac{u_1}{2}\right)^{5 / 2+p_1}\left(\frac{u_2}{2}\right)^{5 / 2+p_2} e^{\mp \mathrm{i}\pi\left(p_1-p_2\right)  / 2}\\
\nonumber &\times \sum_{n_1=0}^{\infty}\left\{\frac{(-1)^{n_1}}{n_{1} !} u_1^{n_1+\i \mu} \Gamma\left[n_1+p_1+\frac{5}{2}+\i \mu, n_1+\i \mu+\frac{1}{2} \mp \i \kappa(k_s),-n_1-2 \i \mu\right]+(\mu \rightarrow-\mu)\right\}\\
&\times \sum_{n_2=0}^{\infty}\left\{\frac{(-1)^{n_2}}{n_{2} !} u_2^{n_2+\i \mu} \Gamma\left[n_2+p_2+\frac{5}{2}+\i \mu, n_2+\i \mu+\frac{1}{2} \pm \i \kappa(k_s),-n_2-2 \i \mu\right]+(\mu \rightarrow-\mu)\right\},
\end{align}
and the summations can be performed explicitly, leading to
\begin{align}
\nonumber \mathcal{I}_{\pm \mp}^{ p_1 p_2}=&\ \frac{e^{\pi\kappa(k_s)}}{2 \pi^2} H^2 \left[\cosh \left(2\pi \kappa(k_s)\right)+\cosh (2 \pi \mu)\right]  e^{\mp \mathrm{i}\pi\left(p_1-p_2\right)  / 2}\\
&\times\left[\mathrm{G}_{\mp \i \kappa(k_s), \mu}^{p_1}\left(u_1\right) u_1^{\mathrm{i} \mu}+\mathrm{G}_{\mp \i \kappa(k_s), -\mu}^{p_1}\left(u_1\right) u_1^{-\mathrm{i} \mu}\right]\left[\mathrm{G}_{\pm \i \kappa(k_s), \mu}^{p_2}\left(u_2\right) u_2^{\mathrm{i} \mu}+\mathrm{G}_{\pm \i \kappa(k_s), -\mu}^{p_2}\left(u_2\right) u_2^{-\mathrm{i} \mu}\right],     
\end{align}
where
\begin{align}
\mathrm{G}_{\pm \i\kappa(k_s), \mu}^p(u) \equiv  \i\pi \operatorname{csch}(2 \pi \mu)\left(\frac{u}{2}\right)^{5 / 2+p} {}_2\mathcal{F}_1\left[\begin{array}{c|c}
\frac{5}{2}+p+\mathrm{i} \mu, \frac{1}{2}+\mathrm{i} \mu \pm \mathrm{i} \kappa(k_s) \\
1+2 \mathrm{i} \mu
\end{array}\ u\right].    \label{def_G}
\end{align}
In particular, in the hierarchical collapsed limit $u_1 \ll u_2 \ll 1$, we obtain
\begin{align}
\lim _{u_1 \ll u_2 \ll 1} \mathcal{I}_{\pm \mp}^{p_1 p_2} =\sum_{\m{a, b}= \pm} \tilde{\mathcal{C}}_{ \pm \mp \mid \m{a b}}\  \tilde{\mathcal{U}}_{ \pm \mid \m{a}}^{p_1}\left(u_1\right) \tilde{\mathcal{U}}_{\mp \mid \mathrm{b}}^{p_2}\left(u_2\right),  \label{Ipm_sq}
\end{align}
where 
\begin{align}
\tilde{\mathcal{C}}_{ \pm \mp \mid ++} =\tilde{\mathcal{C}}_{ \pm \mp \mid +-}=\tilde{\mathcal{C}}_{ \pm \mp \mid -+}=\tilde{\mathcal{C}}_{ \pm \mp \mid --}=\frac{e^{\pi \kappa}}{2 \pi^2} H^2 \left[\cosh \left(2\pi \kappa(k_s)\right)+\cosh (2 \pi \mu)\right]  e^{\mp \mathrm{i}\pi\left(p_1-p_2\right)  / 2} , 
\end{align}
and
\begin{align}
\tilde{\mathcal{U}}_{\mathrm{a} \mid \mathrm{b}}^p(u)  =\operatorname{iab} 2^{\mathrm{iab} \mu} \pi \operatorname{csch}(2 \pi \mu)\left(\frac{u}{2}\right)^{5 / 2+p+\mathrm{iab} \mu} \Gamma\left[\begin{array}{c}
\frac{5}{2}+p+\operatorname{iab} \mu, \frac{1}{2}-\operatorname{ia}\kappa +\operatorname{iab} \mu \\
1+2 \operatorname{iab} \mu
\end{array}\right] . \label{def_util}
\end{align}

\subsubsection*{Same sign seed $\mathcal{I}_{\pm \pm}^{ p_1 p_2}$}

For the same sign seed integral, we adopt a division of the propagators in the same way employed in~\cite{Qin:2022fbv,Qin:2023ejc},
\begin{align}
 D_{\pm \pm}\left(k ; \tau_1, \tau_2\right)=D_{\gtrless}\left(k ; \tau_1, \tau_2\right)+\left[D_{\lessgtr}\left(k ; \tau_1, \tau_2\right)-D_{\gtrless}\left(k ; \tau_1, \tau_2\right)\right] \theta\left(\tau_2-\tau_1\right) ,   
\end{align}
and accordingly, we define
\begin{align}
\mathcal{I}_{\pm \pm}^{p_1 p_2}=\mathcal{I}_{\pm \pm, \mathrm{F},>}^{p_1 p_2}+\mathcal{I}_{\pm \pm, \mathrm{TO},>}^{p_1 p_2} , \quad\left(r_1<r_2\right)  
\end{align}
where the factorized (F) integral $\mathcal{I}_{ \pm \pm, \mathrm{F},>}^{p_1 p_2}$ and the time-ordered (TO) integral $\mathcal{I}_{ \pm \pm,  \mathrm{TO},>}^{p_1 p_2}$ are expressed as
\begin{align}
\mathcal{I}_{ \pm \pm, \mathrm{F},>}^{p_1 p_2} =& -k_s^{5+p_{12}} \int_{-\infty}^0 \mathrm{~d} \tau_1 \mathrm{~d} \tau_2\left(-\tau_1\right)^{p_1}\left(-\tau_2\right)^{p_2} e^{ \pm \mathrm{i}\left(k_{12} \tau_1+k_{34} \tau_2\right)} D_{\gtrless}\left(k_s ; \tau_1, \tau_2\right),\\
\nonumber \mathcal{I}_{ \pm \pm,  \mathrm{TO},>}^{p_1 p_2} =& -k_s^{5+p_{12}} \int_{-\infty}^0 \mathrm{~d} \tau_2 \int_{-\infty}^{\tau_2} \mathrm{~d} \tau_1\left(-\tau_1\right)^{p_1}\left(-\tau_2\right)^{p_2} e^{ \pm \mathrm{i}\left(k_{12} \tau_1+k_{34} \tau_2\right)}\\
&\times\left[D_{\lessgtr}\left(k_s ; \tau_1, \tau_2\right)-D_{\gtrless}\left(k_s ; \tau_1, \tau_2\right)\right].
\end{align}

Let us start from the factorized (F) integral $\mathcal{I}_{ \pm \pm, \mathrm{F},>}$. We rewrite the propagators $D_{\pm\mp}$ using the Mellin--Barnes representations Eqs.~\eqref{W_1} and \eqref{W_2}:
\begin{align}
\nonumber D_{\pm\mp}\left(k ; \tau_1, \tau_2\right)=&\ \frac{e^{\pi\kappa}H^2}{\pi \Gamma\left[\frac{1}{2}-\mathrm{i} \mu\pm\i \kappa, \frac{1}{2}+\mathrm{i} \mu\pm\i \kappa\right]} e^{\mp\mathrm{i} k\left(\tau_1+\tau_2\right)}  \\
\nonumber & \times \int_{-\mathrm{i} \infty}^{\mathrm{i} \infty} \frac{\mathrm{d} s_1}{2 \pi \mathrm{i}} \frac{\mathrm{d} s_2}{2 \pi \mathrm{i}} e^{\mp\mathrm{i} \pi\left(s_1-s_2\right) / 2} \cos \pi\left(s_1\mp\i\kappa\right)\left(2 k\right)^{-s_{12}}\left(-\tau_1\right)^{-s_1+3 / 2}\left(-\tau_2\right)^{-s_2+3 / 2}\\
 &\times \Gamma\left[-s_1+\frac{1}{2}\pm\i \kappa,-s_2+\frac{1}{2}\pm\i \kappa, s_1-\mathrm{i} \mu, s_1+\mathrm{i} \mu, s_2-\mathrm{i} \mu, s_2+\mathrm{i} \mu\right],\label{MB_D_pm_2}
\end{align}
which leads to the subsequent expression of $\mathcal{I}_{ \pm \pm, \mathrm{F},>}$ after the $\tau_i$-integration,
\begin{align}
\nonumber \mathcal{I}_{\pm\pm, \mathrm{F},>}^{ p_1 p_2}=&\ \frac{\pm\i e^{\pi\kappa(k_s)}H^2}{\pi \Gamma\left[\frac{1}{2}-\mathrm{i} \mu\mp\i\kappa(k_s), \frac{1}{2}+\mathrm{i} \mu\mp\i \kappa(k_s)\right]}\left(\frac{u_1}{2}\right)^{5 / 2+p_1}\left(\frac{u_2}{2}\right)^{5/ 2+p_2} e^{\mp\mathrm{i} \pi\left(p_1+p_2\right) / 2}    \\
\nonumber & \times \int_{-\mathrm{i} \infty}^{\mathrm{i} \infty} \frac{\mathrm{d} s_1}{2 \pi \mathrm{i}} \frac{\mathrm{d} s_2}{2 \pi \mathrm{i}} e^{\pm\mathrm{i} \pi s_1 } \cos \pi\left(s_1\pm\i \kappa(k_s)\right)u_1^{-s_1}u_2^{-s_2}\Gamma\left[p_1+\frac{5}{2}-s_1, p_2+\frac{5}{2}-s_2\right]\\
 &\times \Gamma\left[-s_1+\frac{1}{2}\mp\i \kappa(k_s),-s_2+\frac{1}{2}\mp\i\kappa(k_s), s_1-\mathrm{i} \mu, s_1+\mathrm{i} \mu, s_2-\mathrm{i} \mu, s_2+\mathrm{i} \mu\right]. 
\end{align}
The integration over $s_i$ and the summation of the residues can be executed following a similar procedure, and we obtain
\begin{align}
\nonumber  \mathcal{I}_{\pm\pm, \mathrm{F},>}^{ p_1 p_2}=&\ \frac{\pm\i e^{\pi\kappa(k_s)}H^2}{\pi \Gamma\left[\frac{1}{2}-\mathrm{i} \mu\mp\i\kappa(k_s), \frac{1}{2}+\mathrm{i} \mu\mp\i\kappa(k_s)\right]}\left(\frac{u_1}{2}\right)^{5 / 2+p_1}\left(\frac{u_2}{2}\right)^{5/ 2+p_2} e^{\mp\mathrm{i} \pi\left(p_1+p_2\right) / 2}    \\
 \nonumber &\times \sum_{n_1=0}^{\infty}\left\{\frac{(-1)^{n_1}}{n_{1} !}e^{\mp\i \pi(n_1+\i \mu)}\cos \pi\left(-n_1-\i \mu\pm\i\kappa(k_s)\right) u_1^{n_1+\i \mu}\right.\\
\nonumber &\times \left.\Gamma\left[-n_1-2 \i \mu, n_1+\i \mu+\frac{1}{2}\mp\i\kappa(k_s), n_1+\i \mu+p_1+\frac{5}{2}\right]+(\mu \rightarrow-\mu)\right\}\\
&\times \sum_{n_2=0}^{\infty}\left\{\frac{(-1)^{n_2}}{n_{2} !}u_2^{n_2+\i \mu} \Gamma\left[-n_2-2 \i \mu, n_2+\i \mu+\frac{1}{2}\mp\i\kappa(k_s), n_2+\i \mu+p_2+\frac{5}{2}\right]+(\mu \rightarrow-\mu)\right\}  \label{IppF_m}\\  
\nonumber =&\ \frac{\pm\i e^{\pi\kappa(k_s)}H^2}{\pi \Gamma\left[\frac{1}{2}-\mathrm{i} \mu\mp\i\kappa(k_s), \frac{1}{2}+\mathrm{i} \mu\mp\i\kappa(k_s)\right]}e^{\mp\mathrm{i} \pi\left(p_1+p_2\right) / 2}    \\
\nonumber &\times\left[e^{\pm\pi \mu} \cosh \pi\left(\mu\mp\kappa(k_s)\right)\mathrm{G}_{\mp \i \kappa(k_s), \mu}^{p_1}\left(u_1\right) u_1^{\mathrm{i} \mu}+e^{\mp\pi \mu} \cosh \pi\left(\mu\pm\kappa(k_s)\right)\mathrm{G}_{\mp \i \kappa(k_s), -\mu}^{p_1}\left(u_1\right) u_1^{-\mathrm{i} \mu}\right]\\
&\times \left[\mathrm{G}_{\mp\i \kappa(k_s), \mu}^{p_2}\left(u_2\right) u_2^{\mathrm{i} \mu}+\mathrm{G}_{\mp \i \kappa(k_s), -\mu}^{p_2}\left(u_2\right) u_2^{-\mathrm{i} \mu}\right],\label{IppF}    
\end{align}
where the $G$-function is defined in Eq.~\eqref{def_G}.

In the same manner, The time-ordered (TO) integral $\mathcal{I}_{ \pm \pm,  \mathrm{TO},>}^{p_1 p_2}$ can be transformed to
\begin{align}
\nonumber \mathcal{I}_{++, \mathrm{TO}, >}^{ p_1 p_2}= &\ \frac{\mathrm{i} e^{\pi \kappa(k_s)}H^2}{\pi \Gamma\left[\frac{1}{2}-\mathrm{i} \mu-\i\kappa(k_s), \frac{1}{2}+\mathrm{i} \mu-\i\kappa(k_s)\right]} \frac{e^{-\mathrm{i} \pi\left(p_1+p_2\right) / 2}}{2^{5+p_1+p_2}} \int_{-\mathrm{i} \infty}^{+\mathrm{i} \infty} \frac{\mathrm{d} s_1}{2 \pi \mathrm{i}} \frac{\mathrm{d} s_2}{2 \pi \mathrm{i}} \\
\nonumber &\times\left[e^{\mathrm{i} \pi s_2} \cos \pi\left(s_2+\i\kappa(k_s)\right)-e^{\mathrm{i} \pi s_1} \cos \pi\left(s_1+\i\kappa(k_s)\right)\right] u_1^{-s_{12}+5+p_1+p_2}\\
\nonumber &\times \Gamma\left[-s_1+\frac{1}{2}-\i\kappa(k_s),-s_2+\frac{1}{2}-\i\kappa(k_s), s_1-\mathrm{i} \mu, s_1+\mathrm{i} \mu, s_2-\mathrm{i} \mu, s_2+\mathrm{i} \mu\right]\\
&\times {}_2\mathcal{F}_1\left[\begin{array}{c|c}
p_2+\frac{5}{2}-s_2, p_1+p_2+5-s_{12}  \\
p_2+\frac{7}{2}-s_2
\end{array} -\frac{u_1}{u_2}\right]
\end{align}
by utilizing Eq.~\eqref{MB_D_pm_2} and the integration formula
\begin{align}
\nonumber &\int_{-\infty}^0 \mathrm{~d} \tau_2 \int_{-\infty}^{\tau_2} \mathrm{~d} \tau_1 e^{ \pm\left(\mathrm{i} k_{12} \tau_1+\mathrm{i} k_{34} \tau_2\right)}\left(-\tau_1\right)^{p-1}\left(-\tau_2\right)^{q-1}\\
&\qquad\qquad\qquad =e^{\mp \mathrm{i}(p+q) \frac{\pi}{2}} k_{12}^{-p-q} {}_2\mathcal{F}_1\left[\begin{array}{c|c}
q, p+q \\
1+q
\end{array}  -\frac{k_{34}}{k_{12}}\right].
\end{align}
Subsequently, the complex $s_i$-integral leads to
\begin{align}
\nonumber \mathcal{I}_{++, \mathrm{TO}, >}^{ p_1 p_2}=&\ \frac{\i H^2e^{-\mathrm{i} \pi\left(p_1+p_2\right) / 2}}{2^{5+p_1+p_2}} \sum_{n_1, n_2=0}^{\infty} \frac{(-1)^{n_{12}}}{n_{1} ! n_{2} !} u_1^{n_{12}+5+p_1+p_2}\\
\nonumber &\times\left\{\frac{1}{2 \mu}\left(\frac{1}{2}-\mathrm{i} \mu-\i\kappa(k_s)\right)_{n_1}\left(\frac{1}{2}+\mathrm{i} \mu-\i\kappa(k_s)\right)_{n_2}(2 \mathrm{i} \mu)_{-n_1}(-2 \mathrm{i} \mu)_{-n_2}\right.\\
&\left.\times {}_2\mathcal{F}_1\left[\begin{array}{c|c}
n_2+p_2+\frac{5}{2}+\i \mu, n_{12}+p_1+p_2+5 \\
n_2+p_2+\frac{7}{2}+\i\mu
\end{array} -\frac{u_1}{u_2}\right]+(\mu \rightarrow-\mu)\right\},\label{IppTO}    
\end{align}
where $(z)_n\equiv \Gamma[z+n]/\Gamma[z]$ denotes the Pochhammer symbol. The minus sign seed $\mathcal{I}_{--, \mathrm{TO}, >}^{ p_1 p_2}$ is obtained by taking a conjugate of Eq.~\eqref{IppTO}.
In contrast to the previous cases, performing the double summation presented above poses technical challenges. 

The hierarchical collapsed limit ($u_1 \ll u_2 \ll 1$) of $\mathcal{I}_{ \pm \pm}^{p_1 p_2}$ enables us to determine the integration constants appearing in Section~\ref{sec:boot_eq}. Under the limit, the dominant contribution comes from the poles with $n_{1,2}=0$ in Eqs.~\eqref{IppF_m} and \eqref{IppTO}. Additionally, $\mathcal{I}_{++, \mathrm{TO}, >}^{ p_1 p_2}$ is subdominant in comparison to $\mathcal{I}_{\pm\pm, \mathrm{F},>}^{ p_1 p_2}$. Combining these evaluations, we obtain
\begin{align}
\lim _{u_1 \ll u_2 \ll 1} \mathcal{I}_{\pm \pm}^{p_1 p_2}=\sum_{\m{a}, \m{b}= \pm}\tilde{\mathcal{C}}_{ \pm \pm \mid \m{a b}}\  \tilde{\mathcal{U}}_{ \pm \mid \m{a}}^{p_1}\left(u_1\right) \tilde{\mathcal{U}}_{\pm \mid \mathrm{b}}^{p_2}\left(u_2\right),   \label{Ipp_sq} 
\end{align}
where 
\begin{align}
&\tilde{\mathcal{C}}_{ \pm \pm \mid ++} =\tilde{\mathcal{C}}_{ \pm \pm \mid +-}= \frac{\pm\i e^{\pi \left(\kappa(k_s)+\mu\right)}\cosh \pi\left(-\mu+\kappa(k_s)\right)H^2}{\pi \Gamma\left[\frac{1}{2}-\mathrm{i} \mu\mp\i\kappa(k_s), \frac{1}{2}+\mathrm{i} \mu\mp\i\kappa(k_s)\right]}e^{\mp\mathrm{i} \pi\left(p_1+p_2\right) / 2} ,\\
&\tilde{\mathcal{C}}_{ \pm \pm \mid -+} =\tilde{\mathcal{C}}_{ \pm \pm \mid --}= \frac{\pm\i e^{\pi \left(\kappa(k_s)-\mu\right)}\cosh \pi\left(\mu+\kappa(k_s)\right)H^2}{\pi \Gamma\left[\frac{1}{2}-\mathrm{i} \mu\mp\i\kappa(k_s), \frac{1}{2}+\mathrm{i} \mu\mp\i\kappa(k_s)\right]}e^{\mp\mathrm{i} \pi\left(p_1+p_2\right) / 2}. 
\end{align}
Note that we confirmed that our outcomes~\eqref{Ipm}, \eqref{IppF}, and \eqref{IppTO} are consistent with the constant mass results of Refs.~\cite{Qin:2022fbv,Qin:2023ejc} in the limit $\kappa\rightarrow0$, as described in the following appendix in detail.

\section{Constant Mass Limit}\label{CML}
In case where the coupling~\eqref{e_mass} is constant, $g(\phi)=m_0^2$, which corresponds to $\kappa\rightarrow 0$, the seed integrals with single soft limit, Eqs.~\eqref{I_3_pm} and \eqref{I_3_pp}, should reproduce the results for the constant mass scenario presented in Ref.~\cite{Qin:2022fbv,Qin:2023ejc}. Since the mode function of the massive field $\sigma$ for the constant mass is expressed by the Hankel function (see Eq.~\eqref{hankel}) whereas our case is the Whittaker function (see Eq.~\eqref{sol_v}), this limit provides a non-trivial consistency check.

In the subsequent discussion, we frequently use the following formulae for the Gamma function:
\begin{align}
&\Gamma(-z)\Gamma(z+1)=-\frac{\pi}{ \sin \pi z},\\
&\Gamma(2 z)=\frac{2^{2 z-1}}{\sqrt{\pi}} \Gamma(z) \Gamma\left(z+\frac{1}{z}\right).
\end{align}
In the limit $\kappa\rightarrow0$, Eqs.~\eqref{I_3_pm} and \eqref{I_3_pp} are reduced to
\begin{align}
&\mathcal{I}_{ \pm \mp}^{p_1 p_2}(u_1, 1,v(k_3))/H^2=\frac{e^{\mp \mathrm{i} \frac{\pi}{2}\bar{p}_{12} }}{2^{7 / 2+p_2} \pi^{1 / 2}} \Gamma\left[\begin{array}{c}
\frac{5}{2}+p_2-\mathrm{i} \mu, \frac{5}{2}+p_2+\mathrm{i} \mu \\
3+p_2
\end{array}\right]\left[\mathcal{Y}_{+}^{p_1}(u)+\mathcal{Y}_{-}^{p_1}(u)\right],\\    
\nonumber &\mathcal{I}_{ \pm \pm}^{p_1 p_2}(u_1, 1,v(k_3))/H^2\\
\nonumber &\qquad=\frac{e^{\mp \mathrm{i} \frac{\pi}{2}p_{12} } \Gamma\left(5+p_{12}\right) u_1^{5+p_{12}}}{2^{5+p_{12}}\left[\left(\frac{5}{2}+p_2\right)^2+\mu^2\right]}{ }_3 \mathrm{F}_2\left[\begin{array}{c|c}
1,3+p_2, 5+p_{12} \\
\frac{7}{2}+p_2-\mathrm{i} \mu, \frac{7}{2}+p_2+\mathrm{i} \mu
\end{array} \ u_1\right]\\
&\qquad+\frac{ \pm \mathrm{i} e^{\mp \mathrm{i} \frac{\pi}{2}p_{12} }}{2^{7 / 2+p_2} \pi^{1 / 2}} \Gamma\left[\begin{array}{c}
\frac{5}{2}+p_2-\mathrm{i} \mu, \frac{5}{2}+p_2+\mathrm{i} \mu \\
3+p_2
\end{array}\right]\left[e^{\pi \mu} \mathcal{Y}_{ \pm}^{p_1}(u_1)+e^{-\pi \mu} \mathcal{Y}_{\mp}^{p_1}(u_1)\right],
\end{align}
respectively, where
\begin{align}
\mathcal{Y}_{ \pm}^p(u)\equiv 2^{\mp \mathrm{i} \mu}\left(\frac{u}{2}\right)^{5 / 2+p \pm \mathrm{i} \mu} \Gamma\left[\frac{5}{2}+p \pm \mathrm{i} \mu, \mp \mathrm{i} \mu\right]{ }_2 \mathrm{F}_1\left[\begin{array}{c|c}
\frac{5}{2}+p \pm \mathrm{i} \mu, \frac{1}{2} \pm \mathrm{i} \mu \\
1 \pm 2 \mathrm{i} \mu
\end{array} \ u\right].    
\end{align}
These equations are in agreement with the results of Ref.~\cite{Qin:2023ejc}.

In the same way, for the seed integrals with double soft limit (Eqs.\eqref{I_pm_2} and \eqref{I_pp_2}), we can express them in the limit $\kappa\rightarrow 0$ as follows:
\begin{align}
\mathcal{I}_{ \pm \mp}^{p_1 p_2}(1,1,v(k_3))/H^2=\frac{e^{\mp \mathrm{i} \frac{\pi}{2}\bar{p}_{12}}}{2^{5+p_{12}}} \Gamma\left[\begin{array}{c}
\frac{5}{2}+p_1-\mathrm{i} \mu, \frac{5}{2}+p_1+\mathrm{i} \mu, \frac{5}{2}+p_2-\mathrm{i} \mu, \frac{5}{2}+p_2+\mathrm{i} \mu \\
3+p_1, 3+p_2
\end{array}\right] ,    
\end{align}
and 
\begin{align}
\nonumber &\mathcal{I}_{ \pm \pm}^{p_1 p_2}(1,1,v(k_3))/H^2\\
\nonumber &=\frac{ \pm \mathrm{i} e^{\mp \mathrm{i} \frac{\pi}{2}p_{12} } e^{-\pi \mu}}{2^{5+p_{12}}} \Gamma\left[\begin{array}{c}
\frac{5}{2}+p_1-\mathrm{i} \mu, \frac{5}{2}+p_1+\mathrm{i} \mu, \frac{5}{2}+p_2-\mathrm{i} \mu, \frac{5}{2}+p_2+\mathrm{i} \mu \\
3+p_1, 3+p_2
\end{array}\right] \\
&- \frac{e^{\mp \mathrm{i} \frac{\pi}{2}p_{12} }}{2^{p_{12}+5}} \Gamma\left[\begin{array}{c}
\frac{5}{2}+p_2 \pm \i \mu, \frac{5}{2}+p_1 \pm \i \mu \\
\frac{1}{2} \pm \i \mu
\end{array}\right] {}_3\mathcal{F}_2\left[\begin{array}{c|c}
\frac{1}{2} \pm \i \mu, 5+p_{12}, 1 \\
\frac{7}{2}+p_1 \pm \i \mu, \frac{7}{2}+p_2 \pm \i \mu
\end{array}\ 1\right].
\end{align}
Again, these expressions are consistent with the results of Ref.~\cite{Qin:2023ejc}.

\end{appendix}

\addcontentsline{toc}{section}{References}
\printbibliography

\end{fmffile}
\end{document}

%% file: setting.tex
\pdfoutput=1
\usepackage{pifont,cancel,comment}
\usepackage{here}
\usepackage{subfigure}
\usepackage{graphicx,xcolor}
\usepackage{tikz}
\usepackage{feynmp}
\usepackage{slashed}
\usepackage[T1]{fontenc}
\usepackage{lmodern}
\usepackage{bm}
\usepackage[labelsep=quad]{caption}
\usepackage{amsmath,amssymb,mathtools}
\usepackage{physics}
\usepackage{amsthm}
\usepackage{thmtools}
\usepackage{braket}
\usepackage[version=4]{mhchem}
\usepackage{booktabs}
\usepackage{makeidx}
\usepackage{imakeidx}
\usepackage[stable]{footmisc}
\usepackage{fancyhdr}
\usepackage{titleref}
\usepackage{wasysym}
\usepackage{upgreek}
\usepackage{textgreek}
\usepackage{hyperref}
\usepackage{ascmac}
\usepackage{framed}
\usepackage{subfiles}
\usepackage[backend=biber,style=numeric-comp,sorting=none,date=year,giveninits=true]{biblatex}

\hypersetup{
    colorlinks=true,
    linkcolor=blue,
    filecolor=magenta,    
    citecolor=green,
    anchorcolor=yellow,
    urlcolor=cyan
    }

\usetikzlibrary{graphs}

\DeclareFieldFormat*{title}{\mkbibquote{#1\adddot}}
\renewbibmacro{in:}{}

\numberwithin{equation}{section}

\allowdisplaybreaks[1]

\newcommand{\pdiff}[2]{\dfrac{\partial #1}{\partial #2}}

\newcommand{\mc}[1]{\mathcal{#1}}

\makeatletter

\@addtoreset{equation}{section}

\newcommand{\lambdabar}{{\mathchoice
  {\smash@bar\textfont\displaystyle{0.25}{1.2}\lambda}
  {\smash@bar\textfont\textstyle{0.25}{1.2}\lambda}
  {\smash@bar\scriptfont\scriptstyle{0.25}{1.2}\lambda}
  {\smash@bar\scriptscriptfont\scriptscriptstyle{0.25}{1.2}\lambda}
}}
\newcommand{\smash@bar}[4]{
  \smash{\rlap{\raisebox{-#3\fontdimen5#10}{$\m@th#2\mkern#4mu\mathchar'26$}}}%
}

\newcommand{\figcaption}[1]{\def\@captype{figure}\caption{#1}}
\newcommand{\tblcaption}[1]{\def\@captype{table}\caption{#1}}

\makeatother

%% file: ref.bib
@article{Green:2023ids,
    author = "Green, Daniel and Huang, Yiwen and Shen, Chia-Hsien and Baumann, Daniel",
    title = "{Positivity from Cosmological Correlators}",
    eprint = "2310.02490",
    archivePrefix = "arXiv",
    primaryClass = "hep-th",
    month = "10",
    year = "2023"
}

@article{Planck:2018jri,
    author = "Akrami, Y. and others",
    collaboration = "Planck",
    title = "{Planck 2018 results. X. Constraints on inflation}",
    eprint = "1807.06211",
    archivePrefix = "arXiv",
    primaryClass = "astro-ph.CO",
    doi = "10.1051/0004-6361/201833887",
    journal = "Astron. Astrophys.",
    volume = "641",
    pages = "A10",
    year = "2020"
}

@article{Boomerang:2000jdg,
    author = "Jaffe, Andrew H. and others",
    collaboration = "Boomerang",
    title = "{Cosmology from MAXIMA-1, BOOMERANG and COBE / DMR CMB observations}",
    eprint = "astro-ph/0007333",
    archivePrefix = "arXiv",
    doi = "10.1103/PhysRevLett.86.3475",
    journal = "Phys. Rev. Lett.",
    volume = "86",
    pages = "3475--3479",
    year = "2001"
}

@article{WMAP:2012fli,
    author = "Bennett, C. L. and others",
    collaboration = "WMAP",
    title = "{Nine-Year Wilkinson Microwave Anisotropy Probe (WMAP) Observations: Final Maps and Results}",
    eprint = "1212.5225",
    archivePrefix = "arXiv",
    primaryClass = "astro-ph.CO",
    doi = "10.1088/0067-0049/208/2/20",
    journal = "Astrophys. J. Suppl.",
    volume = "208",
    pages = "20",
    year = "2013"
}

@article{Albrecht:1982wi,
    author = "Albrecht, Andreas and Steinhardt, Paul J.",
    editor = "Fang, Li-Zhi and Ruffini, R.",
    title = "{Cosmology for Grand Unified Theories with Radiatively Induced Symmetry Breaking}",
    reportNumber = "UPR-0185T",
    doi = "10.1103/PhysRevLett.48.1220",
    journal = "Phys. Rev. Lett.",
    volume = "48",
    pages = "1220--1223",
    year = "1982"
}

@article{Linde:1981mu,
    author = "Linde, Andrei D.",
    editor = "Fang, Li-Zhi and Ruffini, R.",
    title = "{A New Inflationary Universe Scenario: A Possible Solution of the Horizon, Flatness, Homogeneity, Isotropy and Primordial Monopole Problems}",
    reportNumber = "LEBEDEV-81-229",
    doi = "10.1016/0370-2693(82)91219-9",
    journal = "Phys. Lett. B",
    volume = "108",
    pages = "389--393",
    year = "1982"
}

@article{Guth:1980zm,
    author = "Guth, Alan H.",
    editor = "Fang, Li-Zhi and Ruffini, R.",
    title = "{The Inflationary Universe: A Possible Solution to the Horizon and Flatness Problems}",
    reportNumber = "SLAC-PUB-2576",
    doi = "10.1103/PhysRevD.23.347",
    journal = "Phys. Rev. D",
    volume = "23",
    pages = "347--356",
    year = "1981"
}

@article{Starobinsky:1980te,
    author = "Starobinsky, Alexei A.",
    editor = "Khalatnikov, I. M. and Mineev, V. P.",
    title = "{A New Type of Isotropic Cosmological Models Without Singularity}",
    doi = "10.1016/0370-2693(80)90670-X",
    journal = "Phys. Lett. B",
    volume = "91",
    pages = "99--102",
    year = "1980"
}

@article{Sato:1980yn,
    author = "Sato, K.",
    title = "{First Order Phase Transition of a Vacuum and Expansion of the Universe}",
    reportNumber = "NORDITA-80-29",
    journal = "Mon. Not. Roy. Astron. Soc.",
    volume = "195",
    pages = "467--479",
    year = "1981"
}

@article{Qin:2022fbv,
    author = "Qin, Zhehan and Xianyu, Zhong-Zhi",
    title = "{Helical Inflation Correlators: Partial Mellin-Barnes and Bootstrap Equations}",
    eprint = "2208.13790",
    archivePrefix = "arXiv",
    primaryClass = "hep-th",
    month = "8",
    year = "2022"
}

@article{Qin:2023ejc,
    author = "Qin, Zhehan and Xianyu, Zhong-Zhi",
    title = "{Closed-Form Formulae for Inflation Correlators}",
    eprint = "2301.07047",
    archivePrefix = "arXiv",
    primaryClass = "hep-th",
    month = "1",
    year = "2023"
}

@article{Reece:2022soh,
    author = "Reece, Matthew and Wang, Lian-Tao and Xianyu, Zhong-Zhi",
    title = "{Large-Field Inflation and the Cosmological Collider}",
    eprint = "2204.11869",
    archivePrefix = "arXiv",
    primaryClass = "hep-ph",
    month = "4",
    year = "2022"
}

@article{Chen:2017ryl,
    author = "Chen, Xingang and Wang, Yi and Xianyu, Zhong-Zhi",
    title = "{Schwinger-Keldysh Diagrammatics for Primordial Perturbations}",
    eprint = "1703.10166",
    archivePrefix = "arXiv",
    primaryClass = "hep-th",
    doi = "10.1088/1475-7516/2017/12/006",
    journal = "JCAP",
    volume = "12",
    pages = "006",
    year = "2017"
}

@article{Arkani-Hamed:2015bza,
    author = "Arkani-Hamed, Nima and Maldacena, Juan",
    title = "{Cosmological Collider Physics}",
    eprint = "1503.08043",
    archivePrefix = "arXiv",
    primaryClass = "hep-th",
    month = "3",
    year = "2015"
}

@article{Chen:2012ge,
    author = "Chen, Xingang and Wang, Yi",
    title = "{Quasi-Single Field Inflation with Large Mass}",
    eprint = "1205.0160",
    archivePrefix = "arXiv",
    primaryClass = "hep-th",
    doi = "10.1088/1475-7516/2012/09/021",
    journal = "JCAP",
    volume = "09",
    pages = "021",
    year = "2012"
}

@book{Baumann:2014nda,
    author = "Baumann, Daniel and McAllister, Liam",
    title = "{Inflation and String Theory}",
    eprint = "1404.2601",
    archivePrefix = "arXiv",
    primaryClass = "hep-th",
    doi = "10.1017/CBO9781316105733",
    publisher = "Cambridge University Press",
    series = "Cambridge Monographs on Mathematical Physics",
    month = "5",
    year = "2015"
}

@article{Werth:2023pfl,
    author = "Werth, Denis and Pinol, Lucas and Renaux-Petel, S\'ebastien",
    title = "{Cosmological Flow of Primordial Correlators}",
    eprint = "2302.00655",
    archivePrefix = "arXiv",
    primaryClass = "hep-th",
    month = "2",
    year = "2023"
}

@article{Lee:2023dcy,
    author = "Lee, Hyun Min and Menkara, Adriana",
    title = "{Graceful exit from inflation and reheating with twin waterfalls}",
    eprint = "2304.08686",
    archivePrefix = "arXiv",
    primaryClass = "hep-ph",
    month = "4",
    year = "2023"
}

@article{Lee:2022fkd,
    author = "Lee, Hyun Min and Menkara, Adriana G.",
    title = "{Pseudo-Nambu-Goldstone inflation with twin waterfalls}",
    eprint = "2206.05523",
    archivePrefix = "arXiv",
    primaryClass = "hep-ph",
    doi = "10.1016/j.physletb.2022.137483",
    journal = "Phys. Lett. B",
    volume = "834",
    pages = "137483",
    year = "2022"
}

@article{Deshpande:2020lmf,
    author = "Deshpande, Kaustubh and Kumar, Soubhik and Sundrum, Raman",
    title = "{TwInflation}",
    eprint = "2101.06275",
    archivePrefix = "arXiv",
    primaryClass = "hep-ph",
    reportNumber = "UMD-PP-021-01",
    doi = "10.1007/JHEP07(2021)147",
    journal = "JHEP",
    volume = "21",
    pages = "147",
    year = "2020"
}

@article{Wang:2013zva,
    author = "Wang, Yi",
    title = "{Inflation, Cosmic Perturbations and Non-Gaussianities}",
    eprint = "1303.1523",
    archivePrefix = "arXiv",
    primaryClass = "hep-th",
    doi = "10.1088/0253-6102/62/1/19",
    journal = "Commun. Theor. Phys.",
    volume = "62",
    pages = "109--166",
    year = "2014"
}

@article{Chen:2009zp,
    author = "Chen, Xingang and Wang, Yi",
    title = "{Quasi-Single Field Inflation and Non-Gaussianities}",
    eprint = "0911.3380",
    archivePrefix = "arXiv",
    primaryClass = "hep-th",
    doi = "10.1088/1475-7516/2010/04/027",
    journal = "JCAP",
    volume = "04",
    pages = "027",
    year = "2010"
}

@article{Pinol:2021aun,
    author = "Pinol, Lucas and Aoki, Shuntaro and Renaux-Petel, S\'ebastien and Yamaguchi, Masahide",
    title = "{Inflationary flavor oscillations and the cosmic spectroscopy}",
    eprint = "2112.05710",
    archivePrefix = "arXiv",
    primaryClass = "hep-th",
    month = "12",
    year = "2021"
}

@article{Aoki:2020zbj,
    author = "Aoki, Shuntaro and Yamaguchi, Masahide",
    title = "{Disentangling mass spectra of multiple fields in cosmological collider}",
    eprint = "2012.13667",
    archivePrefix = "arXiv",
    primaryClass = "hep-th",
    reportNumber = "WU-HEP-20-15",
    doi = "10.1007/JHEP04(2021)127",
    journal = "JHEP",
    volume = "04",
    pages = "127",
    year = "2021"
}

@article{Baumann:2011nk,
    author = "Baumann, Daniel and Green, Daniel",
    title = "{Signatures of Supersymmetry from the Early Universe}",
    eprint = "1109.0292",
    archivePrefix = "arXiv",
    primaryClass = "hep-th",
    doi = "10.1103/PhysRevD.85.103520",
    journal = "Phys. Rev. D",
    volume = "85",
    pages = "103520",
    year = "2012"
}

@article{Noumi:2012vr,
    author = "Noumi, Toshifumi and Yamaguchi, Masahide and Yokoyama, Daisuke",
    title = "{Effective field theory approach to quasi-single field inflation and effects of heavy fields}",
    eprint = "1211.1624",
    archivePrefix = "arXiv",
    primaryClass = "hep-th",
    reportNumber = "UT-KOMABA-12-9, TIT-HEP-625",
    doi = "10.1007/JHEP06(2013)051",
    journal = "JHEP",
    volume = "06",
    pages = "051",
    year = "2013"
}

@article{Chen:2009we,
    author = "Chen, Xingang and Wang, Yi",
    title = "{Large non-Gaussianities with Intermediate Shapes from Quasi-Single Field Inflation}",
    eprint = "0909.0496",
    archivePrefix = "arXiv",
    primaryClass = "astro-ph.CO",
    reportNumber = "MIT-CTP-4071",
    doi = "10.1103/PhysRevD.81.063511",
    journal = "Phys. Rev. D",
    volume = "81",
    pages = "063511",
    year = "2010"
}

@article{Assassi:2012zq,
    author = "Assassi, Valentin and Baumann, Daniel and Green, Daniel",
    title = "{On Soft Limits of Inflationary Correlation Functions}",
    eprint = "1204.4207",
    archivePrefix = "arXiv",
    primaryClass = "hep-th",
    doi = "10.1088/1475-7516/2012/11/047",
    journal = "JCAP",
    volume = "11",
    pages = "047",
    year = "2012"
}

@article{Sefusatti:2012ye,
    author = "Sefusatti, Emiliano and Fergusson, James R. and Chen, Xingang and Shellard, E. P. S.",
    title = "{Effects and Detectability of Quasi-Single Field Inflation in the Large-Scale Structure and Cosmic Microwave Background}",
    eprint = "1204.6318",
    archivePrefix = "arXiv",
    primaryClass = "astro-ph.CO",
    doi = "10.1088/1475-7516/2012/08/033",
    journal = "JCAP",
    volume = "08",
    pages = "033",
    year = "2012"
}

@article{Pi:2012gf,
    author = "Pi, Shi and Sasaki, Misao",
    title = "{Curvature Perturbation Spectrum in Two-field Inflation with a Turning Trajectory}",
    eprint = "1205.0161",
    archivePrefix = "arXiv",
    primaryClass = "hep-th",
    reportNumber = "YITP-12-36",
    doi = "10.1088/1475-7516/2012/10/051",
    journal = "JCAP",
    volume = "10",
    pages = "051",
    year = "2012"
}

@article{Cespedes:2013rda,
    author = "C\'espedes, Sebasti\'an and Palma, Gonzalo A.",
    title = "{Cosmic inflation in a landscape of heavy-fields}",
    eprint = "1303.4703",
    archivePrefix = "arXiv",
    primaryClass = "hep-th",
    doi = "10.1088/1475-7516/2013/10/051",
    journal = "JCAP",
    volume = "10",
    pages = "051",
    year = "2013"
}

@article{Gong:2013sma,
    author = "Gong, Jinn-Ouk and Pi, Shi and Sasaki, Misao",
    title = "{Equilateral non-Gaussianity from heavy fields}",
    eprint = "1306.3691",
    archivePrefix = "arXiv",
    primaryClass = "hep-th",
    doi = "10.1088/1475-7516/2013/11/043",
    journal = "JCAP",
    volume = "11",
    pages = "043",
    year = "2013"
}

@article{Emami:2013lma,
    author = "Emami, Razieh",
    title = "{Spectroscopy of Masses and Couplings during Inflation}",
    eprint = "1311.0184",
    archivePrefix = "arXiv",
    primaryClass = "hep-th",
    doi = "10.1088/1475-7516/2014/04/031",
    journal = "JCAP",
    volume = "04",
    pages = "031",
    year = "2014"
}

@article{Kehagias:2015jha,
    author = "Kehagias, Alex and Riotto, Antonio",
    title = "{High Energy Physics Signatures from Inflation and Conformal Symmetry of de Sitter}",
    eprint = "1501.03515",
    archivePrefix = "arXiv",
    primaryClass = "hep-th",
    doi = "10.1002/prop.201500025",
    journal = "Fortsch. Phys.",
    volume = "63",
    pages = "531--542",
    year = "2015"
}

@article{Liu:2015tza,
    author = "Liu, Junyu and Wang, Yi and Zhou, Siyi",
    title = "{Inflation with Massive Vector Fields}",
    eprint = "1502.05138",
    archivePrefix = "arXiv",
    primaryClass = "hep-th",
    doi = "10.1088/1475-7516/2015/08/033",
    journal = "JCAP",
    volume = "08",
    pages = "033",
    year = "2015"
}

@article{Dimastrogiovanni:2015pla,
    author = "Dimastrogiovanni, Emanuela and Fasiello, Matteo and Kamionkowski, Marc",
    title = "{Imprints of Massive Primordial Fields on Large-Scale Structure}",
    eprint = "1504.05993",
    archivePrefix = "arXiv",
    primaryClass = "astro-ph.CO",
    doi = "10.1088/1475-7516/2016/02/017",
    journal = "JCAP",
    volume = "02",
    pages = "017",
    year = "2016"
}

@article{Schmidt:2015xka,
    author = "Schmidt, Fabian and Chisari, Nora Elisa and Dvorkin, Cora",
    title = "{Imprint of inflation on galaxy shape correlations}",
    eprint = "1506.02671",
    archivePrefix = "arXiv",
    primaryClass = "astro-ph.CO",
    doi = "10.1088/1475-7516/2015/10/032",
    journal = "JCAP",
    volume = "10",
    pages = "032",
    year = "2015"
}

@article{Pinol:2023oux,
    author = "Pinol, Lucas and Renaux-Petel, S\'ebastien and Werth, Denis",
    title = "{The Cosmological Flow: A Systematic Approach to Primordial Correlators}",
    eprint = "2312.06559",
    archivePrefix = "arXiv",
    primaryClass = "astro-ph.CO",
    month = "12",
    year = "2023"
}

@article{Chen:2015lza,
    author = "Chen, Xingang and Namjoo, Mohammad Hossein and Wang, Yi",
    title = "{Quantum Primordial Standard Clocks}",
    eprint = "1509.03930",
    archivePrefix = "arXiv",
    primaryClass = "astro-ph.CO",
    doi = "10.1088/1475-7516/2016/02/013",
    journal = "JCAP",
    volume = "02",
    pages = "013",
    year = "2016"
}

@article{Delacretaz:2015edn,
    author = "Delacretaz, Luca V. and Noumi, Toshifumi and Senatore, Leonardo",
    title = "{Boost Breaking in the EFT of Inflation}",
    eprint = "1512.04100",
    archivePrefix = "arXiv",
    primaryClass = "hep-th",
    doi = "10.1088/1475-7516/2017/02/034",
    journal = "JCAP",
    volume = "02",
    pages = "034",
    year = "2017"
}

@article{Bonga:2015urq,
    author = "Bonga, B\'eatrice and Brahma, Suddhasattwa and Deutsch, Anne-Sylvie and Shandera, Sarah",
    title = "{Cosmic variance in inflation with two light scalars}",
    eprint = "1512.05365",
    archivePrefix = "arXiv",
    primaryClass = "astro-ph.CO",
    reportNumber = "IGC-15-12-1",
    doi = "10.1088/1475-7516/2016/05/018",
    journal = "JCAP",
    volume = "05",
    pages = "018",
    year = "2016"
}

@article{Flauger:2016idt,
    author = "Flauger, Raphael and Mirbabayi, Mehrdad and Senatore, Leonardo and Silverstein, Eva",
    title = "{Productive Interactions: heavy particles and non-Gaussianity}",
    eprint = "1606.00513",
    archivePrefix = "arXiv",
    primaryClass = "hep-th",
    reportNumber = "SU-ITP-16-12, UTTG-09-16",
    doi = "10.1088/1475-7516/2017/10/058",
    journal = "JCAP",
    volume = "10",
    pages = "058",
    year = "2017"
}

@article{Lee:2016vti,
    author = "Lee, Hayden and Baumann, Daniel and Pimentel, Guilherme L.",
    title = "{Non-Gaussianity as a Particle Detector}",
    eprint = "1607.03735",
    archivePrefix = "arXiv",
    primaryClass = "hep-th",
    doi = "10.1007/JHEP12(2016)040",
    journal = "JHEP",
    volume = "12",
    pages = "040",
    year = "2016"
}

@article{Delacretaz:2016nhw,
    author = "Delacretaz, Luca V. and Gorbenko, Victor and Senatore, Leonardo",
    title = "{The Supersymmetric Effective Field Theory of Inflation}",
    eprint = "1610.04227",
    archivePrefix = "arXiv",
    primaryClass = "hep-th",
    doi = "10.1007/JHEP03(2017)063",
    journal = "JHEP",
    volume = "03",
    pages = "063",
    year = "2017"
}

@article{Meerburg:2016zdz,
    author = {Meerburg, P. Daniel and M\"unchmeyer, Moritz and Mu\~noz, Julian B. and Chen, Xingang},
    title = "{Prospects for Cosmological Collider Physics}",
    eprint = "1610.06559",
    archivePrefix = "arXiv",
    primaryClass = "astro-ph.CO",
    doi = "10.1088/1475-7516/2017/03/050",
    journal = "JCAP",
    volume = "03",
    pages = "050",
    year = "2017"
}

@article{Chen:2016uwp,
    author = "Chen, Xingang and Wang, Yi and Xianyu, Zhong-Zhi",
    title = "{Standard Model Background of the Cosmological Collider}",
    eprint = "1610.06597",
    archivePrefix = "arXiv",
    primaryClass = "hep-th",
    doi = "10.1103/PhysRevLett.118.261302",
    journal = "Phys. Rev. Lett.",
    volume = "118",
    number = "26",
    pages = "261302",
    year = "2017"
}

@article{Chen:2016hrz,
    author = "Chen, Xingang and Wang, Yi and Xianyu, Zhong-Zhi",
    title = "{Standard Model Mass Spectrum in Inflationary Universe}",
    eprint = "1612.08122",
    archivePrefix = "arXiv",
    primaryClass = "hep-th",
    doi = "10.1007/JHEP04(2017)058",
    journal = "JHEP",
    volume = "04",
    pages = "058",
    year = "2017"
}

@article{Kehagias:2017cym,
    author = "Kehagias, Alex and Riotto, Antonio",
    title = "{On the Inflationary Perturbations of Massive Higher-Spin Fields}",
    eprint = "1705.05834",
    archivePrefix = "arXiv",
    primaryClass = "hep-th",
    doi = "10.1088/1475-7516/2017/07/046",
    journal = "JCAP",
    volume = "07",
    pages = "046",
    year = "2017"
}

@article{An:2017hlx,
    author = "An, Haipeng and McAneny, Michael and Ridgway, Alexander K. and Wise, Mark B.",
    title = "{Quasi Single Field Inflation in the non-perturbative regime}",
    eprint = "1706.09971",
    archivePrefix = "arXiv",
    primaryClass = "hep-ph",
    doi = "10.1007/JHEP06(2018)105",
    journal = "JHEP",
    volume = "06",
    pages = "105",
    year = "2018"
}

@article{Tong:2017iat,
    author = "Tong, Xi and Wang, Yi and Zhou, Siyi",
    title = "{On the Effective Field Theory for Quasi-Single Field Inflation}",
    eprint = "1708.01709",
    archivePrefix = "arXiv",
    primaryClass = "astro-ph.CO",
    doi = "10.1088/1475-7516/2017/11/045",
    journal = "JCAP",
    volume = "11",
    pages = "045",
    year = "2017"
}

@article{Iyer:2017qzw,
    author = "Iyer, Aditya Varna and Pi, Shi and Wang, Yi and Wang, Ziwei and Zhou, Siyi",
    title = "{Strongly Coupled Quasi-Single Field Inflation}",
    eprint = "1710.03054",
    archivePrefix = "arXiv",
    primaryClass = "hep-th",
    doi = "10.1088/1475-7516/2018/01/041",
    journal = "JCAP",
    volume = "01",
    pages = "041",
    year = "2018"
}

@article{An:2017rwo,
    author = "An, Haipeng and McAneny, Michael and Ridgway, Alexander K. and Wise, Mark B.",
    title = "{Non-Gaussian Enhancements of Galactic Halo Correlations in Quasi-Single Field Inflation}",
    eprint = "1711.02667",
    archivePrefix = "arXiv",
    primaryClass = "hep-ph",
    doi = "10.1103/PhysRevD.97.123528",
    journal = "Phys. Rev. D",
    volume = "97",
    number = "12",
    pages = "123528",
    year = "2018"
}

@article{Kumar:2017ecc,
    author = "Kumar, Soubhik and Sundrum, Raman",
    title = "{Heavy-Lifting of Gauge Theories By Cosmic Inflation}",
    eprint = "1711.03988",
    archivePrefix = "arXiv",
    primaryClass = "hep-ph",
    reportNumber = "UMD-PP-017-31",
    doi = "10.1007/JHEP05(2018)011",
    journal = "JHEP",
    volume = "05",
    pages = "011",
    year = "2018"
}

@article{RiquelmeM:2017qhp,
    author = "Riquelme M., Simon",
    title = "{Non-Gaussianities in a two-field generalization of Natural Inflation}",
    eprint = "1711.08549",
    archivePrefix = "arXiv",
    primaryClass = "astro-ph.CO",
    doi = "10.1088/1475-7516/2018/04/027",
    journal = "JCAP",
    volume = "04",
    pages = "027",
    year = "2018"
}

@article{Franciolini:2017ktv,
    author = "Franciolini, Gabriele and Kehagias, Alex and Riotto, Antonio",
    title = "{Imprints of Spinning Particles on Primordial Cosmological Perturbations}",
    eprint = "1712.06626",
    archivePrefix = "arXiv",
    primaryClass = "hep-th",
    doi = "10.1088/1475-7516/2018/02/023",
    journal = "JCAP",
    volume = "02",
    pages = "023",
    year = "2018"
}

@article{Saito:2018omt,
    author = "Saito, Ryo and Kubota, Takahiro",
    title = "{Heavy Particle Signatures in Cosmological Correlation Functions with Tensor Modes}",
    eprint = "1804.06974",
    archivePrefix = "arXiv",
    primaryClass = "hep-th",
    doi = "10.1088/1475-7516/2018/06/009",
    journal = "JCAP",
    volume = "06",
    pages = "009",
    year = "2018"
}

@article{Cabass:2018roz,
    author = "Cabass, Giovanni and Pajer, Enrico and Schmidt, Fabian",
    title = "{Imprints of Oscillatory Bispectra on Galaxy Clustering}",
    eprint = "1804.07295",
    archivePrefix = "arXiv",
    primaryClass = "astro-ph.CO",
    doi = "10.1088/1475-7516/2018/09/003",
    journal = "JCAP",
    volume = "09",
    pages = "003",
    year = "2018"
}

@article{Wang:2018tbf,
    author = "Wang, Yi and Wu, Yi-Peng and Yokoyama, Jun'ichi and Zhou, Siyi",
    title = "{Hybrid Quasi-Single Field Inflation}",
    eprint = "1804.07541",
    archivePrefix = "arXiv",
    primaryClass = "astro-ph.CO",
    reportNumber = "RESCEU 05/18, RESCEU-05-18",
    doi = "10.1088/1475-7516/2018/07/068",
    journal = "JCAP",
    volume = "07",
    pages = "068",
    year = "2018"
}

@article{Chen:2018xck,
    author = "Chen, Xingang and Wang, Yi and Xianyu, Zhong-Zhi",
    title = "{Neutrino Signatures in Primordial Non-Gaussianities}",
    eprint = "1805.02656",
    archivePrefix = "arXiv",
    primaryClass = "hep-ph",
    doi = "10.1007/JHEP09(2018)022",
    journal = "JHEP",
    volume = "09",
    pages = "022",
    year = "2018"
}

@article{Bartolo:2018hjc,
    author = "Bartolo, Nicola and Bianco, Domenico Matteo and Jimenez, Raul and Matarrese, Sabino and Verde, Licia",
    title = "{Supergravity, $\alpha$-attractors and primordial non-Gaussianity}",
    eprint = "1805.04269",
    archivePrefix = "arXiv",
    primaryClass = "astro-ph.CO",
    doi = "10.1088/1475-7516/2018/10/017",
    journal = "JCAP",
    volume = "10",
    pages = "017",
    year = "2018"
}

@article{Dimastrogiovanni:2018uqy,
    author = "Dimastrogiovanni, Emanuela and Fasiello, Matteo and Tasinato, Gianmassimo",
    title = "{Probing the inflationary particle content: extra spin-2 field}",
    eprint = "1806.00850",
    archivePrefix = "arXiv",
    primaryClass = "astro-ph.CO",
    doi = "10.1088/1475-7516/2018/08/016",
    journal = "JCAP",
    volume = "08",
    pages = "016",
    year = "2018"
}

@article{Bordin:2018pca,
    author = "Bordin, Lorenzo and Creminelli, Paolo and Khmelnitsky, Andrei and Senatore, Leonardo",
    title = "{Light Particles with Spin in Inflation}",
    eprint = "1806.10587",
    archivePrefix = "arXiv",
    primaryClass = "hep-th",
    doi = "10.1088/1475-7516/2018/10/013",
    journal = "JCAP",
    volume = "10",
    pages = "013",
    year = "2018"
}

@article{Achucarro:2018ngj,
    author = "Ach\'ucarro, Ana and C\'espedes, Sebasti\'an and Davis, Anne-Christine and Palma, Gonzalo A.",
    title = "{Constraints on Holographic Multifield Inflation and Models Based on the Hamilton-Jacobi Formalism}",
    eprint = "1809.05341",
    archivePrefix = "arXiv",
    primaryClass = "hep-th",
    doi = "10.1103/PhysRevLett.122.191301",
    journal = "Phys. Rev. Lett.",
    volume = "122",
    number = "19",
    pages = "191301",
    year = "2019"
}

@article{Chua:2018dqh,
    author = "Chua, Wan Zhen and Ding, Qianhang and Wang, Yi and Zhou, Siyi",
    title = "{Imprints of Schwinger Effect on Primordial Spectra}",
    eprint = "1810.09815",
    archivePrefix = "arXiv",
    primaryClass = "hep-th",
    doi = "10.1007/JHEP04(2019)066",
    journal = "JHEP",
    volume = "04",
    pages = "066",
    year = "2019"
}

@article{Arkani-Hamed:2018kmz,
    author = "Arkani-Hamed, Nima and Baumann, Daniel and Lee, Hayden and Pimentel, Guilherme L.",
    title = "{The Cosmological Bootstrap: Inflationary Correlators from Symmetries and Singularities}",
    eprint = "1811.00024",
    archivePrefix = "arXiv",
    primaryClass = "hep-th",
    doi = "10.1007/JHEP04(2020)105",
    journal = "JHEP",
    volume = "04",
    pages = "105",
    year = "2020"
}

@article{Kumar:2018jxz,
    author = "Kumar, Soubhik and Sundrum, Raman",
    title = "{Seeing Higher-Dimensional Grand Unification In Primordial Non-Gaussianities}",
    eprint = "1811.11200",
    archivePrefix = "arXiv",
    primaryClass = "hep-ph",
    reportNumber = "UMD-PP-018-09",
    doi = "10.1007/JHEP04(2019)120",
    journal = "JHEP",
    volume = "04",
    pages = "120",
    year = "2019"
}

@article{Goon:2018fyu,
    author = "Goon, Garrett and Hinterbichler, Kurt and Joyce, Austin and Trodden, Mark",
    title = "{Shapes of gravity: Tensor non-Gaussianity and massive spin-2 fields}",
    eprint = "1812.07571",
    archivePrefix = "arXiv",
    primaryClass = "hep-th",
    doi = "10.1007/JHEP10(2019)182",
    journal = "JHEP",
    volume = "10",
    pages = "182",
    year = "2019"
}

@article{Wu:2018lmx,
    author = "Wu, Yi-Peng",
    title = "{Higgs as heavy-lifted physics during inflation}",
    eprint = "1812.10654",
    archivePrefix = "arXiv",
    primaryClass = "hep-ph",
    reportNumber = "RESCEU-17/18",
    doi = "10.1007/JHEP04(2019)125",
    journal = "JHEP",
    volume = "04",
    pages = "125",
    year = "2019"
}

@article{Anninos:2019nib,
    author = "Anninos, D. and De Luca, V. and Franciolini, G. and Kehagias, A. and Riotto, A.",
    title = "{Cosmological Shapes of Higher-Spin Gravity}",
    eprint = "1902.01251",
    archivePrefix = "arXiv",
    primaryClass = "hep-th",
    doi = "10.1088/1475-7516/2019/04/045",
    journal = "JCAP",
    volume = "04",
    pages = "045",
    year = "2019"
}

@article{Li:2019ves,
    author = "Li, Lingfeng and Nakama, Tomohiro and Sou, Chon Man and Wang, Yi and Zhou, Siyi",
    title = "{Gravitational Production of Superheavy Dark Matter and Associated Cosmological Signatures}",
    eprint = "1903.08842",
    archivePrefix = "arXiv",
    primaryClass = "astro-ph.CO",
    doi = "10.1007/JHEP07(2019)067",
    journal = "JHEP",
    volume = "07",
    pages = "067",
    year = "2019"
}

@article{McAneny:2019epy,
    author = "McAneny, Michael and Ridgway, Alexander K.",
    title = "{New Shapes of Primordial Non-Gaussianity from Quasi-Single Field Inflation with Multiple Isocurvatons}",
    eprint = "1903.11607",
    archivePrefix = "arXiv",
    primaryClass = "astro-ph.CO",
    doi = "10.1103/PhysRevD.100.043534",
    journal = "Phys. Rev. D",
    volume = "100",
    number = "4",
    pages = "043534",
    year = "2019"
}

@article{Kim:2019wjo,
    author = "Kim, Suro and Noumi, Toshifumi and Takeuchi, Keito and Zhou, Siyi",
    title = "{Heavy Spinning Particles from Signs of Primordial Non-Gaussianities: Beyond the Positivity Bounds}",
    eprint = "1906.11840",
    archivePrefix = "arXiv",
    primaryClass = "hep-th",
    reportNumber = "KOBE-COSMO-19-09",
    doi = "10.1007/JHEP12(2019)107",
    journal = "JHEP",
    volume = "12",
    pages = "107",
    year = "2019"
}

@article{Lu:2019tjj,
    author = "Lu, Shiyun and Wang, Yi and Xianyu, Zhong-Zhi",
    title = "{A Cosmological Higgs Collider}",
    eprint = "1907.07390",
    archivePrefix = "arXiv",
    primaryClass = "hep-th",
    doi = "10.1007/JHEP02(2020)011",
    journal = "JHEP",
    volume = "02",
    pages = "011",
    year = "2020"
}

@article{Hook:2019zxa,
    author = "Hook, Anson and Huang, Junwu and Racco, Davide",
    title = "{Searches for other vacua. Part II. A new Higgstory at the cosmological collider}",
    eprint = "1907.10624",
    archivePrefix = "arXiv",
    primaryClass = "hep-ph",
    doi = "10.1007/JHEP01(2020)105",
    journal = "JHEP",
    volume = "01",
    pages = "105",
    year = "2020"
}

@article{Hook:2019vcn,
    author = "Hook, Anson and Huang, Junwu and Racco, Davide",
    title = "{Minimal signatures of the Standard Model in non-Gaussianities}",
    eprint = "1908.00019",
    archivePrefix = "arXiv",
    primaryClass = "hep-ph",
    doi = "10.1103/PhysRevD.101.023519",
    journal = "Phys. Rev. D",
    volume = "101",
    number = "2",
    pages = "023519",
    year = "2020"
}

@article{Kumar:2019ebj,
    author = "Kumar, Soubhik and Sundrum, Raman",
    title = "{Cosmological Collider Physics and the Curvaton}",
    eprint = "1908.11378",
    archivePrefix = "arXiv",
    primaryClass = "hep-ph",
    reportNumber = "UMD-PP-019-04",
    doi = "10.1007/JHEP04(2020)077",
    journal = "JHEP",
    volume = "04",
    pages = "077",
    year = "2020"
}

@article{Liu:2019fag,
    author = "Liu, Tao and Tong, Xi and Wang, Yi and Xianyu, Zhong-Zhi",
    title = "{Probing P and CP Violations on the Cosmological Collider}",
    eprint = "1909.01819",
    archivePrefix = "arXiv",
    primaryClass = "hep-ph",
    doi = "10.1007/JHEP04(2020)189",
    journal = "JHEP",
    volume = "04",
    pages = "189",
    year = "2020"
}

@article{Wang:2019gbi,
    author = "Wang, Lian-Tao and Xianyu, Zhong-Zhi",
    title = "{In Search of Large Signals at the Cosmological Collider}",
    eprint = "1910.12876",
    archivePrefix = "arXiv",
    primaryClass = "hep-ph",
    doi = "10.1007/JHEP02(2020)044",
    journal = "JHEP",
    volume = "02",
    pages = "044",
    year = "2020"
}

@article{Wang:2020uic,
    author = "Wang, Yi and Zhu, Yuhang",
    title = "{Cosmological Collider Signatures of Massive Vectors from Non-Gaussian Gravitational Waves}",
    eprint = "2001.03879",
    archivePrefix = "arXiv",
    primaryClass = "astro-ph.CO",
    doi = "10.1088/1475-7516/2020/04/049",
    journal = "JCAP",
    volume = "04",
    pages = "049",
    year = "2020"
}

@article{Li:2020xwr,
    author = "Li, Lingfeng and Lu, Shiyun and Wang, Yi and Zhou, Siyi",
    title = "{Cosmological Signatures of Superheavy Dark Matter}",
    eprint = "2002.01131",
    archivePrefix = "arXiv",
    primaryClass = "hep-ph",
    doi = "10.1007/JHEP07(2020)231",
    journal = "JHEP",
    volume = "07",
    pages = "231",
    year = "2020"
}

@article{DuasoPueyo:2023viy,
    author = "Duaso Pueyo, Carlos and Pajer, Enrico",
    title = "{A Cosmological Bootstrap for Resonant Non-Gaussianity}",
    eprint = "2311.01395",
    archivePrefix = "arXiv",
    primaryClass = "hep-th",
    month = "11",
    year = "2023"
}

@article{Wang:2020ioa,
    author = "Wang, Lian-Tao and Xianyu, Zhong-Zhi",
    title = "{Gauge Boson Signals at the Cosmological Collider}",
    eprint = "2004.02887",
    archivePrefix = "arXiv",
    primaryClass = "hep-ph",
    doi = "10.1007/JHEP11(2020)082",
    journal = "JHEP",
    volume = "11",
    pages = "082",
    year = "2020"
}

@article{Bodas:2020yho,
    author = "Bodas, Arushi and Kumar, Soubhik and Sundrum, Raman",
    title = "{The Scalar Chemical Potential in Cosmological Collider Physics}",
    eprint = "2010.04727",
    archivePrefix = "arXiv",
    primaryClass = "hep-ph",
    reportNumber = "UMD-PP-020-09",
    doi = "10.1007/JHEP02(2021)079",
    journal = "JHEP",
    volume = "02",
    pages = "079",
    year = "2021"
}

@article{Norena:2012yi,
    author = "Norena, Jorge and Verde, Licia and Barenboim, Gabriela and Bosch, Cristian",
    title = "{Prospects for constraining the shape of non-Gaussianity with the scale-dependent bias}",
    eprint = "1204.6324",
    archivePrefix = "arXiv",
    primaryClass = "astro-ph.CO",
    doi = "10.1088/1475-7516/2012/08/019",
    journal = "JCAP",
    volume = "08",
    pages = "019",
    year = "2012"
}

@article{Maru:2021ezc,
    author = "Maru, Nobuhito and Okawa, Akira",
    title = "{Non-Gaussianity from $X, Y$ gauge bosons in Cosmological Collider Physics}",
    eprint = "2101.10634",
    archivePrefix = "arXiv",
    primaryClass = "hep-ph",
    reportNumber = "OCU-PHYS 528, NITEP 86",
    month = "1",
    year = "2021"
}

@article{Lu:2021gso,
    author = "Lu, Shiyun",
    title = "{Axion isocurvature collider}",
    eprint = "2103.05958",
    archivePrefix = "arXiv",
    primaryClass = "hep-th",
    doi = "10.1007/JHEP04(2022)157",
    journal = "JHEP",
    volume = "04",
    pages = "157",
    year = "2022"
}

@article{Sou:2021juh,
    author = "Sou, Chon Man and Tong, Xi and Wang, Yi",
    title = "{Chemical-potential-assisted particle production in FRW spacetimes}",
    eprint = "2104.08772",
    archivePrefix = "arXiv",
    primaryClass = "hep-th",
    doi = "10.1007/JHEP06(2021)129",
    journal = "JHEP",
    volume = "06",
    pages = "129",
    year = "2021"
}

@article{Lu:2021wxu,
    author = "Lu, Qianshu and Reece, Matthew and Xianyu, Zhong-Zhi",
    title = "{Missing scalars at the cosmological collider}",
    eprint = "2108.11385",
    archivePrefix = "arXiv",
    primaryClass = "hep-ph",
    doi = "10.1007/JHEP12(2021)098",
    journal = "JHEP",
    volume = "12",
    pages = "098",
    year = "2021"
}

@article{Cui:2021iie,
    author = "Cui, Yanou and Xianyu, Zhong-Zhi",
    title = "{Probing Leptogenesis with the Cosmological Collider}",
    eprint = "2112.10793",
    archivePrefix = "arXiv",
    primaryClass = "hep-ph",
    doi = "10.1103/PhysRevLett.129.111301",
    journal = "Phys. Rev. Lett.",
    volume = "129",
    number = "11",
    pages = "111301",
    year = "2022"
}

@article{Tong:2022cdz,
    author = "Tong, Xi and Xianyu, Zhong-Zhi",
    title = "{Large spin-2 signals at the cosmological collider}",
    eprint = "2203.06349",
    archivePrefix = "arXiv",
    primaryClass = "hep-ph",
    doi = "10.1007/JHEP10(2022)194",
    journal = "JHEP",
    volume = "10",
    pages = "194",
    year = "2022"
}

@article{Chen:2022vzh,
    author = "Chen, Xingang and Ebadi, Reza and Kumar, Soubhik",
    title = "{Classical cosmological collider physics and primordial features}",
    eprint = "2205.01107",
    archivePrefix = "arXiv",
    primaryClass = "hep-ph",
    doi = "10.1088/1475-7516/2022/08/083",
    journal = "JCAP",
    volume = "08",
    pages = "083",
    year = "2022"
}

@article{Qin:2022lva,
    author = "Qin, Zhehan and Xianyu, Zhong-Zhi",
    title = "{Phase information in cosmological collider signals}",
    eprint = "2205.01692",
    archivePrefix = "arXiv",
    primaryClass = "hep-th",
    doi = "10.1007/JHEP10(2022)192",
    journal = "JHEP",
    volume = "10",
    pages = "192",
    year = "2022"
}

@article{Wang:2021qez,
    author = "Wang, Lian-Tao and Xianyu, Zhong-Zhi and Zhong, Yi-Ming",
    title = "{Precision calculation of inflation correlators at one loop}",
    eprint = "2109.14635",
    archivePrefix = "arXiv",
    primaryClass = "hep-ph",
    doi = "10.1007/JHEP02(2022)085",
    journal = "JHEP",
    volume = "02",
    pages = "085",
    year = "2022"
}

@article{Kim:2021pbr,
    author = "Kim, Suro and Noumi, Toshifumi and Takeuchi, Keito and Zhou, Siyi",
    title = "{Perturbative unitarity in quasi-single field inflation}",
    eprint = "2102.04101",
    archivePrefix = "arXiv",
    primaryClass = "hep-th",
    reportNumber = "KOBE-COSMO-21-01",
    doi = "10.1007/JHEP07(2021)018",
    journal = "JHEP",
    volume = "07",
    pages = "018",
    year = "2021"
}

@article{Chen:2006nt,
    author = "Chen, Xingang and Huang, Min-xin and Kachru, Shamit and Shiu, Gary",
    title = "{Observational signatures and non-Gaussianities of general single field inflation}",
    eprint = "hep-th/0605045",
    archivePrefix = "arXiv",
    reportNumber = "SLAC-PUB-11840, MAD-TH-06-3, UFIFT-HEP-06-9, SU-ITP-06-12, CU-TP-1147",
    doi = "10.1088/1475-7516/2007/01/002",
    journal = "JCAP",
    volume = "01",
    pages = "002",
    year = "2007"
}

@article{Niu:2022quw,
    author = "Niu, Xuce and Rahat, Moinul Hossain and Srinivasan, Karthik and Xue, Wei",
    title = "{Gravitational Wave Probes of Massive Gauge Bosons at the Cosmological Collider}",
    eprint = "2211.14331",
    archivePrefix = "arXiv",
    primaryClass = "hep-ph",
    month = "11",
    year = "2022"
}

@article{Chen:2016nrs,
    author = "Chen, Xingang and Wang, Yi and Xianyu, Zhong-Zhi",
    title = "{Loop Corrections to Standard Model Fields in Inflation}",
    eprint = "1604.07841",
    archivePrefix = "arXiv",
    primaryClass = "hep-th",
    doi = "10.1007/JHEP08(2016)051",
    journal = "JHEP",
    volume = "08",
    pages = "051",
    year = "2016"
}

@article{Tong:2018tqf,
    author = "Tong, Xi and Wang, Yi and Zhou, Siyi",
    title = "{Unsuppressed primordial standard clocks in warm quasi-single field inflation}",
    eprint = "1801.05688",
    archivePrefix = "arXiv",
    primaryClass = "hep-th",
    doi = "10.1088/1475-7516/2018/06/013",
    journal = "JCAP",
    volume = "06",
    pages = "013",
    year = "2018"
}

@article{Chen:2018sce,
    author = "Chen, Xingang and Chua, Wan Zhen and Guo, Yuxun and Wang, Yi and Xianyu, Zhong-Zhi and Xie, Tianyou",
    title = "{Quantum Standard Clocks in the Primordial Trispectrum}",
    eprint = "1803.04412",
    archivePrefix = "arXiv",
    primaryClass = "hep-th",
    doi = "10.1088/1475-7516/2018/05/049",
    journal = "JCAP",
    volume = "05",
    pages = "049",
    year = "2018"
}

@article{Chen:2018cgg,
    author = "Chen, Xingang and Loeb, Abraham and Xianyu, Zhong-Zhi",
    title = "{Unique Fingerprints of Alternatives to Inflation in the Primordial Power Spectrum}",
    eprint = "1809.02603",
    archivePrefix = "arXiv",
    primaryClass = "astro-ph.CO",
    doi = "10.1103/PhysRevLett.122.121301",
    journal = "Phys. Rev. Lett.",
    volume = "122",
    number = "12",
    pages = "121301",
    year = "2019"
}

@article{Alexander:2019vtb,
    author = "Alexander, Stephon and Gates, S. James and Jenks, Leah and Koutrolikos, K. and McDonough, Evan",
    title = "{Higher Spin Supersymmetry at the Cosmological Collider: Sculpting SUSY Rilles in the CMB}",
    eprint = "1907.05829",
    archivePrefix = "arXiv",
    primaryClass = "hep-th",
    doi = "10.1007/JHEP10(2019)156",
    journal = "JHEP",
    volume = "10",
    pages = "156",
    year = "2019"
}

@article{Wang:2019gok,
    author = "Wang, Dong-Gang",
    title = "{On the inflationary massive field with a curved field manifold}",
    eprint = "1911.04459",
    archivePrefix = "arXiv",
    primaryClass = "astro-ph.CO",
    doi = "10.1088/1475-7516/2020/01/046",
    journal = "JCAP",
    volume = "01",
    pages = "046",
    year = "2020"
}

@article{Fan:2020xgh,
    author = "Fan, JiJi and Xianyu, Zhong-Zhi",
    title = "{A Cosmic Microscope for the Preheating Era}",
    eprint = "2005.12278",
    archivePrefix = "arXiv",
    primaryClass = "hep-ph",
    doi = "10.1007/JHEP01(2021)021",
    journal = "JHEP",
    volume = "01",
    pages = "021",
    year = "2021"
}

@article{Cabass:2022rhr,
    author = "Cabass, Giovanni and Jazayeri, Sadra and Pajer, Enrico and Stefanyszyn, David",
    title = "{Parity violation in the scalar trispectrum: no-go theorems and yes-go examples}",
    eprint = "2210.02907",
    archivePrefix = "arXiv",
    primaryClass = "hep-th",
    month = "10",
    year = "2022"
}

@article{Cabass:2022oap,
    author = "Cabass, Giovanni and Ivanov, Mikhail M. and Philcox, Oliver H. E.",
    title = "{Colliding Ghosts: Constraining Inflation with the Parity-Odd Galaxy Four-Point Function}",
    eprint = "2210.16320",
    archivePrefix = "arXiv",
    primaryClass = "astro-ph.CO",
    month = "10",
    year = "2022"
}

@article{Aoki:2023tjm,
    author = "Aoki, Shuntaro",
    title = "{Continuous spectrum on cosmological collider}",
    eprint = "2301.07920",
    archivePrefix = "arXiv",
    primaryClass = "hep-th",
    doi = "10.1088/1475-7516/2023/04/002",
    journal = "JCAP",
    volume = "04",
    pages = "002",
    year = "2023"
}

@article{Tong:2023krn,
    author = "Tong, Xi and Wang, Yi and Zhang, Chen and Zhu, Yuhang",
    title = "{BCS in the Sky: Signatures of Inflationary Fermion Condensation}",
    eprint = "2304.09428",
    archivePrefix = "arXiv",
    primaryClass = "hep-th",
    month = "4",
    year = "2023"
}

@article{Xianyu:2022jwk,
    author = "Xianyu, Zhong-Zhi and Zhang, Hongyu",
    title = "{Bootstrapping one-loop inflation correlators with the spectral decomposition}",
    eprint = "2211.03810",
    archivePrefix = "arXiv",
    primaryClass = "hep-th",
    doi = "10.1007/JHEP04(2023)103",
    journal = "JHEP",
    volume = "04",
    pages = "103",
    year = "2023"
}

@article{Jazayeri:2023xcj,
    author = "Jazayeri, Sadra and Renaux-Petel, S\'ebastien and Werth, Denis",
    title = "{Shapes of the Cosmological Low-Speed Collider}",
    eprint = "2307.01751",
    archivePrefix = "arXiv",
    primaryClass = "hep-th",
    month = "7",
    year = "2023"
}

@article{Niu:2022fki,
    author = "Niu, Xuce and Rahat, Moinul Hossain and Srinivasan, Karthik and Xue, Wei",
    title = "{Parity-Odd and Even Trispectrum from Axion Inflation}",
    eprint = "2211.14324",
    archivePrefix = "arXiv",
    primaryClass = "hep-ph",
    month = "11",
    year = "2022"
}

@article{Baumann:2019oyu,
    author = "Baumann, Daniel and Duaso Pueyo, Carlos and Joyce, Austin and Lee, Hayden and Pimentel, Guilherme L.",
    title = "{The cosmological bootstrap: weight-shifting operators and scalar seeds}",
    eprint = "1910.14051",
    archivePrefix = "arXiv",
    primaryClass = "hep-th",
    doi = "10.1007/JHEP12(2020)204",
    journal = "JHEP",
    volume = "12",
    pages = "204",
    year = "2020"
}

@article{Baumann:2020dch,
    author = "Baumann, Daniel and Duaso Pueyo, Carlos and Joyce, Austin and Lee, Hayden and Pimentel, Guilherme L.",
    title = "{The Cosmological Bootstrap: Spinning Correlators from Symmetries and Factorization}",
    eprint = "2005.04234",
    archivePrefix = "arXiv",
    primaryClass = "hep-th",
    doi = "10.21468/SciPostPhys.11.3.071",
    journal = "SciPost Phys.",
    volume = "11",
    pages = "071",
    year = "2021"
}

@article{Pajer:2020wnj,
    author = "Pajer, Enrico and Stefanyszyn, David and Supe\l{}, Jakub",
    title = "{The Boostless Bootstrap: Amplitudes without Lorentz boosts}",
    eprint = "2007.00027",
    archivePrefix = "arXiv",
    primaryClass = "hep-th",
    doi = "10.1007/JHEP12(2020)198",
    journal = "JHEP",
    volume = "12",
    pages = "198",
    year = "2020",
    note = "[Erratum: JHEP 04, 023 (2022)]"
}

@article{Pajer:2020wxk,
    author = "Pajer, Enrico",
    title = "{Building a Boostless Bootstrap for the Bispectrum}",
    eprint = "2010.12818",
    archivePrefix = "arXiv",
    primaryClass = "hep-th",
    doi = "10.1088/1475-7516/2021/01/023",
    journal = "JCAP",
    volume = "01",
    pages = "023",
    year = "2021"
}

@article{Cabass:2021fnw,
    author = "Cabass, Giovanni and Pajer, Enrico and Stefanyszyn, David and Supe\l{}, Jakub",
    title = "{Bootstrapping large graviton non-Gaussianities}",
    eprint = "2109.10189",
    archivePrefix = "arXiv",
    primaryClass = "hep-th",
    doi = "10.1007/JHEP05(2022)077",
    journal = "JHEP",
    volume = "05",
    pages = "077",
    year = "2022"
}

@article{Pimentel:2022fsc,
    author = "Pimentel, Guilherme L. and Wang, Dong-Gang",
    title = "{Boostless cosmological collider bootstrap}",
    eprint = "2205.00013",
    archivePrefix = "arXiv",
    primaryClass = "hep-th",
    doi = "10.1007/JHEP10(2022)177",
    journal = "JHEP",
    volume = "10",
    pages = "177",
    year = "2022"
}

@article{Jazayeri:2022kjy,
    author = "Jazayeri, Sadra and Renaux-Petel, S\'ebastien",
    title = "{Cosmological bootstrap in slow motion}",
    eprint = "2205.10340",
    archivePrefix = "arXiv",
    primaryClass = "hep-th",
    doi = "10.1007/JHEP12(2022)137",
    journal = "JHEP",
    volume = "12",
    pages = "137",
    year = "2022"
}

@article{Wang:2022eop,
    author = "Wang, Dong-Gang and Pimentel, Guilherme L. and Ach\'ucarro, Ana",
    title = "{Bootstrapping Multi-Field Inflation: non-Gaussianities from light scalars revisited}",
    eprint = "2212.14035",
    archivePrefix = "arXiv",
    primaryClass = "astro-ph.CO",
    month = "12",
    year = "2022"
}

@inproceedings{Baumann:2022jpr,
    author = "Baumann, Daniel and Green, Daniel and Joyce, Austin and Pajer, Enrico and Pimentel, Guilherme L. and Sleight, Charlotte and Taronna, Massimo",
    title = "{Snowmass White Paper: The Cosmological Bootstrap}",
    booktitle = "{2022 Snowmass Summer Study}",
    eprint = "2203.08121",
    archivePrefix = "arXiv",
    primaryClass = "hep-th",
    month = "3",
    year = "2022"
}

@article{Sleight:2019mgd,
    author = "Sleight, Charlotte",
    title = "{A Mellin Space Approach to Cosmological Correlators}",
    eprint = "1906.12302",
    archivePrefix = "arXiv",
    primaryClass = "hep-th",
    doi = "10.1007/JHEP01(2020)090",
    journal = "JHEP",
    volume = "01",
    pages = "090",
    year = "2020"
}

@article{Sleight:2019hfp,
    author = "Sleight, Charlotte and Taronna, Massimo",
    title = "{Bootstrapping Inflationary Correlators in Mellin Space}",
    eprint = "1907.01143",
    archivePrefix = "arXiv",
    primaryClass = "hep-th",
    reportNumber = "PUPT-2590",
    doi = "10.1007/JHEP02(2020)098",
    journal = "JHEP",
    volume = "02",
    pages = "098",
    year = "2020"
}

@article{Sleight:2020obc,
    author = "Sleight, Charlotte and Taronna, Massimo",
    title = "{From AdS to dS exchanges: Spectral representation, Mellin amplitudes, and crossing}",
    eprint = "2007.09993",
    archivePrefix = "arXiv",
    primaryClass = "hep-th",
    doi = "10.1103/PhysRevD.104.L081902",
    journal = "Phys. Rev. D",
    volume = "104",
    number = "8",
    pages = "L081902",
    year = "2021"
}

@article{Sleight:2021iix,
    author = "Sleight, Charlotte and Taronna, Massimo",
    title = "{On the consistency of (partially-)massless matter couplings in de Sitter space}",
    eprint = "2106.00366",
    archivePrefix = "arXiv",
    primaryClass = "hep-th",
    doi = "10.1007/JHEP10(2021)156",
    journal = "JHEP",
    volume = "10",
    pages = "156",
    year = "2021"
}

@article{Sleight:2021plv,
    author = "Sleight, Charlotte and Taronna, Massimo",
    title = "{From dS to AdS and back}",
    eprint = "2109.02725",
    archivePrefix = "arXiv",
    primaryClass = "hep-th",
    doi = "10.1007/JHEP12(2021)074",
    journal = "JHEP",
    volume = "12",
    pages = "074",
    year = "2021"
}

@article{Goodhew:2020hob,
    author = "Goodhew, Harry and Jazayeri, Sadra and Pajer, Enrico",
    title = "{The Cosmological Optical Theorem}",
    eprint = "2009.02898",
    archivePrefix = "arXiv",
    primaryClass = "hep-th",
    doi = "10.1088/1475-7516/2021/04/021",
    journal = "JCAP",
    volume = "04",
    pages = "021",
    year = "2021"
}

@article{Melville:2021lst,
    author = "Melville, Scott and Pajer, Enrico",
    title = "{Cosmological Cutting Rules}",
    eprint = "2103.09832",
    archivePrefix = "arXiv",
    primaryClass = "hep-th",
    doi = "10.1007/JHEP05(2021)249",
    journal = "JHEP",
    volume = "05",
    pages = "249",
    year = "2021"
}

@article{Goodhew:2021oqg,
    author = "Goodhew, Harry and Jazayeri, Sadra and Gordon Lee, Mang Hei and Pajer, Enrico",
    title = "{Cutting cosmological correlators}",
    eprint = "2104.06587",
    archivePrefix = "arXiv",
    primaryClass = "hep-th",
    doi = "10.1088/1475-7516/2021/08/003",
    journal = "JCAP",
    volume = "08",
    pages = "003",
    year = "2021"
}

@article{DiPietro:2021sjt,
    author = "Di Pietro, Lorenzo and Gorbenko, Victor and Komatsu, Shota",
    title = "{Analyticity and unitarity for cosmological correlators}",
    eprint = "2108.01695",
    archivePrefix = "arXiv",
    primaryClass = "hep-th",
    reportNumber = "CERN-TH-2021-118",
    doi = "10.1007/JHEP03(2022)023",
    journal = "JHEP",
    volume = "03",
    pages = "023",
    year = "2022"
}

@article{Tong:2021wai,
    author = "Tong, Xi and Wang, Yi and Zhu, Yuhang",
    title = "{Cutting rule for cosmological collider signals: a bulk evolution perspective}",
    eprint = "2112.03448",
    archivePrefix = "arXiv",
    primaryClass = "hep-th",
    doi = "10.1007/JHEP03(2022)181",
    journal = "JHEP",
    volume = "03",
    pages = "181",
    year = "2022"
}

@article{Bonifacio:2021azc,
    author = "Bonifacio, James and Pajer, Enrico and Wang, Dong-Gang",
    title = "{From amplitudes to contact cosmological correlators}",
    eprint = "2106.15468",
    archivePrefix = "arXiv",
    primaryClass = "hep-th",
    doi = "10.1007/JHEP10(2021)001",
    journal = "JHEP",
    volume = "10",
    pages = "001",
    year = "2021"
}

@article{Hogervorst:2021uvp,
    author = "Hogervorst, Matthijs and Penedones, Jo\~ao and Vaziri, Kamran Salehi",
    title = "{Towards the non-perturbative cosmological bootstrap}",
    eprint = "2107.13871",
    archivePrefix = "arXiv",
    primaryClass = "hep-th",
    month = "7",
    year = "2021"
}

@article{Meltzer:2021zin,
    author = "Meltzer, David",
    title = "{The inflationary wavefunction from analyticity and factorization}",
    eprint = "2107.10266",
    archivePrefix = "arXiv",
    primaryClass = "hep-th",
    reportNumber = "CALT-TH-2021-028",
    doi = "10.1088/1475-7516/2021/12/018",
    journal = "JCAP",
    volume = "12",
    number = "12",
    pages = "018",
    year = "2021"
}

@article{Heckelbacher:2022hbq,
    author = "Heckelbacher, Till and Sachs, Ivo and Skvortsov, Evgeny and Vanhove, Pierre",
    title = "{Analytical evaluation of cosmological correlation functions}",
    eprint = "2204.07217",
    archivePrefix = "arXiv",
    primaryClass = "hep-th",
    reportNumber = "IPhT-t22/02, LMU-ASC 13/22",
    doi = "10.1007/JHEP08(2022)139",
    journal = "JHEP",
    volume = "08",
    pages = "139",
    year = "2022"
}

@article{Gomez:2021qfd,
    author = "Gomez, Humberto and Jusinskas, Renann Lipinski and Lipstein, Arthur",
    title = "{Cosmological Scattering Equations}",
    eprint = "2106.11903",
    archivePrefix = "arXiv",
    primaryClass = "hep-th",
    doi = "10.1103/PhysRevLett.127.251604",
    journal = "Phys. Rev. Lett.",
    volume = "127",
    number = "25",
    pages = "251604",
    year = "2021"
}

@article{Gomez:2021ujt,
    author = "Gomez, Humberto and Lipinski Jusinskas, Renann and Lipstein, Arthur",
    title = "{Cosmological scattering equations at tree-level and one-loop}",
    eprint = "2112.12695",
    archivePrefix = "arXiv",
    primaryClass = "hep-th",
    doi = "10.1007/JHEP07(2022)004",
    journal = "JHEP",
    volume = "07",
    pages = "004",
    year = "2022"
}

@article{Baumann:2021fxj,
    author = "Baumann, Daniel and Chen, Wei-Ming and Duaso Pueyo, Carlos and Joyce, Austin and Lee, Hayden and Pimentel, Guilherme L.",
    title = "{Linking the singularities of cosmological correlators}",
    eprint = "2106.05294",
    archivePrefix = "arXiv",
    primaryClass = "hep-th",
    doi = "10.1007/JHEP09(2022)010",
    journal = "JHEP",
    volume = "09",
    pages = "010",
    year = "2022"
}

@article{Jazayeri:2021fvk,
    author = "Jazayeri, Sadra and Pajer, Enrico and Stefanyszyn, David",
    title = "{From locality and unitarity to cosmological correlators}",
    eprint = "2103.08649",
    archivePrefix = "arXiv",
    primaryClass = "hep-th",
    doi = "10.1007/JHEP10(2021)065",
    journal = "JHEP",
    volume = "10",
    pages = "065",
    year = "2021"
}

@article{Albayrak:2018tam,
    author = "Albayrak, Soner and Kharel, Savan",
    title = "{Towards the higher point holographic momentum space amplitudes}",
    eprint = "1810.12459",
    archivePrefix = "arXiv",
    primaryClass = "hep-th",
    doi = "10.1007/JHEP02(2019)040",
    journal = "JHEP",
    volume = "02",
    pages = "040",
    year = "2019"
}

@article{Albayrak:2019asr,
    author = "Albayrak, Soner and Chowdhury, Chandramouli and Kharel, Savan",
    title = "{New relation for Witten diagrams}",
    eprint = "1904.10043",
    archivePrefix = "arXiv",
    primaryClass = "hep-th",
    doi = "10.1007/JHEP10(2019)274",
    journal = "JHEP",
    volume = "10",
    pages = "274",
    year = "2019"
}

@article{Albayrak:2019yve,
    author = "Albayrak, Soner and Kharel, Savan",
    title = "{Towards the higher point holographic momentum space amplitudes. Part II. Gravitons}",
    eprint = "1908.01835",
    archivePrefix = "arXiv",
    primaryClass = "hep-th",
    doi = "10.1007/JHEP12(2019)135",
    journal = "JHEP",
    volume = "12",
    pages = "135",
    year = "2019"
}

@article{Albayrak:2020bso,
    author = "Albayrak, Soner and Kharel, Savan",
    title = "{Spinning loop amplitudes in anti\textendash{}de Sitter space}",
    eprint = "2006.12540",
    archivePrefix = "arXiv",
    primaryClass = "hep-th",
    doi = "10.1103/PhysRevD.103.026004",
    journal = "Phys. Rev. D",
    volume = "103",
    number = "2",
    pages = "026004",
    year = "2021"
}

@article{Albayrak:2020fyp,
    author = "Albayrak, Soner and Kharel, Savan and Meltzer, David",
    title = "{On duality of color and kinematics in (A)dS momentum space}",
    eprint = "2012.10460",
    archivePrefix = "arXiv",
    primaryClass = "hep-th",
    doi = "10.1007/JHEP03(2021)249",
    journal = "JHEP",
    volume = "03",
    pages = "249",
    year = "2021"
}

@article{Qin:2023nhv,
    author = "Qin, Zhehan and Xianyu, Zhong-Zhi",
    title = "{Nonanalyticity and On-Shell Factorization of Inflation Correlators at All Loop Orders}",
    eprint = "2308.14802",
    archivePrefix = "arXiv",
    primaryClass = "hep-th",
    month = "8",
    year = "2023"
}

@article{Qin:2023bjk,
    author = "Qin, Zhehan and Xianyu, Zhong-Zhi",
    title = "{Inflation Correlators at the One-Loop Order: Nonanalyticity, Factorization, Cutting Rule, and OPE}",
    eprint = "2304.13295",
    archivePrefix = "arXiv",
    primaryClass = "hep-th",
    month = "4",
    year = "2023"
}

@article{Chen:2023txq,
    author = "Chen, Xingang and Fan, JiJi and Li, Lingfeng",
    title = "{New inflationary probes of axion dark matter}",
    eprint = "2303.03406",
    archivePrefix = "arXiv",
    primaryClass = "hep-ph",
    month = "3",
    year = "2023"
}

@article{Albayrak:2023jzl,
    author = "Albayrak, Soner and Kharel, Savan",
    title = "{All plus four point (A)dS graviton function using generalized on-shell recursion relation}",
    eprint = "2302.09089",
    archivePrefix = "arXiv",
    primaryClass = "hep-th",
    doi = "10.1007/JHEP05(2023)151",
    journal = "JHEP",
    volume = "05",
    pages = "151",
    year = "2023"
}

@article{Yin:2023jlv,
    author = "Yin, Yuan",
    title = "{The cosmological collider signal in the non-BD initial states}",
    eprint = "2309.05244",
    archivePrefix = "arXiv",
    primaryClass = "hep-ph",
    month = "9",
    year = "2023"
}

@article{Stefanyszyn:2023qov,
    author = "Stefanyszyn, David and Tong, Xi and Zhu, Yuhang",
    title = "{Cosmological Correlators Through the Looking Glass: Reality, Parity, and Factorisation}",
    eprint = "2309.07769",
    archivePrefix = "arXiv",
    primaryClass = "hep-th",
    month = "9",
    year = "2023"
}

@article{Xianyu:2023ytd,
    author = "Xianyu, Zhong-Zhi and Zang, Jiaju",
    title = "{Inflation Correlators with Multiple Massive Exchanges}",
    eprint = "2309.10849",
    archivePrefix = "arXiv",
    primaryClass = "hep-th",
    month = "9",
    year = "2023"
}

@article{Chakraborty:2023qbp,
    author = "Chakraborty, Priyesh and Stout, John",
    title = "{Light Scalars at the Cosmological Collider}",
    eprint = "2310.01494",
    archivePrefix = "arXiv",
    primaryClass = "hep-th",
    month = "10",
    year = "2023"
}

@article{De:2023xue,
    author = "De, Shounak and Pokraka, Andrzej",
    title = "{Cosmology meets cohomology}",
    eprint = "2308.03753",
    archivePrefix = "arXiv",
    primaryClass = "hep-th",
    month = "8",
    year = "2023"
}
